\documentclass[aps,prl,twocolumn,superscriptaddress,showpacs,longbibliography]{revtex4-1}
\usepackage{amsmath,amssymb,graphicx,subfigure}

\usepackage[utf8]{inputenc}
\usepackage[T1]{fontenc}
\usepackage{xcolor}

\IfFileExists{newtxtext.sty}
   {\usepackage{newtxtext,newtxmath}}
   {\IfFileExists{stix.sty}
      {\usepackage{stix}}
      {\IfFileExists{mathptmx.sty}
      {\usepackage{mathptmx}}{} } }

\usepackage{textcomp}

\usepackage{bm}

\IfFileExists{siunitx.sty}{\usepackage{booktabs,siunitx}}{}

\pdfoutput=1
\usepackage{color}
\definecolor{LinkColor}{rgb}{0.256,0.439,0.588}
\usepackage{hyperref}
\hypersetup{
   colorlinks=true,
   citecolor=LinkColor,
   linkcolor=LinkColor,
   urlcolor=LinkColor
}

\newcommand{\be}{\begin{equation}}
\newcommand{\ee}{\end{equation}}
\newcommand{\bea}{\begin{eqnarray}}
\newcommand{\eea}{\end{eqnarray}}

\usepackage{pifont}

\begin{document}

\title{Intertwined van-Hove Singularities as a Mechanism for Loop Current Order in Kagome Metals  }

\author{Heqiu Li}
	\affiliation{Department of Physics, University of Toronto, Toronto, Ontario M5S 1A7, Canada}

\author{Yong Baek Kim}
\email{ybkim@physics.utoronto.ca}
	\affiliation{Department of Physics, University of Toronto, Toronto, Ontario M5S 1A7, Canada}
	\affiliation{School of Physics, Korea Institute for Advanced Study, Seoul 02455, Korea}

\author{Hae-Young Kee}
\email{hykee@physics.utoronto.ca}
	\affiliation{Department of Physics, University of Toronto, Toronto, Ontario M5S 1A7, Canada}
	\affiliation{Canadian Institute for Advanced Research, CIFAR Program in Quantum Materials, Toronto, Ontario M5G 1M1, Canada}

\begin{abstract}

Recent experiments on kagome metals AV$_3$Sb$_5$ (A=Cs,Rb,K) indicated spontaneous time-reversal symmetry breaking in the charge density wave state in the absence of static magnetization. The loop current order (LCO) is proposed as its cause, but a microscopic model explaining the emergence of LCO through electronic correlations has not been firmly established. We show that the coupling between van-Hove singularities (vHS) with distinct mirror symmetries is a key ingredient to generate LCO ground state. By constructing an effective model, we find that when multiple vHS with opposite mirror eigenvalues are close in energy, the nearest-neighbor electron repulsion favors a ground state with coexisting LCO and charge bond order. It is then demonstrated that this mechanism applies to the kagome metals AV$_3$Sb$_5$. Our findings provide an intriguing mechanism of LCO and pave the way for a deeper understanding of complex quantum phenomena in kagome systems. 


\end{abstract}

\date{\today}

\maketitle


\emph{Introduction}--- The vanadium-based kagome metals AV$_3$Sb$_5$ (A=Cs,Rb,K) have generated considerable interest due to the discovery of exotic phases in this family of materials~\cite{Jiang2022s,Shuo2020,Neupert2022o,Wang2020s,Hu2022n,Oey2022,Christensen2022c,Zhu2022d,Stahl2022r,Wu2022d,Kang2022v,Jiang2021,Li2021u,Wu2021u,Zhao2021,Li2022s,Romer2022,Mertz2022,Ptok2022,Ferrari2022,Uykur2022,Grandi2023,Xu2022n,Saykin2022,Hu2022s,Kang2022,Wang2023dk,wilson2023k}. Superconductivity in these materials emerges at $T_c\sim0.9-2.8K$~\cite{Ortiz2020k,Ortiz2021,Chen2021,Chen2021b}, with magnetoresistance measurements indicating the possibility of novel superconductivity with charge 4e and 6e flux quantization~\cite{Ge2022c}. Additionally, a $2\times 2$ charge density wave (CDW) is detected below $T_{CDW}\sim80-100K$~\cite{Ortiz2019,Ortiz2020k,Shumiya2021,Ortiz2021sm,Si2022,Song2022} with scanning tunneling microscopy, emphasizing the important role of vHS at $M$ point of the Brillouin zone. Intriguingly, these materials exhibit spontaneous time-reversal symmetry breaking (TRSB) after the CDW 
transition, evidenced through techniques such as muon spin relaxation and scanning tunneling microscope ~\cite{Shumiya2021,Jiang2021,Mielke2022} in the CDW phase without evidence of static magnetic order~\cite{Ortiz2019,Kenney_2021,Ortiz2021}. These observations indicate an unconventional CDW order in AV$_3$Sb$_5$.

The observation of TRSB without static magnetic order leads to the hypothesis of loop current order (LCO)~\cite{Jiang2021,Lin2019lco,Yang2022c,Dong2022}, but the mechanism to generate LCO remains unclear. Enormous experimental and theoretical efforts are devoted to determine the properties of CDW in this kagome system~\cite{Nie2022,Li2021u,Labollita2021,Stahl2022r,Tan2021d,Liang2021c,Yu2021cs,Jiang2022e,Ritz2022,Ortiz2021sm,Hu2022t,Yang2022c,Luo2022s,Mu_2022,Luo2022n,Kato2022,Cho2021n,Denner2021,Park2021,Lin2021c,Christensen2021,Jeong2022,Rina2022,Zhou2022f,Wu2022sub,Ferrari2022,Dong2022,Li2022pres,Tsirlin2022s,Ritz2022,Frass2022,Li2023kagome,Le2023c,Wenzel2023p,Scammell2023-wl,Tsirlin2023}. The simplest way to model the system is through a three-band model obtained by assigning a single orbital to each site. When the chemical potential is close to the vHS, incorporating nearest neighbor (NN) electron interactions and electron-phonon coupling leads to a charge bond order (CBO) ground state rather than LCO~\cite{Ferrari2022}. Ref.\onlinecite{Dong2022} shows that LCO can be induced by electron interaction, but this necessitates a substantial next-nearest-neighbor (NNN) interaction, a condition not aligned with realistic scenarios. This poses a critical question: what are the conditions for the emergence of LCO in generic kaogme materials?

Noticing the presence of multiple vHS in AV$_3$Sb$_5$, in this paper we demonstrate that when two vHS with \emph{different mirror symmetry eigenvalues are close to the Fermi level, a simple NN interaction can generate LCO when the coupling between different vHS is taken into account.} This ground state has LCO coexisting with CBO, dubbed loop current charge bond order (LCBO). We apply this analysis to AV$_3$Sb$_5$ by considering a tight binding model with multiple vHS. We find that the ground state of AV$_3$Sb$_5$ is LCBO under the conditions described below. This study unveils a mechanism for generating LCO in kagome systems with multiple vHS.

\begin{figure}
\centering
\includegraphics[width=3.4 in]{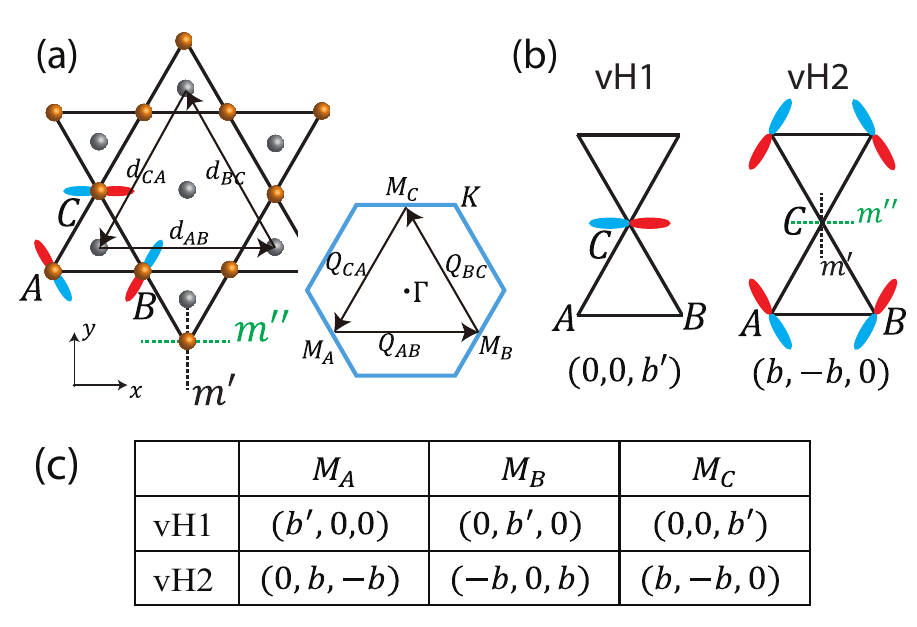}
\caption{(a): Kagome plane of AV$_3$Sb$_5$ (A=Cs, Rb, K). The red (blue) parts denote regions of orbitals with positive (negative) amplitude. The inset shows the Brillouin zone. (b): Real space wave function of vH1 and vH2 at $M_C$ allowed by mirror symmetries. (c): Weight of wave function in (A,B,C) sublattices for vH1 and vH2 at three distinct $M$ points imposed by mirror symmetries, where $b$ and $b’$ are constants.  }
\label{fig_csorder}
\end{figure}

\emph{Conditions imposed by mirror symmetries}---  We first show that mirror symmetries impose important constraints on the wave functions at vHS, which are key ingredients for the emergence of LCBO. Each vHS at momentum $M$ has little group $D_{2h}$ with mutually perpendicular mirror planes $m_z,m',m''$, where $m_z$ coincides with kagome plane, $m'$ and $m''$ are shown in Fig.\ref{fig_csorder}(a).
Consider two vHS near the Fermi level denoted by vH1 and vH2 at the three distinct momenta $M$ denoted by $M_A,M_B,M_C$ as in Fig.\ref{fig_csorder}(a). We show that mirror symmetries will constrain the wave function of vH1 and vH2 at three distinct $M$ points to take the form of Fig.\ref{fig_csorder}(c) as long as the following conditions are satisfied: (1) The wave functions of vH1 and vH2 have opposite eigenvalues under $m'$ and same eigenvalues under $m''$. (2) vH1 and vH2 consist of the same type of orbital at the kagome sites.


{To demonstrate this conclusion explicitly, we consider two vHS made of the colored orbitals in Fig.\ref{fig_csorder}(a), and let the wave function at vH1 (vH2) be odd (even) under $m'$.} Let us inspect the form of symmetry-allowed wave function at momentum $M_c$ as shown in Fig.\ref{fig_csorder}(b). In this case, $m'$ coincides with $m_x$ which maps sublattice A and B to each other and maps sublattice C to itself. Because the wave function of vH2 is even under $m'$ and the orbital at sublattice C is odd under $m'$, the weight of wave function must vanish at $m'$-invariant sublattice C. Furthermore,  wave function components of vH2 at sublattice A and B must have opposite signs to make the wave function even under $m'$. Therefore, the wave function of vH2 at momentum $M_C$ must take the form $(b,-b,0)$ at A,B,C sublattices respectively, where $b$ is a constant. A similar analysis can be applied to vH1, which gives the form of $(0,0,b')$ instead, where $b'$ is another constant. Finally, the threefold rotation symmetry leads to the wave function structure at the three distinct $M$ points given in Fig.\ref{fig_csorder}(c). Similarly, this wave function structure can also be obtained for orbitals with other mirror eigenvalues, as shown in Sec.I of the supplementary material (SM)~\cite{supp}.

\begin{figure}
\centering
\includegraphics[width=3.5 in]{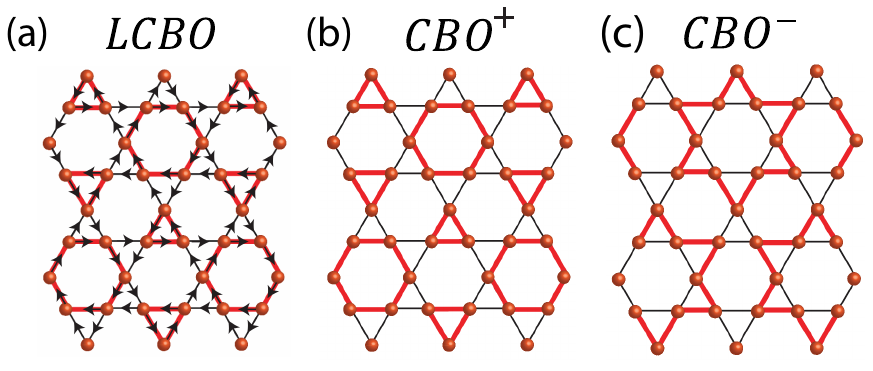}
\caption{ (a): Coexisting loop current order and charge bond order (LCBO). The red bonds represent modulations of $\langle c^\dagger_{\mathbf r}c_{\mathbf r'}\rangle$ at NN bonds and the arrows represent current $I\sim \langle ic^\dagger_{\mathbf r}c_{\mathbf r'}-ic^\dagger_{\mathbf r'}c_{\mathbf r}\rangle$. (b): Charge bond order with $\Delta>0$.  (c): Charge bond order with $\Delta<0$. }
\label{fig_LCBO}
\end{figure}

\emph{Effective model for coupled vHS}---  We construct an effective model that describes the coupling between different vHS. The order parameter for a complex CDW with $2\times 2$ periodicity is written as:
\be
\Delta_{\alpha\beta}=\frac{V}{2N_c}\sum_{\mathbf R}\left(\langle c^\dagger_{\mathbf R,\alpha}c_{\mathbf R,\beta} \rangle -\langle c^\dagger_{\mathbf R,\alpha}c_{\mathbf R-\mathbf d_{\alpha\beta},\beta} \rangle \right)\cos(\mathbf Q_{\alpha\beta} \cdot \mathbf R),
\label{Deldef}
\ee
Here $\mathbf R$ labels unit cells, $V$ is the NN interaction strength, $N_c$ is the number of unit cells, $\alpha,\beta=A,B,C$ denote the kagome sublattices and $\mathbf Q_{\alpha\beta}$ connects different momenta $M$ as in Fig.\ref{fig_csorder}(a). In phases that preserve threefold rotation symmetry, the order parameters satisfy $\Delta_{AB}=\Delta_{BC}=\Delta_{CA}\equiv \Delta$. The real part of $\Delta$ represents CBO, the imaginary part represents LCO and a complex value of $\Delta$ represents the coexisting phase of LCO and CBO, denoted as LCBO in Fig.\ref{fig_LCBO}(a). The phase with real $\Delta>0$ ($\Delta<0$) is denoted as CBO$^+$ (CBO$^-$) as shown in Fig.\ref{fig_LCBO}(b,c). There are other possible LCO phases distinct from the ones in Fig.\ref{fig_LCBO}, but those phases have higher free energy as shown in Sec.IV of SM, hence we mainly focus on the phases in Fig.\ref{fig_LCBO}.

{We can construct an effective model on patches near the three $M$ points. This model can effectively capture the coupling between different vHS, which is the dominant contribution leading to the LCBO ground state.} The coupling between vHS at different $M$ points is proportional to the order parameter with coupling strength determined by the wave function components at vHS. We choose the basis $u_1(M_A),u_1(M_B),u_1(M_C),u_2(M_A),u_2(M_B),u_2(M_C)$ where $u_1,u_2$ denotes the wave function for vH1 and vH2 respectively. Let $\mathbf k$ denote the small deviation from $M$ with $|\mathbf k|<k_{cut}$. Given the form of wave functions in Fig.\ref{fig_csorder}(c) and the order parameter in Eq.\eqref{Deldef}, the effective Hamiltonian with leading terms in $\mathbf k$ is derived in Sec.II of SM to be:
\bea
&&H_{\rm{eff}}(\mathbf k,\Delta)=\begin{pmatrix}
\epsilon_1 &  s_1 \Delta & s_1 \Delta^* &  \lambda^* k_1 &0 &0\\
s_1 \Delta^* & \epsilon_1  & s_1 \Delta &0 &  \lambda^* k_2 & 0\\
s_1 \Delta & s_1 \Delta^* & \epsilon_1 &0 &0 &  \lambda^* k_3 \\
 \lambda k_1 &0 & 0 & \epsilon_2 & s_2 \Delta^* & s_2 \Delta\\ 
0&  \lambda k_2  & 0  & s_2 \Delta & \epsilon_2 & s_2 \Delta^*\\
0  & 0 &  \lambda k_3 & s_2 \Delta^* & s_2 \Delta & \epsilon_2
\end{pmatrix}, \nonumber\\
&&\ \ \ \ \ \ \ \ \ \ \ \ \ \ \ \ \   \equiv\begin{pmatrix}
P_1 & Q^\dagger \\
Q & P_2  
\end{pmatrix}. 
\label{Heff}
\eea
Here $P_1,P_2,Q$ are $3\times 3$ matrices, $k_1=-\frac{1}{2}k_x+\frac{\sqrt{3}}{2}k_y,k_2=-\frac{1}{2}k_x-\frac{\sqrt{3}}{2}k_y,k_3=k_x$. $\epsilon_1$ and $\epsilon_2$ denote the energies of vH1 and vH2 respectively. The chemical potential $\mu$ is set between $\epsilon_1$ and $\epsilon_2$. {Let's first inspect the form of $P$ matrices imposed by symmetries.} The matrix $P_1$ ($P_2$) describes the effect of CDW order on vH1 (vH2) at momenta $M_A,M_B,M_C$. The threefold rotation symmetry permutes the three $M$ points, which requires $(P_n)_{12}=(P_n)_{23}=(P_n)_{31}$ for $n=1,2$, and whether these matrix elements are related to $\Delta$ or $\Delta^*$ is determined by the wave function at vHS. {The coefficients $s_1$ and $s_2$ are determined by wave functions at the vHS, which are found to be $s_1=-2|b'|^2$ and $s_2=2|b|^2$. The relative sign difference in these coefficients comes from the $-b$ term in Fig.\ref{fig_csorder}(c) as shown in Sec.III of SM, which is a consequence of the mirror symmetries.}

{Another important consequence of the mirror symmetries is that they enforce the $Q$ matrix in the off-diagonal block to be a diagonal matrix. With the wave function structure in Fig.\ref{fig_csorder}(c), the off-diagonal elements of $Q$ must vanish because they are multiplied by the zeros of wave function components from either vH1 or vH2, as shown in Sec.III of SM. Therefore, only the diagonal terms remain in the $Q$ matrix, which describes the coupling between two vHS at the same $M$ point. These terms are linear in $\mathbf k$ because $\epsilon_1$ and $\epsilon_2$ are exact eigenvalues when both $\mathbf k$ and $\Delta$ are zero, hence at $\mathbf k=0$ the diagonal terms should vanish and the leading order is linear in $\mathbf k$.  }



\emph{Mechanism to generate LCBO}---  We now discuss the last condition for LCBO to be the ground state of a system described by Eq.\eqref{Heff}. To derive this, we start from the limit with $\lambda=0$ and do a perturbation theory on $\lambda$. When $\lambda=0$, $H_{\rm{eff}}(\mathbf k,\Delta)$ and $H_{\rm{eff}}(\mathbf k,\Delta e^{\frac{2\pi i}{3}})$ have the same eigenvalues because they are related by a gauge transformation $\mathcal{U}=diag\{1,\omega,\omega^*,1,\omega^*,\omega\}$ with $\omega=e^{\frac{2\pi i}{3}}$. Hence when $\lambda=0$ the free energy $F$ is invariant under $\Delta\rightarrow\Delta e^{\frac{2\pi}{3}i}$, and $F$ has degenerate minima at $\Delta=-|\Delta|$ and $\Delta= |\Delta|e^{\pm\frac{\pi}{3}i}$ corresponding to CBO$^-$ and LCBO respectively. The eigenvalues of $H_{\rm{eff}}-\mu$ at both minima are the same, which are given by:
\bea
&&E_1=\epsilon_2-\mu-4|b|^2 |\Delta|,\ E_2=E_3=\epsilon_1-\mu-2|b'|^2 |\Delta|,\nonumber\\
&&E_4=E_5=\epsilon_2-\mu+2|b|^2 |\Delta|,\ E_6=\epsilon_1-\mu+4|b'|^2 |\Delta|
\label{xi16}
\eea
When the energy separation between vH1 and vH2 is small, the sign of each eigenvalue is determined by the $\Delta$ term, hence the negative eigenvalues are $E_1,E_2,E_3$. In the low-temperature limit the sum of them determines the free energy. When $\lambda$ becomes finite, the degenerate minima of $F$ at $\Delta=-|\Delta|$ and $|\Delta|e^{\pm\frac{\pi}{3}i}$ corresponding to CBO$^-$ and LCBO splits. The amount of splitting can be computed by degenerate perturbation theory that captures the evolution of $E_{1\sim 3}$ with $\lambda$. Define $\delta\epsilon\equiv\epsilon_2-\epsilon_1$ as the separation between vH1 and vH2 and denote $A$ as the system area. We find that the difference in free energy density $f=F/A$ between CBO$^-$ and LCBO is given by:
\bea
&&f_{CBO^-}-f_{LCBO}=  \nonumber\\
&&\sum_{|\mathbf k|\lesssim k_{cut}}\frac{-2|\lambda|^2(k_1 k_2+k_2k_3+k_1k_3)|\Delta|(|b|^2+|b'|^2)}{A(2|\Delta|(|b|^2+|b'|^2)+\delta\epsilon)(4|\Delta|(|b|^2+|b'|^2)-\delta\epsilon)}  \nonumber\\
&&=\frac{3}{16\pi}\frac{|\lambda|^2 k_{cut}^4 |\Delta|(|b|^2+|b'|^2)}{(2|\Delta|(|b|^2+|b'|^2)+\delta\epsilon)(4|\Delta|(|b|^2+|b'|^2)-\delta\epsilon)}>0. \nonumber\\
\label{Fdif}
\eea
Eq.\eqref{Fdif} shows that for small energy separation $\delta\epsilon<4(|b|^2+|b'|^2)|\Delta|$, a finite coupling $\lambda$ between the two vHS will make LCBO have lower energy and be more favorable than the competing phase CBO$^-$. This is the mechanism to generate LCBO in kagome systems.

{Importantly, we find that the LCBO phase cannot be described by a small order parameter expansion of the Ginzburg-Landau (GL) theory, hence it is not sensitive to the dispersion near vHS. In Sec.V of SM, we demonstrate that the solution corresponding to LCBO cannot be obtained from GL free energy even if high power terms of $\Delta$ are included. We also reiterate that the free energy analysis in Eq.\eqref{Fdif} leading to an LCBO ground state is a result of the coupling between two vHS with distinct mirror symmetries.   }

\begin{figure}
\centering
\includegraphics[width=3.5 in]{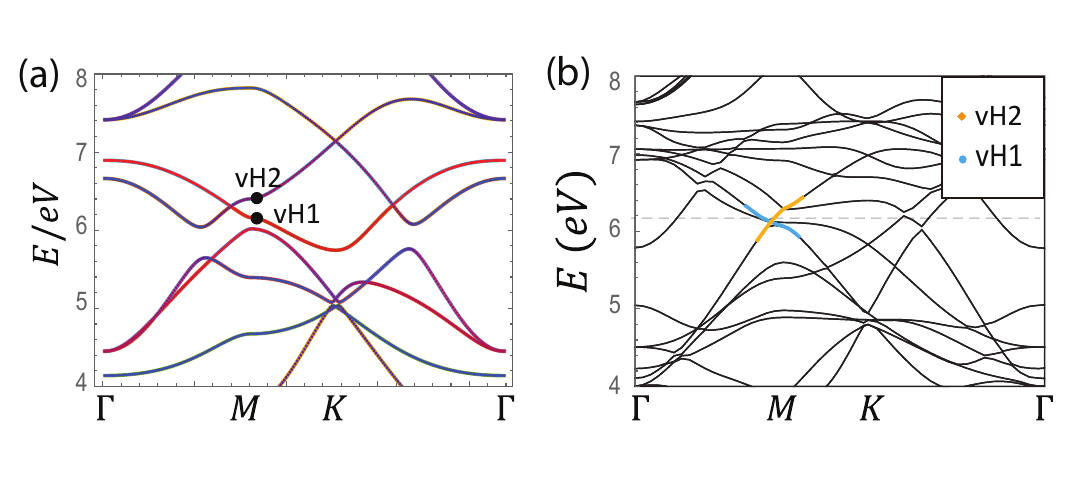}
\caption{ (a): Band structure of the $9\times 9$ tight-binding model $H_{TB}(\mathbf k)$ that can reproduce vH1 and vH2. The red color represents the weight of $\tilde d$ orbitals in the wave function. (b): Band structure obtained from DFT with vH1 and vH2 highlighted. The figure is adapted from Ref.\onlinecite{Li2023kagome}. }
\label{fig_band}
\end{figure}

\emph{Application to AV$_3$Sb$_5$}--- We apply the above analysis to AV$_3$Sb$_5$ and explicitly construct the effective Hamiltonian $H_{\rm{eff}}$. We start from a tight binding model that captures multiple vHS near the Fermi level. The bands close to the Fermi level in AV$_3$Sb$_5$ are mainly made of $d$ orbitals at V sites and $p$ orbitals at Sb sites. We consider the tight binding model introduced in Ref.~\cite{Li2023kagome}. This model includes three $p$ orbitals at each out-of-plane Sb site and one $d$ orbital at each V site. This $d$ orbital is made of a specific linear combination of $d_{xz},d_{yz}$ orbitals as indicated by the colored orbitals in Fig.\ref{fig_csorder}(a), which is odd (even) under $m'$ ($m''$), denoted as $\tilde d$ orbitals. Hence there are three $\tilde d$ orbitals and six $p$ orbitals in each unit cell, leading to a 9-band model $H_{TB}(\mathbf k)$. The hopping parameters in this model are obtained from DFT band structure~\cite{Li2023kagome}. The band structure of $H_{TB}(\mathbf k)$ is shown in Fig.\ref{fig_band}(a). Compared with the DFT band structure in Fig.\ref{fig_band}(b), the 9-band model reproduces two vHS at momentum $M$ denoted by vH1 and vH2. vH1 is odd (even) under $m'$ ($m''$) and is mainly made of $\tilde d$ orbitals. vH2 is even under both $m'$ and $m''$ and is a superposition of $\tilde d$ and $p$ orbitals. Compared with commonly used three-band models that can only describe vH1, this 9-band model can capture both vH1 and vH2, hence it provides a useful platform to study the interplay between different vHS.

Next we consider the NN electron interaction given by
\be
H_V=V\sum_{\langle \mathbf R\alpha;\mathbf R'\beta\rangle} c^\dagger_{\mathbf R,\alpha}c_{\mathbf R,\alpha}c^\dagger_{\mathbf R',\beta}c_{\mathbf R',\beta},
\ee
where $\langle \mathbf R\alpha;\mathbf R'\beta\rangle$ denotes NN bonds. With the order parameter $\Delta_{\alpha\beta}$ defined in Eq.\eqref{Deldef}, the NN interaction can be mean-field decoupled as~\cite{Li2023kagome}:
\begin{eqnarray}
H_V^{MF}&=& -\sum_{\mathbf k} \left(\Delta_{\alpha\beta}(1-e^{i\mathbf k \cdot \mathbf d_{\alpha\beta}})c^\dagger_{\mathbf k-\mathbf Q_{\alpha\beta},\beta}c_{\mathbf k,\alpha}+h.c. \right)\nonumber \\
 &&+2N_c \frac{|\Delta_{\alpha\beta}|^2}{V} ,\label{Hvmf} 
\end{eqnarray}
We can write down a mean field Hamiltonian that includes all bands in Fig.\ref{fig_band}(a) and the CDW order parameter in Eq.\eqref{Hvmf} with $\Delta_{AB}=\Delta_{BC}=\Delta_{CA}\equiv\Delta$. To construct the effective patch model $H_{\rm{eff}}$, we focus on momenta near the $M$ points and perform a unitary transformation into the band basis in which the basis functions at $M$ points are eigenfunctions of the tight binding model. Then we keep only the matrix elements corresponding to the energies and couplings between vH1 and vH2. This leads to a $6\times 6$ matrix $H_{\rm{eff}}(\mathbf k,\Delta)$ corresponding to the six patches at vH1 and vH2 near the three $M$ points. By performing a Taylor expansion in $\mathbf k$ and keeping leading order terms, we obtain $H_{\rm{eff}}$ in Eq.\eqref{Heff} with parameters $\epsilon_1=6.16eV,\epsilon_2=6.40eV,b=0.52,b'=0.96,\lambda=0.35eV\cdot a$, where $a=5.48$\AA$\ $is the lattice constant. Because the wave functions at both vHS have significant weight on $\tilde d$ orbitals, the coupling $\lambda$ between the two vHS receives major contribution from the hopping amplitude $t_{dd}$ between nearest-neighbor $\tilde d$ orbitals hence $\lambda$ is generally nonzero. With a finite $\lambda$, the above theory for LCBO is applicable to AV$_3$Sb$_5$, indicating that the LCBO phase can be stabilized as the ground state.



\begin{figure}
\centering
\includegraphics[width=3.4 in]{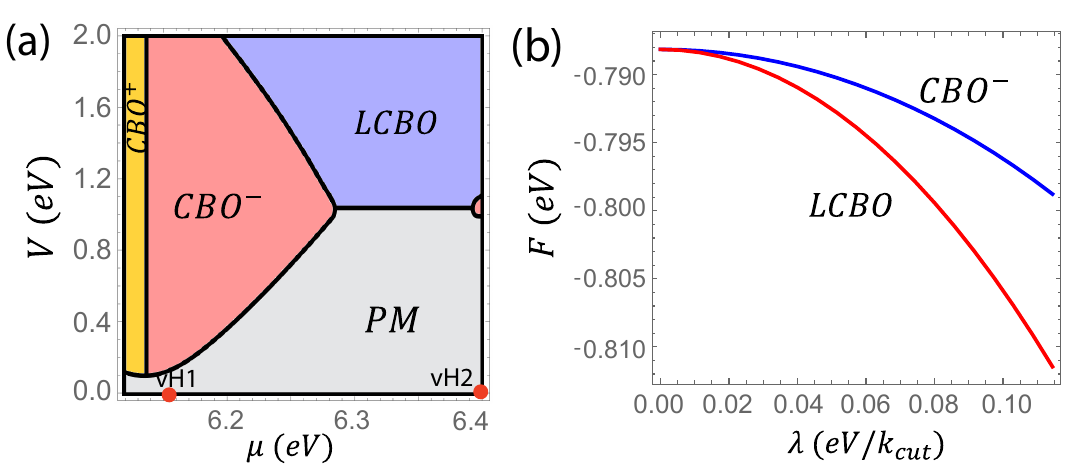}
\caption{ (a): Phase diagram of $H_{\rm{eff}}$ at different interaction strength and chemical potential with parameters: $\epsilon_1=6.16eV,\epsilon_2=6.40eV,b=0.52,b'=0.96,\lambda k_{cut}=0.1eV$ and temperature is 90K. $PM$ refers to the pristine metal without any CDW order. (b): Free energy of LCBO and CBO$^-$ as a function of coupling $\lambda$ at a fixed interaction strength. It shows LCBO is favored at finite $\lambda$. 
}
\label{fig_engRI}
\end{figure}

\emph{Phase diagram of CDW orders}--- The phase diagram of $H_{\rm{eff}}$ obtained by minimizing the free energy with respect to $\Delta$ at different chemical potential and interaction strength is shown in Fig.\ref{fig_engRI}(a). The LCBO phase is more pronounced near vH2 due to the difference in wave function structures at vH1 and vH2. Eq.\eqref{Fdif} requires the eigenvalues $E_{1\sim 3}$ be negative and $E_{4\sim 6}$ be positive. Based on Eq.\eqref{xi16}, these conditions lead to $4|b'|^2 |\Delta|>\delta\epsilon$ when $\mu\sim\epsilon_2$, while when $\mu\sim\epsilon_1$ they lead to $4|b|^2 |\Delta|>\delta\epsilon$. Since $|b'|>|b|$ due to the larger weight of $\tilde d$ orbital at vH1, when $\mu\sim\epsilon_2$ it requires smaller $|\Delta|$ and smaller interaction to realize LCBO. This leads to the smaller critical interaction strength near vH2 as shown in the phase diagram. The competition between CBO$^-$ and LCBO depends on the strength of $\lambda$. The free energy of the CBO$^-$ and LCBO phases at $\mu=\epsilon_2, V=1.3eV$ as a function of coupling strength $\lambda$  is shown in Fig.\ref{fig_engRI}(b). It shows LCBO and CBO$^-$ are degenerate when $\lambda=0$, and a finite $\lambda$ makes the free energy of LCBO lower than CBO$^-$, consistent with Eq.\eqref{Fdif}.



\begin{figure}
\centering
\includegraphics[width=3.4 in]{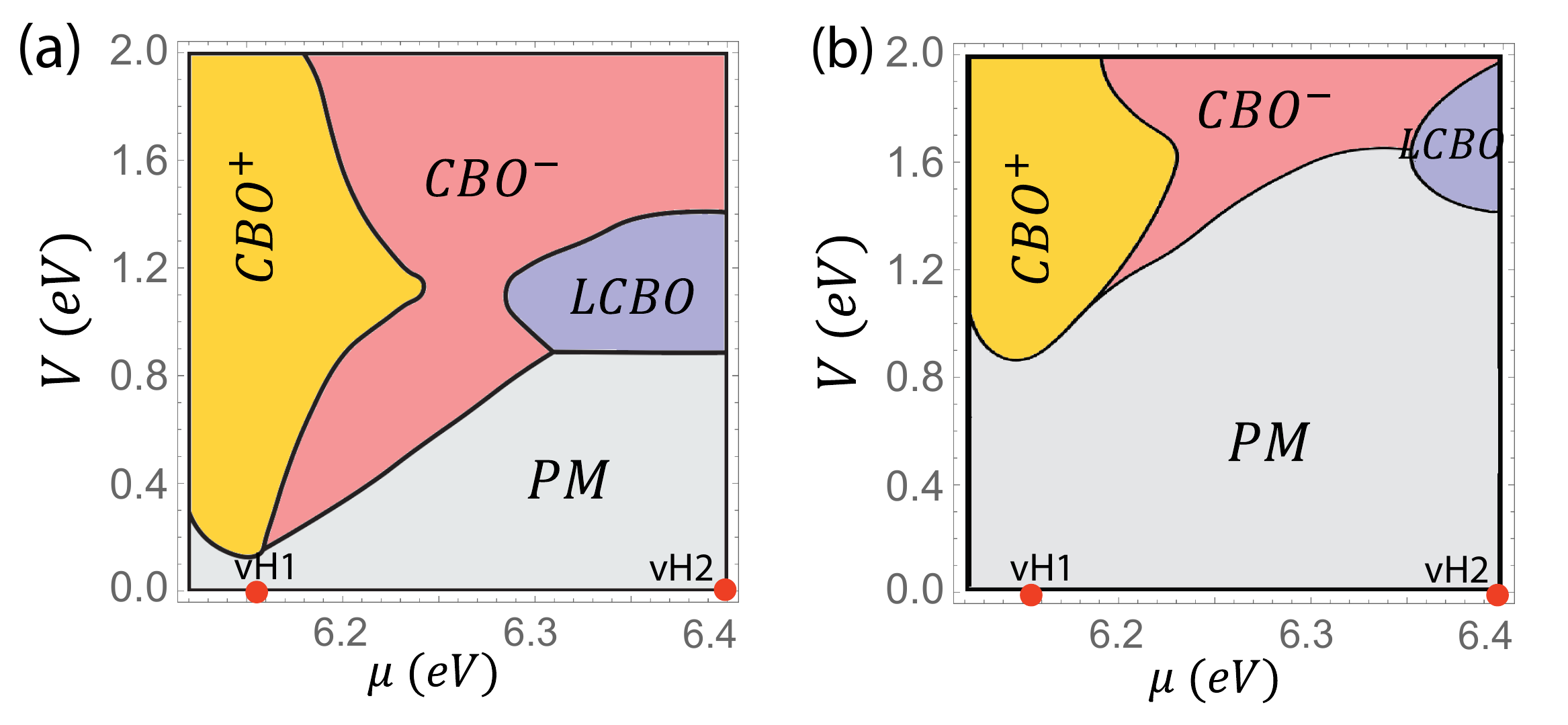}
\caption{ (a): Phase diagram of the effective patch model obtained by including vH1, vH2 and $\epsilon_3$. $PM$ refers to the pristine metal without any CDW order. (b): Phase diagram that takes into account all bands in $H_{TB}$ and the momentum summation is over the Brillouin zone. The LCBO phase still exists near vH2. 
}
\label{fig_phaseallbz}
\end{figure}

\emph{Effects of the other bands}--- In AV$_3$Sb$_5$ there are other bands near the Fermi level and their effects need to be investigated. For this purpose, we consider an effective patch model obtained by adding one more band below vH1 (denoted as $\epsilon_3$) in Fig.\ref{fig_band}(a) to $H_{\rm{eff}}$, which expands it to a $9\times 9$ matrix near the $M$ points. This model includes vH1, vH2 and $\epsilon_3$, and its phase diagram is shown in Fig.\ref{fig_phaseallbz}(a). {Compared with Fig.\ref{fig_engRI}(a), which only includes vH1 and vH2,  the key variation in Fig.\ref{fig_phaseallbz}(a) occurs near vH1, which is close to the additional band at $\epsilon_3$. However, in regions around vH2 that are more distant from $\epsilon_3$, the two phase diagrams resemble each other, with LCBO emerging in both scenarios.} We further demonstrate that the emergence of LCBO inferred from the patch model remains valid when all the bands in the tight-binding model are considered and the momentum cutoff is removed. The phase diagram obtained with all bands in $H_{TB}(\mathbf k)$ included is shown in Fig.\ref{fig_phaseallbz}(b). The real-space configuration of these phases are shown in Fig.\ref{fig_LCBO} with the order parameter given in Eq.\eqref{Deldef}. The summation of momentum in computing the free energy is taken over the whole Brillouin zone. The LCBO phase exists near vH2, whereas near vH1 the ground state is CBO, due to the effect of band structure away from $M$ points and the other bands that are not taken into account in the patch models. This comparison suggests despite the quantitative difference in these phase diagrams, our main finding of LCBO remains valid in the full-band model as long as the chemical potential is near vH2.



\emph{Discussion}--- We provide a mechanism to realize LCBO in kagome systems based on the coupling between multiple vHS with different mirror symmetries. The imaginary part of LCBO breaks time-reversal symmetry, whereas the real part of LCBO can induce lattice distortion with star of David or tri-hexagonal patterns. Experiments on AV$_3$Sb$_5$ have observed staggered patterns of lattice distortion among different kagome layers~\cite{Kang2022}. If the ground state is described by LCBO, we expect the loop current order to be staggered along the c axis as well. Our theory shows LCBO is more favorable when the energy difference $\delta\epsilon$ between vHS is small. Experiments and first-principle computations suggest that pressure can lead to an increase of $\delta\epsilon$~\cite{Labollita2021,Tsirlin2023}, hence we expect LCBO to disappear under high pressure, which is consistent with the disappearance of CDW under high pressure observed in experiments~\cite{Chen2021,Yu2021cs,Li2022pres,Wenzel2023p}. The phase diagram of AV$_3$Sb$_5$ in Fig.\ref{fig_phaseallbz}(b) suggests that LCBO emerges when the chemical potential is close to vH2. Thus we predict that electron-doping the material is more likely to induce the LCBO phase. Given the general applicability of our theory to kagome systems having vHS with distinct mirror eigenvalues, we hope this work will inspire further exploration of kagome materials with mirror symmetries.


\emph{Acknowledgements}--- This work is supported by the Natural Sciences and Engineering Research Council of Canada (NSERC) Discovery Grant No. 2022-04601 and the Center for Quantum Materials at the University of Toronto. H.Y.K acknowledges the support by the Canadian Institute for Advanced Research (CIFAR) and the Canada Research Chairs Program. Y.B.K. is supported by the Simons Fellowship from the Simons Foundation and the Guggenheim Fellowship from the John Simon Guggenheim Memorial Foundation. Computations were performed on the Niagara supercomputer at the SciNet HPC Consortium. SciNet is funded by: the Canada Foundation for Innovation under the auspices of Compute Canada; the Government of Ontario; Ontario Research Fund - Research Excellence; and the University of Toronto.


\begin{thebibliography}{73}%
\makeatletter
\providecommand \@ifxundefined [1]{%
 \@ifx{#1\undefined}
}%
\providecommand \@ifnum [1]{%
 \ifnum #1\expandafter \@firstoftwo
 \else \expandafter \@secondoftwo
 \fi
}%
\providecommand \@ifx [1]{%
 \ifx #1\expandafter \@firstoftwo
 \else \expandafter \@secondoftwo
 \fi
}%
\providecommand \natexlab [1]{#1}%
\providecommand \enquote  [1]{``#1''}%
\providecommand \bibnamefont  [1]{#1}%
\providecommand \bibfnamefont [1]{#1}%
\providecommand \citenamefont [1]{#1}%
\providecommand \href@noop [0]{\@secondoftwo}%
\providecommand \href [0]{\begingroup \@sanitize@url \@href}%
\providecommand \@href[1]{\@@startlink{#1}\@@href}%
\providecommand \@@href[1]{\endgroup#1\@@endlink}%
\providecommand \@sanitize@url [0]{\catcode `\\12\catcode `\$12\catcode
  `\&12\catcode `\#12\catcode `\^12\catcode `\_12\catcode `\%12\relax}%
\providecommand \@@startlink[1]{}%
\providecommand \@@endlink[0]{}%
\providecommand \url  [0]{\begingroup\@sanitize@url \@url }%
\providecommand \@url [1]{\endgroup\@href {#1}{\urlprefix }}%
\providecommand \urlprefix  [0]{URL }%
\providecommand \Eprint [0]{\href }%
\providecommand \doibase [0]{http://dx.doi.org/}%
\providecommand \selectlanguage [0]{\@gobble}%
\providecommand \bibinfo  [0]{\@secondoftwo}%
\providecommand \bibfield  [0]{\@secondoftwo}%
\providecommand \translation [1]{[#1]}%
\providecommand \BibitemOpen [0]{}%
\providecommand \bibitemStop [0]{}%
\providecommand \bibitemNoStop [0]{.\EOS\space}%
\providecommand \EOS [0]{\spacefactor3000\relax}%
\providecommand \BibitemShut  [1]{\csname bibitem#1\endcsname}%
\let\auto@bib@innerbib\@empty
\bibitem [{\citenamefont {Jiang}\ \emph {et~al.}(2022)\citenamefont {Jiang},
  \citenamefont {Wu}, \citenamefont {Yin}, \citenamefont {Wang}, \citenamefont
  {Hasan}, \citenamefont {Wilson}, \citenamefont {Chen},\ and\ \citenamefont
  {Hu}}]{Jiang2022s}%
  \BibitemOpen
  \bibfield  {author} {\bibinfo {author} {\bibfnamefont {Kun}\ \bibnamefont
  {Jiang}}, \bibinfo {author} {\bibfnamefont {Tao}\ \bibnamefont {Wu}},
  \bibinfo {author} {\bibfnamefont {Jia-Xin}\ \bibnamefont {Yin}}, \bibinfo
  {author} {\bibfnamefont {Zhenyu}\ \bibnamefont {Wang}}, \bibinfo {author}
  {\bibfnamefont {M~Zahid}\ \bibnamefont {Hasan}}, \bibinfo {author}
  {\bibfnamefont {Stephen~D}\ \bibnamefont {Wilson}}, \bibinfo {author}
  {\bibfnamefont {Xianhui}\ \bibnamefont {Chen}}, \ and\ \bibinfo {author}
  {\bibfnamefont {Jiangping}\ \bibnamefont {Hu}},\ }\bibfield  {title}
  {\enquote {\bibinfo {title} {{Kagome superconductors AV3Sb5 (A=K, Rb,
  Cs)}},}\ }\href {\doibase 10.1093/nsr/nwac199} {\bibfield  {journal}
  {\bibinfo  {journal} {National Science Review}\ } (\bibinfo {year} {2022}),\
  10.1093/nsr/nwac199},\ \bibinfo {note} {nwac199}\BibitemShut {NoStop}%
\bibitem [{\citenamefont {Yang}\ \emph {et~al.}(2020)\citenamefont {Yang},
  \citenamefont {Wang}, \citenamefont {Ortiz}, \citenamefont {Liu},
  \citenamefont {Gayles}, \citenamefont {Derunova}, \citenamefont
  {Gonzalez-Hernandez}, \citenamefont {Šmejkal}, \citenamefont {Chen},
  \citenamefont {Parkin}, \citenamefont {Wilson}, \citenamefont {Toberer},
  \citenamefont {McQueen},\ and\ \citenamefont {Ali}}]{Shuo2020}%
  \BibitemOpen
  \bibfield  {author} {\bibinfo {author} {\bibfnamefont {Shuo-Ying}\
  \bibnamefont {Yang}}, \bibinfo {author} {\bibfnamefont {Yaojia}\ \bibnamefont
  {Wang}}, \bibinfo {author} {\bibfnamefont {Brenden~R.}\ \bibnamefont
  {Ortiz}}, \bibinfo {author} {\bibfnamefont {Defa}\ \bibnamefont {Liu}},
  \bibinfo {author} {\bibfnamefont {Jacob}\ \bibnamefont {Gayles}}, \bibinfo
  {author} {\bibfnamefont {Elena}\ \bibnamefont {Derunova}}, \bibinfo {author}
  {\bibfnamefont {Rafael}\ \bibnamefont {Gonzalez-Hernandez}}, \bibinfo
  {author} {\bibfnamefont {Libor}\ \bibnamefont {Šmejkal}}, \bibinfo {author}
  {\bibfnamefont {Yulin}\ \bibnamefont {Chen}}, \bibinfo {author}
  {\bibfnamefont {Stuart S.~P.}\ \bibnamefont {Parkin}}, \bibinfo {author}
  {\bibfnamefont {Stephen~D.}\ \bibnamefont {Wilson}}, \bibinfo {author}
  {\bibfnamefont {Eric~S.}\ \bibnamefont {Toberer}}, \bibinfo {author}
  {\bibfnamefont {Tyrel}\ \bibnamefont {McQueen}}, \ and\ \bibinfo {author}
  {\bibfnamefont {Mazhar~N.}\ \bibnamefont {Ali}},\ }\bibfield  {title}
  {\enquote {\bibinfo {title} {Giant, unconventional anomalous hall effect in
  the metallic frustrated magnet candidate, kv3sb5},}\ }\href {\doibase
  10.1126/sciadv.abb6003} {\bibfield  {journal} {\bibinfo  {journal} {Science
  Advances}\ }\textbf {\bibinfo {volume} {6}},\ \bibinfo {pages} {eabb6003}
  (\bibinfo {year} {2020})}\BibitemShut {NoStop}%
\bibitem [{\citenamefont {Neupert}\ \emph {et~al.}(2022)\citenamefont
  {Neupert}, \citenamefont {Denner}, \citenamefont {Yin}, \citenamefont
  {Thomale},\ and\ \citenamefont {Hasan}}]{Neupert2022o}%
  \BibitemOpen
  \bibfield  {author} {\bibinfo {author} {\bibfnamefont {Titus}\ \bibnamefont
  {Neupert}}, \bibinfo {author} {\bibfnamefont {M.~Michael}\ \bibnamefont
  {Denner}}, \bibinfo {author} {\bibfnamefont {Jia-Xin}\ \bibnamefont {Yin}},
  \bibinfo {author} {\bibfnamefont {Ronny}\ \bibnamefont {Thomale}}, \ and\
  \bibinfo {author} {\bibfnamefont {M.~Zahid}\ \bibnamefont {Hasan}},\
  }\bibfield  {title} {\enquote {\bibinfo {title} {Charge order and
  superconductivity in kagome materials},}\ }\href {\doibase
  10.1038/s41567-021-01404-y} {\bibfield  {journal} {\bibinfo  {journal}
  {Nature Physics}\ }\textbf {\bibinfo {volume} {18}},\ \bibinfo {pages}
  {137--143} (\bibinfo {year} {2022})}\BibitemShut {NoStop}%
\bibitem [{\citenamefont {Wang}\ \emph {et~al.}(2020)\citenamefont {Wang},
  \citenamefont {Yang}, \citenamefont {Sivakumar}, \citenamefont {Ortiz},
  \citenamefont {Teicher}, \citenamefont {Wu}, \citenamefont {Srivastava},
  \citenamefont {Garg}, \citenamefont {Liu}, \citenamefont {Parkin},
  \citenamefont {Toberer}, \citenamefont {McQueen}, \citenamefont {Wilson},\
  and\ \citenamefont {Ali}}]{Wang2020s}%
  \BibitemOpen
  \bibfield  {author} {\bibinfo {author} {\bibfnamefont {Yaojia}\ \bibnamefont
  {Wang}}, \bibinfo {author} {\bibfnamefont {Shuoying}\ \bibnamefont {Yang}},
  \bibinfo {author} {\bibfnamefont {Pranava~K.}\ \bibnamefont {Sivakumar}},
  \bibinfo {author} {\bibfnamefont {Brenden~R.}\ \bibnamefont {Ortiz}},
  \bibinfo {author} {\bibfnamefont {Samuel M.~L.}\ \bibnamefont {Teicher}},
  \bibinfo {author} {\bibfnamefont {Heng}\ \bibnamefont {Wu}}, \bibinfo
  {author} {\bibfnamefont {Abhay~K.}\ \bibnamefont {Srivastava}}, \bibinfo
  {author} {\bibfnamefont {Chirag}\ \bibnamefont {Garg}}, \bibinfo {author}
  {\bibfnamefont {Defa}\ \bibnamefont {Liu}}, \bibinfo {author} {\bibfnamefont
  {Stuart S.~P.}\ \bibnamefont {Parkin}}, \bibinfo {author} {\bibfnamefont
  {Eric~S.}\ \bibnamefont {Toberer}}, \bibinfo {author} {\bibfnamefont {Tyrel}\
  \bibnamefont {McQueen}}, \bibinfo {author} {\bibfnamefont {Stephen~D.}\
  \bibnamefont {Wilson}}, \ and\ \bibinfo {author} {\bibfnamefont {Mazhar~N.}\
  \bibnamefont {Ali}},\ }\href@noop {} {\enquote {\bibinfo {title}
  {Proximity-induced spin-triplet superconductivity and edge supercurrent in
  the topological kagome metal, $\mathrm{K_{1-x}V_3Sb_5}$},}\ } (\bibinfo
  {year} {2020}),\ \Eprint {http://arxiv.org/abs/2012.05898} {arXiv:2012.05898}
  \BibitemShut {NoStop}%
\bibitem [{\citenamefont {Hu}\ \emph {et~al.}(2022{\natexlab{a}})\citenamefont
  {Hu}, \citenamefont {Wu}, \citenamefont {Ortiz}, \citenamefont {Ju},
  \citenamefont {Han}, \citenamefont {Ma}, \citenamefont {Plumb}, \citenamefont
  {Radovic}, \citenamefont {Thomale}, \citenamefont {Wilson}, \citenamefont
  {Schnyder},\ and\ \citenamefont {Shi}}]{Hu2022n}%
  \BibitemOpen
  \bibfield  {author} {\bibinfo {author} {\bibfnamefont {Yong}\ \bibnamefont
  {Hu}}, \bibinfo {author} {\bibfnamefont {Xianxin}\ \bibnamefont {Wu}},
  \bibinfo {author} {\bibfnamefont {Brenden~R.}\ \bibnamefont {Ortiz}},
  \bibinfo {author} {\bibfnamefont {Sailong}\ \bibnamefont {Ju}}, \bibinfo
  {author} {\bibfnamefont {Xinloong}\ \bibnamefont {Han}}, \bibinfo {author}
  {\bibfnamefont {Junzhang}\ \bibnamefont {Ma}}, \bibinfo {author}
  {\bibfnamefont {Nicholas~C.}\ \bibnamefont {Plumb}}, \bibinfo {author}
  {\bibfnamefont {Milan}\ \bibnamefont {Radovic}}, \bibinfo {author}
  {\bibfnamefont {Ronny}\ \bibnamefont {Thomale}}, \bibinfo {author}
  {\bibfnamefont {Stephen~D.}\ \bibnamefont {Wilson}}, \bibinfo {author}
  {\bibfnamefont {Andreas~P.}\ \bibnamefont {Schnyder}}, \ and\ \bibinfo
  {author} {\bibfnamefont {Ming}\ \bibnamefont {Shi}},\ }\bibfield  {title}
  {\enquote {\bibinfo {title} {Rich nature of van hove singularities in kagome
  superconductor csv3sb5},}\ }\href {\doibase 10.1038/s41467-022-29828-x  }
  {\bibfield  {journal} {\bibinfo  {journal} {Nature Communications}\ }\textbf
  {\bibinfo {volume} {13}},\ \bibinfo {pages} {2220} (\bibinfo {year}
  {2022}{\natexlab{a}})}\BibitemShut {NoStop}%
\bibitem [{\citenamefont {Oey}\ \emph {et~al.}(2022)\citenamefont {Oey},
  \citenamefont {Ortiz}, \citenamefont {Kaboudvand}, \citenamefont
  {Frassineti}, \citenamefont {Garcia}, \citenamefont {Cong}, \citenamefont
  {Sanna}, \citenamefont {Mitrovi\ifmmode~\acute{c}\else \'{c}\fi{}},
  \citenamefont {Seshadri},\ and\ \citenamefont {Wilson}}]{Oey2022}%
  \BibitemOpen
  \bibfield  {author} {\bibinfo {author} {\bibfnamefont {Yuzki~M.}\
  \bibnamefont {Oey}}, \bibinfo {author} {\bibfnamefont {Brenden~R.}\
  \bibnamefont {Ortiz}}, \bibinfo {author} {\bibfnamefont {Farnaz}\
  \bibnamefont {Kaboudvand}}, \bibinfo {author} {\bibfnamefont {Jonathan}\
  \bibnamefont {Frassineti}}, \bibinfo {author} {\bibfnamefont {Erick}\
  \bibnamefont {Garcia}}, \bibinfo {author} {\bibfnamefont {Rong}\ \bibnamefont
  {Cong}}, \bibinfo {author} {\bibfnamefont {Samuele}\ \bibnamefont {Sanna}},
  \bibinfo {author} {\bibfnamefont {Vesna~F.}\ \bibnamefont
  {Mitrovi\ifmmode~\acute{c}\else \'{c}\fi{}}}, \bibinfo {author}
  {\bibfnamefont {Ram}\ \bibnamefont {Seshadri}}, \ and\ \bibinfo {author}
  {\bibfnamefont {Stephen~D.}\ \bibnamefont {Wilson}},\ }\bibfield  {title}
  {\enquote {\bibinfo {title} {Fermi level tuning and double-dome
  superconductivity in the kagome metal
  ${\mathrm{csv}}_{3}{\mathrm{sb}}_{5\ensuremath{-}x}{\mathrm{sn}}_{x}$},}\
  }\href {\doibase 10.1103/PhysRevMaterials.6.L041801 } {\bibfield  {journal}
  {\bibinfo  {journal} {Phys. Rev. Mater.}\ }\textbf {\bibinfo {volume} {6}},\
  \bibinfo {pages} {L041801} (\bibinfo {year} {2022})}\BibitemShut {NoStop}%
\bibitem [{\citenamefont {Christensen}\ \emph {et~al.}(2022)\citenamefont
  {Christensen}, \citenamefont {Birol}, \citenamefont {Andersen},\ and\
  \citenamefont {Fernandes}}]{Christensen2022c}%
  \BibitemOpen
  \bibfield  {author} {\bibinfo {author} {\bibfnamefont {Morten~H.}\
  \bibnamefont {Christensen}}, \bibinfo {author} {\bibfnamefont {Turan}\
  \bibnamefont {Birol}}, \bibinfo {author} {\bibfnamefont {Brian~M.}\
  \bibnamefont {Andersen}}, \ and\ \bibinfo {author} {\bibfnamefont
  {Rafael~M.}\ \bibnamefont {Fernandes}},\ }\bibfield  {title} {\enquote
  {\bibinfo {title} {Loop currents in $a{\mathrm{v}}_{3}{\mathrm{sb}}_{5}$
  kagome metals: Multipolar and toroidal magnetic orders},}\ }\href {\doibase
  10.1103/PhysRevB.106.144504 } {\bibfield  {journal} {\bibinfo  {journal}
  {Phys. Rev. B}\ }\textbf {\bibinfo {volume} {106}},\ \bibinfo {pages}
  {144504} (\bibinfo {year} {2022})}\BibitemShut {NoStop}%
\bibitem [{\citenamefont {Zhu}\ \emph {et~al.}(2022)\citenamefont {Zhu},
  \citenamefont {Yang}, \citenamefont {Xia}, \citenamefont {Yin}, \citenamefont
  {Wang}, \citenamefont {Zhao}, \citenamefont {Dai}, \citenamefont {Tu},
  \citenamefont {Song}, \citenamefont {Tao}, \citenamefont {Tu}, \citenamefont
  {Gong}, \citenamefont {Lei}, \citenamefont {Guo},\ and\ \citenamefont
  {Li}}]{Zhu2022d}%
  \BibitemOpen
  \bibfield  {author} {\bibinfo {author} {\bibfnamefont {C.~C.}\ \bibnamefont
  {Zhu}}, \bibinfo {author} {\bibfnamefont {X.~F.}\ \bibnamefont {Yang}},
  \bibinfo {author} {\bibfnamefont {W.}~\bibnamefont {Xia}}, \bibinfo {author}
  {\bibfnamefont {Q.~W.}\ \bibnamefont {Yin}}, \bibinfo {author} {\bibfnamefont
  {L.~S.}\ \bibnamefont {Wang}}, \bibinfo {author} {\bibfnamefont {C.~C.}\
  \bibnamefont {Zhao}}, \bibinfo {author} {\bibfnamefont {D.~Z.}\ \bibnamefont
  {Dai}}, \bibinfo {author} {\bibfnamefont {C.~P.}\ \bibnamefont {Tu}},
  \bibinfo {author} {\bibfnamefont {B.~Q.}\ \bibnamefont {Song}}, \bibinfo
  {author} {\bibfnamefont {Z.~C.}\ \bibnamefont {Tao}}, \bibinfo {author}
  {\bibfnamefont {Z.~J.}\ \bibnamefont {Tu}}, \bibinfo {author} {\bibfnamefont
  {C.~S.}\ \bibnamefont {Gong}}, \bibinfo {author} {\bibfnamefont {H.~C.}\
  \bibnamefont {Lei}}, \bibinfo {author} {\bibfnamefont {Y.~F.}\ \bibnamefont
  {Guo}}, \ and\ \bibinfo {author} {\bibfnamefont {S.~Y.}\ \bibnamefont {Li}},\
  }\bibfield  {title} {\enquote {\bibinfo {title} {Double-dome
  superconductivity under pressure in the v-based kagome metals
  $a{\mathrm{v}}_{3}{\mathrm{sb}}_{5}$ ($a=\mathrm{Rb}$ and k)},}\ }\href
  {\doibase 10.1103/PhysRevB.105.094507} {\bibfield  {journal} {\bibinfo
  {journal} {Phys. Rev. B}\ }\textbf {\bibinfo {volume} {105}},\ \bibinfo
  {pages} {094507} (\bibinfo {year} {2022})}\BibitemShut {NoStop}%
\bibitem [{\citenamefont {Stahl}\ \emph {et~al.}(2022)\citenamefont {Stahl},
  \citenamefont {Chen}, \citenamefont {Ritschel}, \citenamefont {Shekhar},
  \citenamefont {Sadrollahi}, \citenamefont {Rahn}, \citenamefont {Ivashko},
  \citenamefont {Zimmermann}, \citenamefont {Felser},\ and\ \citenamefont
  {Geck}}]{Stahl2022r}%
  \BibitemOpen
  \bibfield  {author} {\bibinfo {author} {\bibfnamefont {Q.}~\bibnamefont
  {Stahl}}, \bibinfo {author} {\bibfnamefont {D.}~\bibnamefont {Chen}},
  \bibinfo {author} {\bibfnamefont {T.}~\bibnamefont {Ritschel}}, \bibinfo
  {author} {\bibfnamefont {C.}~\bibnamefont {Shekhar}}, \bibinfo {author}
  {\bibfnamefont {E.}~\bibnamefont {Sadrollahi}}, \bibinfo {author}
  {\bibfnamefont {M.~C.}\ \bibnamefont {Rahn}}, \bibinfo {author}
  {\bibfnamefont {O.}~\bibnamefont {Ivashko}}, \bibinfo {author} {\bibfnamefont
  {M.~v.}\ \bibnamefont {Zimmermann}}, \bibinfo {author} {\bibfnamefont
  {C.}~\bibnamefont {Felser}}, \ and\ \bibinfo {author} {\bibfnamefont
  {J.}~\bibnamefont {Geck}},\ }\bibfield  {title} {\enquote {\bibinfo {title}
  {Temperature-driven reorganization of electronic order in
  ${\mathrm{csv}}_{3}{\mathrm{sb}}_{5}$},}\ }\href {\doibase
  10.1103/PhysRevB.105.195136} {\bibfield  {journal} {\bibinfo  {journal}
  {Phys. Rev. B}\ }\textbf {\bibinfo {volume} {105}},\ \bibinfo {pages}
  {195136} (\bibinfo {year} {2022})}\BibitemShut {NoStop}%
\bibitem [{\citenamefont {Wu}\ \emph {et~al.}(2022{\natexlab{a}})\citenamefont
  {Wu}, \citenamefont {Ortiz}, \citenamefont {Tan}, \citenamefont {Wilson},
  \citenamefont {Yan}, \citenamefont {Birol},\ and\ \citenamefont
  {Blumberg}}]{Wu2022d}%
  \BibitemOpen
  \bibfield  {author} {\bibinfo {author} {\bibfnamefont {Shangfei}\
  \bibnamefont {Wu}}, \bibinfo {author} {\bibfnamefont {Brenden~R.}\
  \bibnamefont {Ortiz}}, \bibinfo {author} {\bibfnamefont {Hengxin}\
  \bibnamefont {Tan}}, \bibinfo {author} {\bibfnamefont {Stephen~D.}\
  \bibnamefont {Wilson}}, \bibinfo {author} {\bibfnamefont {Binghai}\
  \bibnamefont {Yan}}, \bibinfo {author} {\bibfnamefont {Turan}\ \bibnamefont
  {Birol}}, \ and\ \bibinfo {author} {\bibfnamefont {Girsh}\ \bibnamefont
  {Blumberg}},\ }\bibfield  {title} {\enquote {\bibinfo {title} {Charge density
  wave order in the kagome metal $a{\mathrm{v}}_{3}{\mathrm{sb}}_{5}$
  $(a=\mathrm{Cs},\mathrm{Rb},\mathrm{K})$},}\ }\href {\doibase
  10.1103/PhysRevB.105.155106} {\bibfield  {journal} {\bibinfo  {journal}
  {Phys. Rev. B}\ }\textbf {\bibinfo {volume} {105}},\ \bibinfo {pages}
  {155106} (\bibinfo {year} {2022}{\natexlab{a}})}\BibitemShut {NoStop}%
\bibitem [{\citenamefont {Kang}\ \emph
  {et~al.}(2022{\natexlab{a}})\citenamefont {Kang}, \citenamefont {Fang},
  \citenamefont {Kim}, \citenamefont {Ortiz}, \citenamefont {Ryu},
  \citenamefont {Kim}, \citenamefont {Yoo}, \citenamefont {Sangiovanni},
  \citenamefont {Di~Sante}, \citenamefont {Park}, \citenamefont {Jozwiak},
  \citenamefont {Bostwick}, \citenamefont {Rotenberg}, \citenamefont {Kaxiras},
  \citenamefont {Wilson}, \citenamefont {Park},\ and\ \citenamefont
  {Comin}}]{Kang2022v}%
  \BibitemOpen
  \bibfield  {author} {\bibinfo {author} {\bibfnamefont {Mingu}\ \bibnamefont
  {Kang}}, \bibinfo {author} {\bibfnamefont {Shiang}\ \bibnamefont {Fang}},
  \bibinfo {author} {\bibfnamefont {Jeong-Kyu}\ \bibnamefont {Kim}}, \bibinfo
  {author} {\bibfnamefont {Brenden~R.}\ \bibnamefont {Ortiz}}, \bibinfo
  {author} {\bibfnamefont {Sae~Hee}\ \bibnamefont {Ryu}}, \bibinfo {author}
  {\bibfnamefont {Jimin}\ \bibnamefont {Kim}}, \bibinfo {author} {\bibfnamefont
  {Jonggyu}\ \bibnamefont {Yoo}}, \bibinfo {author} {\bibfnamefont {Giorgio}\
  \bibnamefont {Sangiovanni}}, \bibinfo {author} {\bibfnamefont {Domenico}\
  \bibnamefont {Di~Sante}}, \bibinfo {author} {\bibfnamefont {Byeong-Gyu}\
  \bibnamefont {Park}}, \bibinfo {author} {\bibfnamefont {Chris}\ \bibnamefont
  {Jozwiak}}, \bibinfo {author} {\bibfnamefont {Aaron}\ \bibnamefont
  {Bostwick}}, \bibinfo {author} {\bibfnamefont {Eli}\ \bibnamefont
  {Rotenberg}}, \bibinfo {author} {\bibfnamefont {Efthimios}\ \bibnamefont
  {Kaxiras}}, \bibinfo {author} {\bibfnamefont {Stephen~D.}\ \bibnamefont
  {Wilson}}, \bibinfo {author} {\bibfnamefont {Jae-Hoon}\ \bibnamefont {Park}},
  \ and\ \bibinfo {author} {\bibfnamefont {Riccardo}\ \bibnamefont {Comin}},\
  }\bibfield  {title} {\enquote {\bibinfo {title} {Twofold van hove singularity
  and origin of charge order in topological kagome superconductor csv3sb5},}\
  }\href {\doibase 10.1038/s41567-021-01451-5} {\bibfield  {journal} {\bibinfo
  {journal} {Nature Physics}\ }\textbf {\bibinfo {volume} {18}},\ \bibinfo
  {pages} {301--308} (\bibinfo {year} {2022}{\natexlab{a}})}\BibitemShut
  {NoStop}%
\bibitem [{\citenamefont {Jiang}\ \emph {et~al.}(2021)\citenamefont {Jiang},
  \citenamefont {Yin}, \citenamefont {Denner}, \citenamefont {Shumiya},
  \citenamefont {Ortiz}, \citenamefont {Xu}, \citenamefont {Guguchia},
  \citenamefont {He}, \citenamefont {Hossain}, \citenamefont {Liu},
  \citenamefont {Ruff}, \citenamefont {Kautzsch}, \citenamefont {Zhang},
  \citenamefont {Chang}, \citenamefont {Belopolski}, \citenamefont {Zhang},
  \citenamefont {Cochran}, \citenamefont {Multer}, \citenamefont {Litskevich},
  \citenamefont {Cheng}, \citenamefont {Yang}, \citenamefont {Wang},
  \citenamefont {Thomale}, \citenamefont {Neupert}, \citenamefont {Wilson},\
  and\ \citenamefont {Hasan}}]{Jiang2021}%
  \BibitemOpen
  \bibfield  {author} {\bibinfo {author} {\bibfnamefont {Yu-Xiao}\ \bibnamefont
  {Jiang}}, \bibinfo {author} {\bibfnamefont {Jia-Xin}\ \bibnamefont {Yin}},
  \bibinfo {author} {\bibfnamefont {M.~Michael}\ \bibnamefont {Denner}},
  \bibinfo {author} {\bibfnamefont {Nana}\ \bibnamefont {Shumiya}}, \bibinfo
  {author} {\bibfnamefont {Brenden~R.}\ \bibnamefont {Ortiz}}, \bibinfo
  {author} {\bibfnamefont {Gang}\ \bibnamefont {Xu}}, \bibinfo {author}
  {\bibfnamefont {Zurab}\ \bibnamefont {Guguchia}}, \bibinfo {author}
  {\bibfnamefont {Junyi}\ \bibnamefont {He}}, \bibinfo {author} {\bibfnamefont
  {Md~Shafayat}\ \bibnamefont {Hossain}}, \bibinfo {author} {\bibfnamefont
  {Xiaoxiong}\ \bibnamefont {Liu}}, \bibinfo {author} {\bibfnamefont {Jacob}\
  \bibnamefont {Ruff}}, \bibinfo {author} {\bibfnamefont {Linus}\ \bibnamefont
  {Kautzsch}}, \bibinfo {author} {\bibfnamefont {Songtian~S.}\ \bibnamefont
  {Zhang}}, \bibinfo {author} {\bibfnamefont {Guoqing}\ \bibnamefont {Chang}},
  \bibinfo {author} {\bibfnamefont {Ilya}\ \bibnamefont {Belopolski}}, \bibinfo
  {author} {\bibfnamefont {Qi}~\bibnamefont {Zhang}}, \bibinfo {author}
  {\bibfnamefont {Tyler~A.}\ \bibnamefont {Cochran}}, \bibinfo {author}
  {\bibfnamefont {Daniel}\ \bibnamefont {Multer}}, \bibinfo {author}
  {\bibfnamefont {Maksim}\ \bibnamefont {Litskevich}}, \bibinfo {author}
  {\bibfnamefont {Zi-Jia}\ \bibnamefont {Cheng}}, \bibinfo {author}
  {\bibfnamefont {Xian~P.}\ \bibnamefont {Yang}}, \bibinfo {author}
  {\bibfnamefont {Ziqiang}\ \bibnamefont {Wang}}, \bibinfo {author}
  {\bibfnamefont {Ronny}\ \bibnamefont {Thomale}}, \bibinfo {author}
  {\bibfnamefont {Titus}\ \bibnamefont {Neupert}}, \bibinfo {author}
  {\bibfnamefont {Stephen~D.}\ \bibnamefont {Wilson}}, \ and\ \bibinfo {author}
  {\bibfnamefont {M.~Zahid}\ \bibnamefont {Hasan}},\ }\bibfield  {title}
  {\enquote {\bibinfo {title} {Unconventional chiral charge order in kagome
  superconductor kv3sb5},}\ }\href {\doibase 10.1038/s41563-021-01034-y}
  {\bibfield  {journal} {\bibinfo  {journal} {Nature Materials}\ }\textbf
  {\bibinfo {volume} {20}},\ \bibinfo {pages} {1353--1357} (\bibinfo {year}
  {2021})}\BibitemShut {NoStop}%
\bibitem [{\citenamefont {Li}\ \emph {et~al.}(2021)\citenamefont {Li},
  \citenamefont {Zhang}, \citenamefont {Yilmaz}, \citenamefont {Pai},
  \citenamefont {Marvinney}, \citenamefont {Said}, \citenamefont {Yin},
  \citenamefont {Gong}, \citenamefont {Tu}, \citenamefont {Vescovo},
  \citenamefont {Nelson}, \citenamefont {Moore}, \citenamefont {Murakami},
  \citenamefont {Lei}, \citenamefont {Lee}, \citenamefont {Lawrie},\ and\
  \citenamefont {Miao}}]{Li2021u}%
  \BibitemOpen
  \bibfield  {author} {\bibinfo {author} {\bibfnamefont {Haoxiang}\
  \bibnamefont {Li}}, \bibinfo {author} {\bibfnamefont {T.~T.}\ \bibnamefont
  {Zhang}}, \bibinfo {author} {\bibfnamefont {T.}~\bibnamefont {Yilmaz}},
  \bibinfo {author} {\bibfnamefont {Y.~Y.}\ \bibnamefont {Pai}}, \bibinfo
  {author} {\bibfnamefont {C.~E.}\ \bibnamefont {Marvinney}}, \bibinfo {author}
  {\bibfnamefont {A.}~\bibnamefont {Said}}, \bibinfo {author} {\bibfnamefont
  {Q.~W.}\ \bibnamefont {Yin}}, \bibinfo {author} {\bibfnamefont {C.~S.}\
  \bibnamefont {Gong}}, \bibinfo {author} {\bibfnamefont {Z.~J.}\ \bibnamefont
  {Tu}}, \bibinfo {author} {\bibfnamefont {E.}~\bibnamefont {Vescovo}},
  \bibinfo {author} {\bibfnamefont {C.~S.}\ \bibnamefont {Nelson}}, \bibinfo
  {author} {\bibfnamefont {R.~G.}\ \bibnamefont {Moore}}, \bibinfo {author}
  {\bibfnamefont {S.}~\bibnamefont {Murakami}}, \bibinfo {author}
  {\bibfnamefont {H.~C.}\ \bibnamefont {Lei}}, \bibinfo {author} {\bibfnamefont
  {H.~N.}\ \bibnamefont {Lee}}, \bibinfo {author} {\bibfnamefont {B.~J.}\
  \bibnamefont {Lawrie}}, \ and\ \bibinfo {author} {\bibfnamefont
  {H.}~\bibnamefont {Miao}},\ }\bibfield  {title} {\enquote {\bibinfo {title}
  {Observation of unconventional charge density wave without acoustic phonon
  anomaly in kagome superconductors ${A\mathrm{V}}_{3}{\mathrm{sb}}_{5}$
  ($a=\mathrm{Rb}$, cs)},}\ }\href {\doibase 10.1103/PhysRevX.11.031050}
  {\bibfield  {journal} {\bibinfo  {journal} {Phys. Rev. X}\ }\textbf {\bibinfo
  {volume} {11}},\ \bibinfo {pages} {031050} (\bibinfo {year}
  {2021})}\BibitemShut {NoStop}%
\bibitem [{\citenamefont {Wu}\ \emph {et~al.}(2021)\citenamefont {Wu},
  \citenamefont {Schwemmer}, \citenamefont {M\"uller}, \citenamefont
  {Consiglio}, \citenamefont {Sangiovanni}, \citenamefont {Di~Sante},
  \citenamefont {Iqbal}, \citenamefont {Hanke}, \citenamefont {Schnyder},
  \citenamefont {Denner}, \citenamefont {Fischer}, \citenamefont {Neupert},\
  and\ \citenamefont {Thomale}}]{Wu2021u}%
  \BibitemOpen
  \bibfield  {author} {\bibinfo {author} {\bibfnamefont {Xianxin}\ \bibnamefont
  {Wu}}, \bibinfo {author} {\bibfnamefont {Tilman}\ \bibnamefont {Schwemmer}},
  \bibinfo {author} {\bibfnamefont {Tobias}\ \bibnamefont {M\"uller}}, \bibinfo
  {author} {\bibfnamefont {Armando}\ \bibnamefont {Consiglio}}, \bibinfo
  {author} {\bibfnamefont {Giorgio}\ \bibnamefont {Sangiovanni}}, \bibinfo
  {author} {\bibfnamefont {Domenico}\ \bibnamefont {Di~Sante}}, \bibinfo
  {author} {\bibfnamefont {Yasir}\ \bibnamefont {Iqbal}}, \bibinfo {author}
  {\bibfnamefont {Werner}\ \bibnamefont {Hanke}}, \bibinfo {author}
  {\bibfnamefont {Andreas~P.}\ \bibnamefont {Schnyder}}, \bibinfo {author}
  {\bibfnamefont {M.~Michael}\ \bibnamefont {Denner}}, \bibinfo {author}
  {\bibfnamefont {Mark~H.}\ \bibnamefont {Fischer}}, \bibinfo {author}
  {\bibfnamefont {Titus}\ \bibnamefont {Neupert}}, \ and\ \bibinfo {author}
  {\bibfnamefont {Ronny}\ \bibnamefont {Thomale}},\ }\bibfield  {title}
  {\enquote {\bibinfo {title} {Nature of unconventional pairing in the kagome
  superconductors $a{\mathrm{v}}_{3}{\mathrm{sb}}_{5}$
  ($a=\mathrm{K},\mathrm{Rb},\mathrm{Cs}$)},}\ }\href {\doibase
  10.1103/PhysRevLett.127.177001} {\bibfield  {journal} {\bibinfo  {journal}
  {Phys. Rev. Lett.}\ }\textbf {\bibinfo {volume} {127}},\ \bibinfo {pages}
  {177001} (\bibinfo {year} {2021})}\BibitemShut {NoStop}%
\bibitem [{\citenamefont {Zhao}\ \emph {et~al.}(2021)\citenamefont {Zhao},
  \citenamefont {Li}, \citenamefont {Ortiz}, \citenamefont {Teicher},
  \citenamefont {Park}, \citenamefont {Ye}, \citenamefont {Wang}, \citenamefont
  {Balents}, \citenamefont {Wilson},\ and\ \citenamefont
  {Zeljkovic}}]{Zhao2021}%
  \BibitemOpen
  \bibfield  {author} {\bibinfo {author} {\bibfnamefont {He}~\bibnamefont
  {Zhao}}, \bibinfo {author} {\bibfnamefont {Hong}\ \bibnamefont {Li}},
  \bibinfo {author} {\bibfnamefont {Brenden~R.}\ \bibnamefont {Ortiz}},
  \bibinfo {author} {\bibfnamefont {Samuel M.~L.}\ \bibnamefont {Teicher}},
  \bibinfo {author} {\bibfnamefont {Takamori}\ \bibnamefont {Park}}, \bibinfo
  {author} {\bibfnamefont {Mengxing}\ \bibnamefont {Ye}}, \bibinfo {author}
  {\bibfnamefont {Ziqiang}\ \bibnamefont {Wang}}, \bibinfo {author}
  {\bibfnamefont {Leon}\ \bibnamefont {Balents}}, \bibinfo {author}
  {\bibfnamefont {Stephen~D.}\ \bibnamefont {Wilson}}, \ and\ \bibinfo {author}
  {\bibfnamefont {Ilija}\ \bibnamefont {Zeljkovic}},\ }\bibfield  {title}
  {\enquote {\bibinfo {title} {Cascade of correlated electron states in the
  kagome superconductor csv3sb5},}\ }\href {\doibase
  10.1038/s41586-021-03946-w} {\bibfield  {journal} {\bibinfo  {journal}
  {Nature}\ }\textbf {\bibinfo {volume} {599}},\ \bibinfo {pages} {216--221}
  (\bibinfo {year} {2021})}\BibitemShut {NoStop}%
\bibitem [{\citenamefont {Li}\ \emph {et~al.}(2022{\natexlab{a}})\citenamefont
  {Li}, \citenamefont {Zhao}, \citenamefont {Ortiz}, \citenamefont {Park},
  \citenamefont {Ye}, \citenamefont {Balents}, \citenamefont {Wang},
  \citenamefont {Wilson},\ and\ \citenamefont {Zeljkovic}}]{Li2022s}%
  \BibitemOpen
  \bibfield  {author} {\bibinfo {author} {\bibfnamefont {Hong}\ \bibnamefont
  {Li}}, \bibinfo {author} {\bibfnamefont {He}~\bibnamefont {Zhao}}, \bibinfo
  {author} {\bibfnamefont {Brenden~R.}\ \bibnamefont {Ortiz}}, \bibinfo
  {author} {\bibfnamefont {Takamori}\ \bibnamefont {Park}}, \bibinfo {author}
  {\bibfnamefont {Mengxing}\ \bibnamefont {Ye}}, \bibinfo {author}
  {\bibfnamefont {Leon}\ \bibnamefont {Balents}}, \bibinfo {author}
  {\bibfnamefont {Ziqiang}\ \bibnamefont {Wang}}, \bibinfo {author}
  {\bibfnamefont {Stephen~D.}\ \bibnamefont {Wilson}}, \ and\ \bibinfo {author}
  {\bibfnamefont {Ilija}\ \bibnamefont {Zeljkovic}},\ }\bibfield  {title}
  {\enquote {\bibinfo {title} {Rotation symmetry breaking in the normal state
  of a kagome superconductor kv3sb5},}\ }\href {\doibase
  10.1038/s41567-021-01479-7} {\bibfield  {journal} {\bibinfo  {journal}
  {Nature Physics}\ }\textbf {\bibinfo {volume} {18}},\ \bibinfo {pages}
  {265--270} (\bibinfo {year} {2022}{\natexlab{a}})}\BibitemShut {NoStop}%
\bibitem [{\citenamefont {R\o{}mer}\ \emph {et~al.}(2022)\citenamefont
  {R\o{}mer}, \citenamefont {Bhattacharyya}, \citenamefont {Valent\'{\i}},
  \citenamefont {Christensen},\ and\ \citenamefont {Andersen}}]{Romer2022}%
  \BibitemOpen
  \bibfield  {author} {\bibinfo {author} {\bibfnamefont {Astrid~T.}\
  \bibnamefont {R\o{}mer}}, \bibinfo {author} {\bibfnamefont {Shinibali}\
  \bibnamefont {Bhattacharyya}}, \bibinfo {author} {\bibfnamefont {Roser}\
  \bibnamefont {Valent\'{\i}}}, \bibinfo {author} {\bibfnamefont {Morten~H.}\
  \bibnamefont {Christensen}}, \ and\ \bibinfo {author} {\bibfnamefont
  {Brian~M.}\ \bibnamefont {Andersen}},\ }\bibfield  {title} {\enquote
  {\bibinfo {title} {Superconductivity from repulsive interactions on the
  kagome lattice},}\ }\href {\doibase 10.1103/PhysRevB.106.174514} {\bibfield
  {journal} {\bibinfo  {journal} {Phys. Rev. B}\ }\textbf {\bibinfo {volume}
  {106}},\ \bibinfo {pages} {174514} (\bibinfo {year} {2022})}\BibitemShut
  {NoStop}%
\bibitem [{\citenamefont {Mertz}\ \emph {et~al.}(2022)\citenamefont {Mertz},
  \citenamefont {Wunderlich}, \citenamefont {Bhattacharyya}, \citenamefont
  {Ferrari},\ and\ \citenamefont {Valent{\'i}}}]{Mertz2022}%
  \BibitemOpen
  \bibfield  {author} {\bibinfo {author} {\bibfnamefont {Thomas}\ \bibnamefont
  {Mertz}}, \bibinfo {author} {\bibfnamefont {Paul}\ \bibnamefont
  {Wunderlich}}, \bibinfo {author} {\bibfnamefont {Shinibali}\ \bibnamefont
  {Bhattacharyya}}, \bibinfo {author} {\bibfnamefont {Francesco}\ \bibnamefont
  {Ferrari}}, \ and\ \bibinfo {author} {\bibfnamefont {Roser}\ \bibnamefont
  {Valent{\'i}}},\ }\bibfield  {title} {\enquote {\bibinfo {title} {Statistical
  learning of engineered topological phases in the kagome superlattice of
  av3sb5},}\ }\href {\doibase 10.1038/s41524-022-00745-3} {\bibfield  {journal}
  {\bibinfo  {journal} {npj Computational Materials}\ }\textbf {\bibinfo
  {volume} {8}},\ \bibinfo {pages} {66} (\bibinfo {year} {2022})}\BibitemShut
  {NoStop}%
\bibitem [{\citenamefont {Ptok}\ \emph {et~al.}(2022)\citenamefont {Ptok},
  \citenamefont {Kobia\l{}ka}, \citenamefont {Sternik}, \citenamefont
  {\L{}a\ifmmode~\dot{z}\else \.{z}\fi{}ewski}, \citenamefont {Jochym},
  \citenamefont {Ole\ifmmode~\acute{s}\else \'{s}\fi{}},\ and\ \citenamefont
  {Piekarz}}]{Ptok2022}%
  \BibitemOpen
  \bibfield  {author} {\bibinfo {author} {\bibfnamefont {Andrzej}\ \bibnamefont
  {Ptok}}, \bibinfo {author} {\bibfnamefont {Aksel}\ \bibnamefont
  {Kobia\l{}ka}}, \bibinfo {author} {\bibfnamefont {Ma\l{}gorzata}\
  \bibnamefont {Sternik}}, \bibinfo {author} {\bibfnamefont {Jan}\ \bibnamefont
  {\L{}a\ifmmode~\dot{z}\else \.{z}\fi{}ewski}}, \bibinfo {author}
  {\bibfnamefont {Pawe\l{}~T.}\ \bibnamefont {Jochym}}, \bibinfo {author}
  {\bibfnamefont {Andrzej~M.}\ \bibnamefont {Ole\ifmmode~\acute{s}\else
  \'{s}\fi{}}}, \ and\ \bibinfo {author} {\bibfnamefont {Przemys\l{}aw}\
  \bibnamefont {Piekarz}},\ }\bibfield  {title} {\enquote {\bibinfo {title}
  {Dynamical study of the origin of the charge density wave in
  $a{\mathrm{v}}_{3}{\mathrm{sb}}_{5}$ ($a=\mathrm{K}$, rb, cs) compounds},}\
  }\href {\doibase 10.1103/PhysRevB.105.235134} {\bibfield  {journal} {\bibinfo
   {journal} {Phys. Rev. B}\ }\textbf {\bibinfo {volume} {105}},\ \bibinfo
  {pages} {235134} (\bibinfo {year} {2022})}\BibitemShut {NoStop}%
\bibitem [{\citenamefont {Ferrari}\ \emph {et~al.}(2022)\citenamefont
  {Ferrari}, \citenamefont {Becca},\ and\ \citenamefont
  {Valent\'{\i}}}]{Ferrari2022}%
  \BibitemOpen
  \bibfield  {author} {\bibinfo {author} {\bibfnamefont {Francesco}\
  \bibnamefont {Ferrari}}, \bibinfo {author} {\bibfnamefont {Federico}\
  \bibnamefont {Becca}}, \ and\ \bibinfo {author} {\bibfnamefont {Roser}\
  \bibnamefont {Valent\'{\i}}},\ }\bibfield  {title} {\enquote {\bibinfo
  {title} {Charge density waves in kagome-lattice extended hubbard models at
  the van hove filling},}\ }\href {\doibase 10.1103/PhysRevB.106.L081107}
  {\bibfield  {journal} {\bibinfo  {journal} {Phys. Rev. B}\ }\textbf {\bibinfo
  {volume} {106}},\ \bibinfo {pages} {L081107} (\bibinfo {year}
  {2022})}\BibitemShut {NoStop}%
\bibitem [{\citenamefont {Uykur}\ \emph {et~al.}(2022)\citenamefont {Uykur},
  \citenamefont {Ortiz}, \citenamefont {Wilson}, \citenamefont {Dressel},\ and\
  \citenamefont {Tsirlin}}]{Uykur2022}%
  \BibitemOpen
  \bibfield  {author} {\bibinfo {author} {\bibfnamefont {Ece}\ \bibnamefont
  {Uykur}}, \bibinfo {author} {\bibfnamefont {Brenden~R.}\ \bibnamefont
  {Ortiz}}, \bibinfo {author} {\bibfnamefont {Stephen~D.}\ \bibnamefont
  {Wilson}}, \bibinfo {author} {\bibfnamefont {Martin}\ \bibnamefont
  {Dressel}}, \ and\ \bibinfo {author} {\bibfnamefont {Alexander~A.}\
  \bibnamefont {Tsirlin}},\ }\bibfield  {title} {\enquote {\bibinfo {title}
  {Optical detection of the density-wave instability in the kagome metal
  kv3sb5},}\ }\href {\doibase 10.1038/s41535-021-00420-8} {\bibfield  {journal}
  {\bibinfo  {journal} {npj Quantum Materials}\ }\textbf {\bibinfo {volume}
  {7}},\ \bibinfo {pages} {16} (\bibinfo {year} {2022})}\BibitemShut {NoStop}%
\bibitem [{\citenamefont {Grandi}\ \emph {et~al.}(2023)\citenamefont {Grandi},
  \citenamefont {Consiglio}, \citenamefont {Sentef}, \citenamefont {Thomale},\
  and\ \citenamefont {Kennes}}]{Grandi2023}%
  \BibitemOpen
  \bibfield  {author} {\bibinfo {author} {\bibfnamefont {Francesco}\
  \bibnamefont {Grandi}}, \bibinfo {author} {\bibfnamefont {Armando}\
  \bibnamefont {Consiglio}}, \bibinfo {author} {\bibfnamefont {Michael~A.}\
  \bibnamefont {Sentef}}, \bibinfo {author} {\bibfnamefont {Ronny}\
  \bibnamefont {Thomale}}, \ and\ \bibinfo {author} {\bibfnamefont {Dante~M.}\
  \bibnamefont {Kennes}},\ }\href@noop {} {\enquote {\bibinfo {title} {Theory
  of nematic charge orders in kagome metals},}\ } (\bibinfo {year} {2023}),\
  \Eprint {http://arxiv.org/abs/2302.01615} {arXiv:2302.01615} \BibitemShut
  {NoStop}%
\bibitem [{\citenamefont {Xu}\ \emph {et~al.}(2022)\citenamefont {Xu},
  \citenamefont {Ni}, \citenamefont {Liu}, \citenamefont {Ortiz}, \citenamefont
  {Deng}, \citenamefont {Wilson}, \citenamefont {Yan}, \citenamefont
  {Balents},\ and\ \citenamefont {Wu}}]{Xu2022n}%
  \BibitemOpen
  \bibfield  {author} {\bibinfo {author} {\bibfnamefont {Yishuai}\ \bibnamefont
  {Xu}}, \bibinfo {author} {\bibfnamefont {Zhuoliang}\ \bibnamefont {Ni}},
  \bibinfo {author} {\bibfnamefont {Yizhou}\ \bibnamefont {Liu}}, \bibinfo
  {author} {\bibfnamefont {Brenden~R.}\ \bibnamefont {Ortiz}}, \bibinfo
  {author} {\bibfnamefont {Qinwen}\ \bibnamefont {Deng}}, \bibinfo {author}
  {\bibfnamefont {Stephen~D.}\ \bibnamefont {Wilson}}, \bibinfo {author}
  {\bibfnamefont {Binghai}\ \bibnamefont {Yan}}, \bibinfo {author}
  {\bibfnamefont {Leon}\ \bibnamefont {Balents}}, \ and\ \bibinfo {author}
  {\bibfnamefont {Liang}\ \bibnamefont {Wu}},\ }\bibfield  {title} {\enquote
  {\bibinfo {title} {Three-state nematicity and magneto-optical kerr effect in
  the charge density waves in kagome superconductors},}\ }\href {\doibase
  10.1038/s41567-022-01805-7 } {\bibfield  {journal} {\bibinfo  {journal}
  {Nature Physics}\ }\textbf {\bibinfo {volume} {18}},\ \bibinfo {pages}
  {1470--1475} (\bibinfo {year} {2022})}\BibitemShut {NoStop}%
\bibitem [{\citenamefont {Saykin}\ \emph {et~al.}(2023)\citenamefont {Saykin},
  \citenamefont {Farhang}, \citenamefont {Kountz}, \citenamefont {Chen},
  \citenamefont {Ortiz}, \citenamefont {Shekhar}, \citenamefont {Felser},
  \citenamefont {Wilson}, \citenamefont {Thomale}, \citenamefont {Xia},\ and\
  \citenamefont {Kapitulnik}}]{Saykin2022}%
  \BibitemOpen
  \bibfield  {author} {\bibinfo {author} {\bibfnamefont {David~R.}\
  \bibnamefont {Saykin}}, \bibinfo {author} {\bibfnamefont {Camron}\
  \bibnamefont {Farhang}}, \bibinfo {author} {\bibfnamefont {Erik~D.}\
  \bibnamefont {Kountz}}, \bibinfo {author} {\bibfnamefont {Dong}\ \bibnamefont
  {Chen}}, \bibinfo {author} {\bibfnamefont {Brenden~R.}\ \bibnamefont
  {Ortiz}}, \bibinfo {author} {\bibfnamefont {Chandra}\ \bibnamefont
  {Shekhar}}, \bibinfo {author} {\bibfnamefont {Claudia}\ \bibnamefont
  {Felser}}, \bibinfo {author} {\bibfnamefont {Stephen~D.}\ \bibnamefont
  {Wilson}}, \bibinfo {author} {\bibfnamefont {Ronny}\ \bibnamefont {Thomale}},
  \bibinfo {author} {\bibfnamefont {Jing}\ \bibnamefont {Xia}}, \ and\ \bibinfo
  {author} {\bibfnamefont {Aharon}\ \bibnamefont {Kapitulnik}},\ }\bibfield
  {title} {\enquote {\bibinfo {title} {High resolution polar kerr effect
  studies of ${\mathrm{csv}}_{3}{\mathrm{sb}}_{5}$: Tests for time-reversal
  symmetry breaking below the charge-order transition},}\ }\href {\doibase
  10.1103/PhysRevLett.131.016901 } {\bibfield  {journal} {\bibinfo  {journal}
  {Phys. Rev. Lett.}\ }\textbf {\bibinfo {volume} {131}},\ \bibinfo {pages}
  {016901} (\bibinfo {year} {2023})}\BibitemShut {NoStop}%
\bibitem [{\citenamefont {Hu}\ \emph {et~al.}(2022{\natexlab{b}})\citenamefont
  {Hu}, \citenamefont {Yamane}, \citenamefont {Mattoni}, \citenamefont {Yada},
  \citenamefont {Obata}, \citenamefont {Li}, \citenamefont {Yao}, \citenamefont
  {Wang}, \citenamefont {Wang}, \citenamefont {Farhang}, \citenamefont {Xia},
  \citenamefont {Maeno},\ and\ \citenamefont {Yonezawa}}]{Hu2022s}%
  \BibitemOpen
  \bibfield  {author} {\bibinfo {author} {\bibfnamefont {Yajian}\ \bibnamefont
  {Hu}}, \bibinfo {author} {\bibfnamefont {Soichiro}\ \bibnamefont {Yamane}},
  \bibinfo {author} {\bibfnamefont {Giordano}\ \bibnamefont {Mattoni}},
  \bibinfo {author} {\bibfnamefont {Kanae}\ \bibnamefont {Yada}}, \bibinfo
  {author} {\bibfnamefont {Keito}\ \bibnamefont {Obata}}, \bibinfo {author}
  {\bibfnamefont {Yongkai}\ \bibnamefont {Li}}, \bibinfo {author}
  {\bibfnamefont {Yugui}\ \bibnamefont {Yao}}, \bibinfo {author} {\bibfnamefont
  {Zhiwei}\ \bibnamefont {Wang}}, \bibinfo {author} {\bibfnamefont {Jingyuan}\
  \bibnamefont {Wang}}, \bibinfo {author} {\bibfnamefont {Camron}\ \bibnamefont
  {Farhang}}, \bibinfo {author} {\bibfnamefont {Jing}\ \bibnamefont {Xia}},
  \bibinfo {author} {\bibfnamefont {Yoshiteru}\ \bibnamefont {Maeno}}, \ and\
  \bibinfo {author} {\bibfnamefont {Shingo}\ \bibnamefont {Yonezawa}},\
  }\href@noop {} {\enquote {\bibinfo {title} {Time-reversal symmetry breaking
  in charge density wave of csv$_3$sb$_5$ detected by polar kerr effect},}\ }
  (\bibinfo {year} {2022}{\natexlab{b}}),\ \Eprint
  {http://arxiv.org/abs/2208.08036} {arXiv:2208.08036} \BibitemShut {NoStop}%
\bibitem [{\citenamefont {Kang}\ \emph
  {et~al.}(2022{\natexlab{b}})\citenamefont {Kang}, \citenamefont {Fang},
  \citenamefont {Yoo}, \citenamefont {Ortiz}, \citenamefont {Oey},
  \citenamefont {Choi}, \citenamefont {Ryu}, \citenamefont {Kim}, \citenamefont
  {Jozwiak}, \citenamefont {Bostwick}, \citenamefont {Rotenberg}, \citenamefont
  {Kaxiras}, \citenamefont {Checkelsky}, \citenamefont {Wilson}, \citenamefont
  {Park},\ and\ \citenamefont {Comin}}]{Kang2022}%
  \BibitemOpen
  \bibfield  {author} {\bibinfo {author} {\bibfnamefont {Mingu}\ \bibnamefont
  {Kang}}, \bibinfo {author} {\bibfnamefont {Shiang}\ \bibnamefont {Fang}},
  \bibinfo {author} {\bibfnamefont {Jonggyu}\ \bibnamefont {Yoo}}, \bibinfo
  {author} {\bibfnamefont {Brenden~R.}\ \bibnamefont {Ortiz}}, \bibinfo
  {author} {\bibfnamefont {Yuzki~M.}\ \bibnamefont {Oey}}, \bibinfo {author}
  {\bibfnamefont {Jonghyeok}\ \bibnamefont {Choi}}, \bibinfo {author}
  {\bibfnamefont {Sae~Hee}\ \bibnamefont {Ryu}}, \bibinfo {author}
  {\bibfnamefont {Jimin}\ \bibnamefont {Kim}}, \bibinfo {author} {\bibfnamefont
  {Chris}\ \bibnamefont {Jozwiak}}, \bibinfo {author} {\bibfnamefont {Aaron}\
  \bibnamefont {Bostwick}}, \bibinfo {author} {\bibfnamefont {Eli}\
  \bibnamefont {Rotenberg}}, \bibinfo {author} {\bibfnamefont {Efthimios}\
  \bibnamefont {Kaxiras}}, \bibinfo {author} {\bibfnamefont {Joseph~G.}\
  \bibnamefont {Checkelsky}}, \bibinfo {author} {\bibfnamefont {Stephen~D.}\
  \bibnamefont {Wilson}}, \bibinfo {author} {\bibfnamefont {Jae-Hoon}\
  \bibnamefont {Park}}, \ and\ \bibinfo {author} {\bibfnamefont {Riccardo}\
  \bibnamefont {Comin}},\ }\bibfield  {title} {\enquote {\bibinfo {title}
  {Charge order landscape and competition with superconductivity in kagome
  metals},}\ }\href {\doibase 10.1038/s41563-022-01375-2 l} {\bibfield
  {journal} {\bibinfo  {journal} {Nature Materials}\ } (\bibinfo {year}
  {2022}{\natexlab{b}}),\ 10.1038/s41563-022-01375-2 l}\BibitemShut {NoStop}%
\bibitem [{\citenamefont {Wang}\ \emph {et~al.}(2023)\citenamefont {Wang},
  \citenamefont {Farhang}, \citenamefont {Ortiz}, \citenamefont {Wilson},\ and\
  \citenamefont {Xia}}]{Wang2023dk}%
  \BibitemOpen
  \bibfield  {author} {\bibinfo {author} {\bibfnamefont {Jingyuan}\
  \bibnamefont {Wang}}, \bibinfo {author} {\bibfnamefont {Camron}\ \bibnamefont
  {Farhang}}, \bibinfo {author} {\bibfnamefont {Brenden~R.}\ \bibnamefont
  {Ortiz}}, \bibinfo {author} {\bibfnamefont {Stephen~D.}\ \bibnamefont
  {Wilson}}, \ and\ \bibinfo {author} {\bibfnamefont {Jing}\ \bibnamefont
  {Xia}},\ }\href@noop {} {\enquote {\bibinfo {title} {Resolving the
  discrepancy between moke measurements at 1550-nm wavelength on kagome metal
  csv3sb5},}\ } (\bibinfo {year} {2023}),\ \Eprint
  {http://arxiv.org/abs/2301.08853} {arXiv:2301.08853 } \BibitemShut {NoStop}%
\bibitem [{\citenamefont {Wilson}\ and\ \citenamefont
  {Ortiz}(2023)}]{wilson2023k}%
  \BibitemOpen
  \bibfield  {author} {\bibinfo {author} {\bibfnamefont {Stephen~D.}\
  \bibnamefont {Wilson}}\ and\ \bibinfo {author} {\bibfnamefont {Brenden~R.}\
  \bibnamefont {Ortiz}},\ }\href@noop {} {\enquote {\bibinfo {title}
  {Av$_3$sb$_5$ kagome superconductors: Progress and future directions},}\ }
  (\bibinfo {year} {2023}),\ \Eprint {http://arxiv.org/abs/2311.05946}
  {arXiv:2311.05946 } \BibitemShut {NoStop}%
\bibitem [{\citenamefont {Ortiz}\ \emph {et~al.}(2020)\citenamefont {Ortiz},
  \citenamefont {Teicher}, \citenamefont {Hu}, \citenamefont {Zuo},
  \citenamefont {Sarte}, \citenamefont {Schueller}, \citenamefont {Abeykoon},
  \citenamefont {Krogstad}, \citenamefont {Rosenkranz}, \citenamefont {Osborn},
  \citenamefont {Seshadri}, \citenamefont {Balents}, \citenamefont {He},\ and\
  \citenamefont {Wilson}}]{Ortiz2020k}%
  \BibitemOpen
  \bibfield  {author} {\bibinfo {author} {\bibfnamefont {Brenden~R.}\
  \bibnamefont {Ortiz}}, \bibinfo {author} {\bibfnamefont {Samuel M.~L.}\
  \bibnamefont {Teicher}}, \bibinfo {author} {\bibfnamefont {Yong}\
  \bibnamefont {Hu}}, \bibinfo {author} {\bibfnamefont {Julia~L.}\ \bibnamefont
  {Zuo}}, \bibinfo {author} {\bibfnamefont {Paul~M.}\ \bibnamefont {Sarte}},
  \bibinfo {author} {\bibfnamefont {Emily~C.}\ \bibnamefont {Schueller}},
  \bibinfo {author} {\bibfnamefont {A.~M.~Milinda}\ \bibnamefont {Abeykoon}},
  \bibinfo {author} {\bibfnamefont {Matthew~J.}\ \bibnamefont {Krogstad}},
  \bibinfo {author} {\bibfnamefont {Stephan}\ \bibnamefont {Rosenkranz}},
  \bibinfo {author} {\bibfnamefont {Raymond}\ \bibnamefont {Osborn}}, \bibinfo
  {author} {\bibfnamefont {Ram}\ \bibnamefont {Seshadri}}, \bibinfo {author}
  {\bibfnamefont {Leon}\ \bibnamefont {Balents}}, \bibinfo {author}
  {\bibfnamefont {Junfeng}\ \bibnamefont {He}}, \ and\ \bibinfo {author}
  {\bibfnamefont {Stephen~D.}\ \bibnamefont {Wilson}},\ }\bibfield  {title}
  {\enquote {\bibinfo {title} {$\mathrm{Cs}{\mathrm{v}}_{3}{\mathrm{sb}}_{5}$:
  A ${\mathbb{z}}_{2}$ topological kagome metal with a superconducting ground
  state},}\ }\href {\doibase 10.1103/PhysRevLett.125.247002 } {\bibfield
  {journal} {\bibinfo  {journal} {Phys. Rev. Lett.}\ }\textbf {\bibinfo
  {volume} {125}},\ \bibinfo {pages} {247002} (\bibinfo {year}
  {2020})}\BibitemShut {NoStop}%
\bibitem [{\citenamefont {Ortiz}\ \emph
  {et~al.}(2021{\natexlab{a}})\citenamefont {Ortiz}, \citenamefont {Sarte},
  \citenamefont {Kenney}, \citenamefont {Graf}, \citenamefont {Teicher},
  \citenamefont {Seshadri},\ and\ \citenamefont {Wilson}}]{Ortiz2021}%
  \BibitemOpen
  \bibfield  {author} {\bibinfo {author} {\bibfnamefont {Brenden~R.}\
  \bibnamefont {Ortiz}}, \bibinfo {author} {\bibfnamefont {Paul~M.}\
  \bibnamefont {Sarte}}, \bibinfo {author} {\bibfnamefont {Eric~M.}\
  \bibnamefont {Kenney}}, \bibinfo {author} {\bibfnamefont {Michael~J.}\
  \bibnamefont {Graf}}, \bibinfo {author} {\bibfnamefont {Samuel M.~L.}\
  \bibnamefont {Teicher}}, \bibinfo {author} {\bibfnamefont {Ram}\ \bibnamefont
  {Seshadri}}, \ and\ \bibinfo {author} {\bibfnamefont {Stephen~D.}\
  \bibnamefont {Wilson}},\ }\bibfield  {title} {\enquote {\bibinfo {title}
  {Superconductivity in the ${\mathbb{z}}_{2}$ kagome metal
  ${\mathrm{kv}}_{3}{\mathrm{sb}}_{5}$},}\ }\href {\doibase
  10.1103/PhysRevMaterials.5.034801} {\bibfield  {journal} {\bibinfo  {journal}
  {Phys. Rev. Mater.}\ }\textbf {\bibinfo {volume} {5}},\ \bibinfo {pages}
  {034801} (\bibinfo {year} {2021}{\natexlab{a}})}\BibitemShut {NoStop}%
\bibitem [{\citenamefont {Chen}\ \emph
  {et~al.}(2021{\natexlab{a}})\citenamefont {Chen}, \citenamefont {Wang},
  \citenamefont {Yin}, \citenamefont {Gu}, \citenamefont {Jiang}, \citenamefont
  {Tu}, \citenamefont {Gong}, \citenamefont {Uwatoko}, \citenamefont {Sun},
  \citenamefont {Lei}, \citenamefont {Hu},\ and\ \citenamefont
  {Cheng}}]{Chen2021}%
  \BibitemOpen
  \bibfield  {author} {\bibinfo {author} {\bibfnamefont {K.~Y.}\ \bibnamefont
  {Chen}}, \bibinfo {author} {\bibfnamefont {N.~N.}\ \bibnamefont {Wang}},
  \bibinfo {author} {\bibfnamefont {Q.~W.}\ \bibnamefont {Yin}}, \bibinfo
  {author} {\bibfnamefont {Y.~H.}\ \bibnamefont {Gu}}, \bibinfo {author}
  {\bibfnamefont {K.}~\bibnamefont {Jiang}}, \bibinfo {author} {\bibfnamefont
  {Z.~J.}\ \bibnamefont {Tu}}, \bibinfo {author} {\bibfnamefont {C.~S.}\
  \bibnamefont {Gong}}, \bibinfo {author} {\bibfnamefont {Y.}~\bibnamefont
  {Uwatoko}}, \bibinfo {author} {\bibfnamefont {J.~P.}\ \bibnamefont {Sun}},
  \bibinfo {author} {\bibfnamefont {H.~C.}\ \bibnamefont {Lei}}, \bibinfo
  {author} {\bibfnamefont {J.~P.}\ \bibnamefont {Hu}}, \ and\ \bibinfo {author}
  {\bibfnamefont {J.-G.}\ \bibnamefont {Cheng}},\ }\bibfield  {title} {\enquote
  {\bibinfo {title} {Double superconducting dome and triple enhancement of
  ${T}_{c}$ in the kagome superconductor ${\mathrm{csv}}_{3}{\mathrm{sb}}_{5}$
  under high pressure},}\ }\href {\doibase 10.1103/PhysRevLett.126.247001}
  {\bibfield  {journal} {\bibinfo  {journal} {Phys. Rev. Lett.}\ }\textbf
  {\bibinfo {volume} {126}},\ \bibinfo {pages} {247001} (\bibinfo {year}
  {2021}{\natexlab{a}})}\BibitemShut {NoStop}%
\bibitem [{\citenamefont {Chen}\ \emph
  {et~al.}(2021{\natexlab{b}})\citenamefont {Chen}, \citenamefont {Yang},
  \citenamefont {Hu}, \citenamefont {Zhao}, \citenamefont {Yuan}, \citenamefont
  {Xing}, \citenamefont {Qian}, \citenamefont {Huang}, \citenamefont {Li},
  \citenamefont {Ye}, \citenamefont {Ma}, \citenamefont {Ni}, \citenamefont
  {Zhang}, \citenamefont {Yin}, \citenamefont {Gong}, \citenamefont {Tu},
  \citenamefont {Lei}, \citenamefont {Tan}, \citenamefont {Zhou}, \citenamefont
  {Shen}, \citenamefont {Dong}, \citenamefont {Yan}, \citenamefont {Wang},\
  and\ \citenamefont {Gao}}]{Chen2021b}%
  \BibitemOpen
  \bibfield  {author} {\bibinfo {author} {\bibfnamefont {Hui}\ \bibnamefont
  {Chen}}, \bibinfo {author} {\bibfnamefont {Haitao}\ \bibnamefont {Yang}},
  \bibinfo {author} {\bibfnamefont {Bin}\ \bibnamefont {Hu}}, \bibinfo {author}
  {\bibfnamefont {Zhen}\ \bibnamefont {Zhao}}, \bibinfo {author} {\bibfnamefont
  {Jie}\ \bibnamefont {Yuan}}, \bibinfo {author} {\bibfnamefont {Yuqing}\
  \bibnamefont {Xing}}, \bibinfo {author} {\bibfnamefont {Guojian}\
  \bibnamefont {Qian}}, \bibinfo {author} {\bibfnamefont {Zihao}\ \bibnamefont
  {Huang}}, \bibinfo {author} {\bibfnamefont {Geng}\ \bibnamefont {Li}},
  \bibinfo {author} {\bibfnamefont {Yuhan}\ \bibnamefont {Ye}}, \bibinfo
  {author} {\bibfnamefont {Sheng}\ \bibnamefont {Ma}}, \bibinfo {author}
  {\bibfnamefont {Shunli}\ \bibnamefont {Ni}}, \bibinfo {author} {\bibfnamefont
  {Hua}\ \bibnamefont {Zhang}}, \bibinfo {author} {\bibfnamefont {Qiangwei}\
  \bibnamefont {Yin}}, \bibinfo {author} {\bibfnamefont {Chunsheng}\
  \bibnamefont {Gong}}, \bibinfo {author} {\bibfnamefont {Zhijun}\ \bibnamefont
  {Tu}}, \bibinfo {author} {\bibfnamefont {Hechang}\ \bibnamefont {Lei}},
  \bibinfo {author} {\bibfnamefont {Hengxin}\ \bibnamefont {Tan}}, \bibinfo
  {author} {\bibfnamefont {Sen}\ \bibnamefont {Zhou}}, \bibinfo {author}
  {\bibfnamefont {Chengmin}\ \bibnamefont {Shen}}, \bibinfo {author}
  {\bibfnamefont {Xiaoli}\ \bibnamefont {Dong}}, \bibinfo {author}
  {\bibfnamefont {Binghai}\ \bibnamefont {Yan}}, \bibinfo {author}
  {\bibfnamefont {Ziqiang}\ \bibnamefont {Wang}}, \ and\ \bibinfo {author}
  {\bibfnamefont {Hong-Jun}\ \bibnamefont {Gao}},\ }\bibfield  {title}
  {\enquote {\bibinfo {title} {Roton pair density wave in a strong-coupling
  kagome superconductor},}\ }\href {\doibase 10.1038/s41586-021-03983-5 }
  {\bibfield  {journal} {\bibinfo  {journal} {Nature}\ }\textbf {\bibinfo
  {volume} {599}},\ \bibinfo {pages} {222--228} (\bibinfo {year}
  {2021}{\natexlab{b}})}\BibitemShut {NoStop}%
\bibitem [{\citenamefont {Ge}\ \emph {et~al.}(2022)\citenamefont {Ge},
  \citenamefont {Wang}, \citenamefont {Xing}, \citenamefont {Yin},
  \citenamefont {Lei}, \citenamefont {Wang},\ and\ \citenamefont
  {Wang}}]{Ge2022c}%
  \BibitemOpen
  \bibfield  {author} {\bibinfo {author} {\bibfnamefont {Jun}\ \bibnamefont
  {Ge}}, \bibinfo {author} {\bibfnamefont {Pinyuan}\ \bibnamefont {Wang}},
  \bibinfo {author} {\bibfnamefont {Ying}\ \bibnamefont {Xing}}, \bibinfo
  {author} {\bibfnamefont {Qiangwei}\ \bibnamefont {Yin}}, \bibinfo {author}
  {\bibfnamefont {Hechang}\ \bibnamefont {Lei}}, \bibinfo {author}
  {\bibfnamefont {Ziqiang}\ \bibnamefont {Wang}}, \ and\ \bibinfo {author}
  {\bibfnamefont {Jian}\ \bibnamefont {Wang}},\ }\href@noop {} {\enquote
  {\bibinfo {title} {Discovery of charge-4e and charge-6e superconductivity in
  kagome superconductor csv3sb5},}\ } (\bibinfo {year} {2022}),\ \Eprint
  {http://arxiv.org/abs/2201.10352} {arXiv:2201.10352 } \BibitemShut {NoStop}%
\bibitem [{\citenamefont {Ortiz}\ \emph {et~al.}(2019)\citenamefont {Ortiz},
  \citenamefont {Gomes}, \citenamefont {Morey}, \citenamefont {Winiarski},
  \citenamefont {Bordelon}, \citenamefont {Mangum}, \citenamefont {Oswald},
  \citenamefont {Rodriguez-Rivera}, \citenamefont {Neilson}, \citenamefont
  {Wilson}, \citenamefont {Ertekin}, \citenamefont {McQueen},\ and\
  \citenamefont {Toberer}}]{Ortiz2019}%
  \BibitemOpen
  \bibfield  {author} {\bibinfo {author} {\bibfnamefont {Brenden~R.}\
  \bibnamefont {Ortiz}}, \bibinfo {author} {\bibfnamefont {L\'{\i}dia~C.}\
  \bibnamefont {Gomes}}, \bibinfo {author} {\bibfnamefont {Jennifer~R.}\
  \bibnamefont {Morey}}, \bibinfo {author} {\bibfnamefont {Michal}\
  \bibnamefont {Winiarski}}, \bibinfo {author} {\bibfnamefont {Mitchell}\
  \bibnamefont {Bordelon}}, \bibinfo {author} {\bibfnamefont {John~S.}\
  \bibnamefont {Mangum}}, \bibinfo {author} {\bibfnamefont {Iain W.~H.}\
  \bibnamefont {Oswald}}, \bibinfo {author} {\bibfnamefont {Jose~A.}\
  \bibnamefont {Rodriguez-Rivera}}, \bibinfo {author} {\bibfnamefont
  {James~R.}\ \bibnamefont {Neilson}}, \bibinfo {author} {\bibfnamefont
  {Stephen~D.}\ \bibnamefont {Wilson}}, \bibinfo {author} {\bibfnamefont
  {Elif}\ \bibnamefont {Ertekin}}, \bibinfo {author} {\bibfnamefont {Tyrel~M.}\
  \bibnamefont {McQueen}}, \ and\ \bibinfo {author} {\bibfnamefont {Eric~S.}\
  \bibnamefont {Toberer}},\ }\bibfield  {title} {\enquote {\bibinfo {title}
  {New kagome prototype materials: discovery of
  ${\mathrm{kv}}_{3}{\mathrm{sb}}_{5},{\mathrm{rbv}}_{3}{\mathrm{sb}}_{5}$, and
  ${\mathrm{csv}}_{3}{\mathrm{sb}}_{5}$},}\ }\href {\doibase
  10.1103/PhysRevMaterials.3.094407} {\bibfield  {journal} {\bibinfo  {journal}
  {Phys. Rev. Mater.}\ }\textbf {\bibinfo {volume} {3}},\ \bibinfo {pages}
  {094407} (\bibinfo {year} {2019})}\BibitemShut {NoStop}%
\bibitem [{\citenamefont {Shumiya}\ \emph {et~al.}(2021)\citenamefont
  {Shumiya}, \citenamefont {Hossain}, \citenamefont {Yin}, \citenamefont
  {Jiang}, \citenamefont {Ortiz}, \citenamefont {Liu}, \citenamefont {Shi},
  \citenamefont {Yin}, \citenamefont {Lei}, \citenamefont {Zhang},
  \citenamefont {Chang}, \citenamefont {Zhang}, \citenamefont {Cochran},
  \citenamefont {Multer}, \citenamefont {Litskevich}, \citenamefont {Cheng},
  \citenamefont {Yang}, \citenamefont {Guguchia}, \citenamefont {Wilson},\ and\
  \citenamefont {Hasan}}]{Shumiya2021}%
  \BibitemOpen
  \bibfield  {author} {\bibinfo {author} {\bibfnamefont {Nana}\ \bibnamefont
  {Shumiya}}, \bibinfo {author} {\bibfnamefont {Md.~Shafayat}\ \bibnamefont
  {Hossain}}, \bibinfo {author} {\bibfnamefont {Jia-Xin}\ \bibnamefont {Yin}},
  \bibinfo {author} {\bibfnamefont {Yu-Xiao}\ \bibnamefont {Jiang}}, \bibinfo
  {author} {\bibfnamefont {Brenden~R.}\ \bibnamefont {Ortiz}}, \bibinfo
  {author} {\bibfnamefont {Hongxiong}\ \bibnamefont {Liu}}, \bibinfo {author}
  {\bibfnamefont {Youguo}\ \bibnamefont {Shi}}, \bibinfo {author}
  {\bibfnamefont {Qiangwei}\ \bibnamefont {Yin}}, \bibinfo {author}
  {\bibfnamefont {Hechang}\ \bibnamefont {Lei}}, \bibinfo {author}
  {\bibfnamefont {Songtian~S.}\ \bibnamefont {Zhang}}, \bibinfo {author}
  {\bibfnamefont {Guoqing}\ \bibnamefont {Chang}}, \bibinfo {author}
  {\bibfnamefont {Qi}~\bibnamefont {Zhang}}, \bibinfo {author} {\bibfnamefont
  {Tyler~A.}\ \bibnamefont {Cochran}}, \bibinfo {author} {\bibfnamefont
  {Daniel}\ \bibnamefont {Multer}}, \bibinfo {author} {\bibfnamefont {Maksim}\
  \bibnamefont {Litskevich}}, \bibinfo {author} {\bibfnamefont {Zi-Jia}\
  \bibnamefont {Cheng}}, \bibinfo {author} {\bibfnamefont {Xian~P.}\
  \bibnamefont {Yang}}, \bibinfo {author} {\bibfnamefont {Zurab}\ \bibnamefont
  {Guguchia}}, \bibinfo {author} {\bibfnamefont {Stephen~D.}\ \bibnamefont
  {Wilson}}, \ and\ \bibinfo {author} {\bibfnamefont {M.~Zahid}\ \bibnamefont
  {Hasan}},\ }\bibfield  {title} {\enquote {\bibinfo {title} {Intrinsic nature
  of chiral charge order in the kagome superconductor
  $\mathrm{Rb}{\mathrm{v}}_{3}{\mathrm{sb}}_{5}$},}\ }\href {\doibase
  10.1103/PhysRevB.104.035131} {\bibfield  {journal} {\bibinfo  {journal}
  {Phys. Rev. B}\ }\textbf {\bibinfo {volume} {104}},\ \bibinfo {pages}
  {035131} (\bibinfo {year} {2021})}\BibitemShut {NoStop}%
\bibitem [{\citenamefont {Ortiz}\ \emph
  {et~al.}(2021{\natexlab{b}})\citenamefont {Ortiz}, \citenamefont {Teicher},
  \citenamefont {Kautzsch}, \citenamefont {Sarte}, \citenamefont {Ratcliff},
  \citenamefont {Harter}, \citenamefont {Ruff}, \citenamefont {Seshadri},\ and\
  \citenamefont {Wilson}}]{Ortiz2021sm}%
  \BibitemOpen
  \bibfield  {author} {\bibinfo {author} {\bibfnamefont {Brenden~R.}\
  \bibnamefont {Ortiz}}, \bibinfo {author} {\bibfnamefont {Samuel M.~L.}\
  \bibnamefont {Teicher}}, \bibinfo {author} {\bibfnamefont {Linus}\
  \bibnamefont {Kautzsch}}, \bibinfo {author} {\bibfnamefont {Paul~M.}\
  \bibnamefont {Sarte}}, \bibinfo {author} {\bibfnamefont {Noah}\ \bibnamefont
  {Ratcliff}}, \bibinfo {author} {\bibfnamefont {John}\ \bibnamefont {Harter}},
  \bibinfo {author} {\bibfnamefont {Jacob P.~C.}\ \bibnamefont {Ruff}},
  \bibinfo {author} {\bibfnamefont {Ram}\ \bibnamefont {Seshadri}}, \ and\
  \bibinfo {author} {\bibfnamefont {Stephen~D.}\ \bibnamefont {Wilson}},\
  }\bibfield  {title} {\enquote {\bibinfo {title} {Fermi surface mapping and
  the nature of charge-density-wave order in the kagome superconductor
  ${\mathrm{csv}}_{3}{\mathrm{sb}}_{5}$},}\ }\href {\doibase
  10.1103/PhysRevX.11.041030} {\bibfield  {journal} {\bibinfo  {journal} {Phys.
  Rev. X}\ }\textbf {\bibinfo {volume} {11}},\ \bibinfo {pages} {041030}
  (\bibinfo {year} {2021}{\natexlab{b}})}\BibitemShut {NoStop}%
\bibitem [{\citenamefont {Si}\ \emph {et~al.}(2022)\citenamefont {Si},
  \citenamefont {Lu}, \citenamefont {Sun}, \citenamefont {Liu},\ and\
  \citenamefont {Wang}}]{Si2022}%
  \BibitemOpen
  \bibfield  {author} {\bibinfo {author} {\bibfnamefont {Jian-Guo}\
  \bibnamefont {Si}}, \bibinfo {author} {\bibfnamefont {Wen-Jian}\ \bibnamefont
  {Lu}}, \bibinfo {author} {\bibfnamefont {Yu-Ping}\ \bibnamefont {Sun}},
  \bibinfo {author} {\bibfnamefont {Peng-Fei}\ \bibnamefont {Liu}}, \ and\
  \bibinfo {author} {\bibfnamefont {Bao-Tian}\ \bibnamefont {Wang}},\
  }\bibfield  {title} {\enquote {\bibinfo {title} {Charge density wave and
  pressure-dependent superconductivity in the kagome metal
  ${\mathrm{csv}}_{3}{\mathrm{sb}}_{5}$: A first-principles study},}\ }\href
  {\doibase 10.1103/PhysRevB.105.024517} {\bibfield  {journal} {\bibinfo
  {journal} {Phys. Rev. B}\ }\textbf {\bibinfo {volume} {105}},\ \bibinfo
  {pages} {024517} (\bibinfo {year} {2022})}\BibitemShut {NoStop}%
\bibitem [{\citenamefont {Song}\ \emph {et~al.}(2022)\citenamefont {Song},
  \citenamefont {Zheng}, \citenamefont {Yu}, \citenamefont {Li}, \citenamefont
  {Nie}, \citenamefont {Shan}, \citenamefont {Zhao}, \citenamefont {Li},
  \citenamefont {Kang}, \citenamefont {Wu}, \citenamefont {Zhou}, \citenamefont
  {Sun}, \citenamefont {Liu}, \citenamefont {Luo}, \citenamefont {Wang},
  \citenamefont {Ying}, \citenamefont {Wan}, \citenamefont {Wu},\ and\
  \citenamefont {Chen}}]{Song2022}%
  \BibitemOpen
  \bibfield  {author} {\bibinfo {author} {\bibfnamefont {DianWu}\ \bibnamefont
  {Song}}, \bibinfo {author} {\bibfnamefont {LiXuan}\ \bibnamefont {Zheng}},
  \bibinfo {author} {\bibfnamefont {FangHang}\ \bibnamefont {Yu}}, \bibinfo
  {author} {\bibfnamefont {Jian}\ \bibnamefont {Li}}, \bibinfo {author}
  {\bibfnamefont {LinPeng}\ \bibnamefont {Nie}}, \bibinfo {author}
  {\bibfnamefont {Min}\ \bibnamefont {Shan}}, \bibinfo {author} {\bibfnamefont
  {Dan}\ \bibnamefont {Zhao}}, \bibinfo {author} {\bibfnamefont {ShunJiao}\
  \bibnamefont {Li}}, \bibinfo {author} {\bibfnamefont {BaoLei}\ \bibnamefont
  {Kang}}, \bibinfo {author} {\bibfnamefont {ZhiMian}\ \bibnamefont {Wu}},
  \bibinfo {author} {\bibfnamefont {YanBing}\ \bibnamefont {Zhou}}, \bibinfo
  {author} {\bibfnamefont {KuangLv}\ \bibnamefont {Sun}}, \bibinfo {author}
  {\bibfnamefont {Kai}\ \bibnamefont {Liu}}, \bibinfo {author} {\bibfnamefont
  {XiGang}\ \bibnamefont {Luo}}, \bibinfo {author} {\bibfnamefont {ZhenYu}\
  \bibnamefont {Wang}}, \bibinfo {author} {\bibfnamefont {JianJun}\
  \bibnamefont {Ying}}, \bibinfo {author} {\bibfnamefont {XianGang}\
  \bibnamefont {Wan}}, \bibinfo {author} {\bibfnamefont {Tao}\ \bibnamefont
  {Wu}}, \ and\ \bibinfo {author} {\bibfnamefont {XianHui}\ \bibnamefont
  {Chen}},\ }\bibfield  {title} {\enquote {\bibinfo {title} {Orbital ordering
  and fluctuations in a kagome superconductor csv3sb5},}\ }\href {\doibase
  10.1007/s11433-021-1826-1} {\bibfield  {journal} {\bibinfo  {journal}
  {Science China Physics, Mechanics {\&} Astronomy}\ }\textbf {\bibinfo
  {volume} {65}},\ \bibinfo {pages} {247462} (\bibinfo {year}
  {2022})}\BibitemShut {NoStop}%
\bibitem [{\citenamefont {Mielke}\ \emph {et~al.}(2022)\citenamefont {Mielke},
  \citenamefont {Das}, \citenamefont {Yin}, \citenamefont {Liu}, \citenamefont
  {Gupta}, \citenamefont {Jiang}, \citenamefont {Medarde}, \citenamefont {Wu},
  \citenamefont {Lei}, \citenamefont {Chang}, \citenamefont {Dai},
  \citenamefont {Si}, \citenamefont {Miao}, \citenamefont {Thomale},
  \citenamefont {Neupert}, \citenamefont {Shi}, \citenamefont {Khasanov},
  \citenamefont {Hasan}, \citenamefont {Luetkens},\ and\ \citenamefont
  {Guguchia}}]{Mielke2022}%
  \BibitemOpen
  \bibfield  {author} {\bibinfo {author} {\bibfnamefont {C.}~\bibnamefont
  {Mielke}}, \bibinfo {author} {\bibfnamefont {D.}~\bibnamefont {Das}},
  \bibinfo {author} {\bibfnamefont {J.-X.}\ \bibnamefont {Yin}}, \bibinfo
  {author} {\bibfnamefont {H.}~\bibnamefont {Liu}}, \bibinfo {author}
  {\bibfnamefont {R.}~\bibnamefont {Gupta}}, \bibinfo {author} {\bibfnamefont
  {Y.-X.}\ \bibnamefont {Jiang}}, \bibinfo {author} {\bibfnamefont
  {M.}~\bibnamefont {Medarde}}, \bibinfo {author} {\bibfnamefont
  {X.}~\bibnamefont {Wu}}, \bibinfo {author} {\bibfnamefont {H.~C.}\
  \bibnamefont {Lei}}, \bibinfo {author} {\bibfnamefont {J.}~\bibnamefont
  {Chang}}, \bibinfo {author} {\bibfnamefont {Pengcheng}\ \bibnamefont {Dai}},
  \bibinfo {author} {\bibfnamefont {Q.}~\bibnamefont {Si}}, \bibinfo {author}
  {\bibfnamefont {H.}~\bibnamefont {Miao}}, \bibinfo {author} {\bibfnamefont
  {R.}~\bibnamefont {Thomale}}, \bibinfo {author} {\bibfnamefont
  {T.}~\bibnamefont {Neupert}}, \bibinfo {author} {\bibfnamefont
  {Y.}~\bibnamefont {Shi}}, \bibinfo {author} {\bibfnamefont {R.}~\bibnamefont
  {Khasanov}}, \bibinfo {author} {\bibfnamefont {M.~Z.}\ \bibnamefont {Hasan}},
  \bibinfo {author} {\bibfnamefont {H.}~\bibnamefont {Luetkens}}, \ and\
  \bibinfo {author} {\bibfnamefont {Z.}~\bibnamefont {Guguchia}},\ }\bibfield
  {title} {\enquote {\bibinfo {title} {Time-reversal symmetry-breaking charge
  order in a kagome superconductor},}\ }\href {\doibase
  10.1038/s41586-021-04327-z} {\bibfield  {journal} {\bibinfo  {journal}
  {Nature}\ }\textbf {\bibinfo {volume} {602}},\ \bibinfo {pages} {245--250}
  (\bibinfo {year} {2022})}\BibitemShut {NoStop}%
\bibitem [{\citenamefont {Kenney}\ \emph {et~al.}(2021)\citenamefont {Kenney},
  \citenamefont {Ortiz}, \citenamefont {Wang}, \citenamefont {Wilson},\ and\
  \citenamefont {Graf}}]{Kenney_2021}%
  \BibitemOpen
  \bibfield  {author} {\bibinfo {author} {\bibfnamefont {Eric~M}\ \bibnamefont
  {Kenney}}, \bibinfo {author} {\bibfnamefont {Brenden~R}\ \bibnamefont
  {Ortiz}}, \bibinfo {author} {\bibfnamefont {Chennan}\ \bibnamefont {Wang}},
  \bibinfo {author} {\bibfnamefont {Stephen~D}\ \bibnamefont {Wilson}}, \ and\
  \bibinfo {author} {\bibfnamefont {Michael~J}\ \bibnamefont {Graf}},\
  }\bibfield  {title} {\enquote {\bibinfo {title} {Absence of local moments in
  the kagome metal kv3sb5 as determined by muon spin spectroscopy},}\ }\href
  {\doibase 10.1088/1361-648X/ abe8f9} {\bibfield  {journal} {\bibinfo
  {journal} {Journal of Physics: Condensed Matter}\ }\textbf {\bibinfo {volume}
  {33}},\ \bibinfo {pages} {235801} (\bibinfo {year} {2021})}\BibitemShut
  {NoStop}%
\bibitem [{\citenamefont {Lin}\ and\ \citenamefont
  {Nandkishore}(2019)}]{Lin2019lco}%
  \BibitemOpen
  \bibfield  {author} {\bibinfo {author} {\bibfnamefont {Yu-Ping}\ \bibnamefont
  {Lin}}\ and\ \bibinfo {author} {\bibfnamefont {Rahul~M.}\ \bibnamefont
  {Nandkishore}},\ }\bibfield  {title} {\enquote {\bibinfo {title} {Chiral
  twist on the high-${T}_{c}$ phase diagram in moir\'e heterostructures},}\
  }\href {\doibase 10.1103/PhysRevB.100.085136} {\bibfield  {journal} {\bibinfo
   {journal} {Phys. Rev. B}\ }\textbf {\bibinfo {volume} {100}},\ \bibinfo
  {pages} {085136} (\bibinfo {year} {2019})}\BibitemShut {NoStop}%
\bibitem [{\citenamefont {Yang}\ \emph {et~al.}(2022)\citenamefont {Yang},
  \citenamefont {Kim}, \citenamefont {Jeong}, \citenamefont {Kim},
  \citenamefont {Han},\ and\ \citenamefont {Lee}}]{Yang2022c}%
  \BibitemOpen
  \bibfield  {author} {\bibinfo {author} {\bibfnamefont {Hyeok-Jun}\
  \bibnamefont {Yang}}, \bibinfo {author} {\bibfnamefont {Hee~Seung}\
  \bibnamefont {Kim}}, \bibinfo {author} {\bibfnamefont {Min~Yong}\
  \bibnamefont {Jeong}}, \bibinfo {author} {\bibfnamefont {Yong~Baek}\
  \bibnamefont {Kim}}, \bibinfo {author} {\bibfnamefont {Myung~Joon}\
  \bibnamefont {Han}}, \ and\ \bibinfo {author} {\bibfnamefont {SungBin}\
  \bibnamefont {Lee}},\ }\href@noop {} {\enquote {\bibinfo {title}
  {Intertwining orbital current order and superconductivity in kagome metal},}\
  } (\bibinfo {year} {2022}),\ \Eprint {http://arxiv.org/abs/2203.07365}
  {arXiv:2203.07365} \BibitemShut {NoStop}%
\bibitem [{\citenamefont {Dong}\ \emph {et~al.}(2023)\citenamefont {Dong},
  \citenamefont {Wang},\ and\ \citenamefont {Zhou}}]{Dong2022}%
  \BibitemOpen
  \bibfield  {author} {\bibinfo {author} {\bibfnamefont {Jin-Wei}\ \bibnamefont
  {Dong}}, \bibinfo {author} {\bibfnamefont {Ziqiang}\ \bibnamefont {Wang}}, \
  and\ \bibinfo {author} {\bibfnamefont {Sen}\ \bibnamefont {Zhou}},\
  }\bibfield  {title} {\enquote {\bibinfo {title} {Loop-current charge density
  wave driven by long-range coulomb repulsion on the kagom\'e lattice},}\
  }\href {\doibase 10.1103/PhysRevB.107.045127  } {\bibfield  {journal} {\bibinfo
   {journal} {Phys. Rev. B}\ }\textbf {\bibinfo {volume} {107}},\ \bibinfo
  {pages} {045127} (\bibinfo {year} {2023})}\BibitemShut {NoStop}%
\bibitem [{\citenamefont {Nie}\ \emph {et~al.}(2022)\citenamefont {Nie},
  \citenamefont {Sun}, \citenamefont {Ma}, \citenamefont {Song}, \citenamefont
  {Zheng}, \citenamefont {Liang}, \citenamefont {Wu}, \citenamefont {Yu},
  \citenamefont {Li}, \citenamefont {Shan}, \citenamefont {Zhao}, \citenamefont
  {Li}, \citenamefont {Kang}, \citenamefont {Wu}, \citenamefont {Zhou},
  \citenamefont {Liu}, \citenamefont {Xiang}, \citenamefont {Ying},
  \citenamefont {Wang}, \citenamefont {Wu},\ and\ \citenamefont
  {Chen}}]{Nie2022}%
  \BibitemOpen
  \bibfield  {author} {\bibinfo {author} {\bibfnamefont {Linpeng}\ \bibnamefont
  {Nie}}, \bibinfo {author} {\bibfnamefont {Kuanglv}\ \bibnamefont {Sun}},
  \bibinfo {author} {\bibfnamefont {Wanru}\ \bibnamefont {Ma}}, \bibinfo
  {author} {\bibfnamefont {Dianwu}\ \bibnamefont {Song}}, \bibinfo {author}
  {\bibfnamefont {Lixuan}\ \bibnamefont {Zheng}}, \bibinfo {author}
  {\bibfnamefont {Zuowei}\ \bibnamefont {Liang}}, \bibinfo {author}
  {\bibfnamefont {Ping}\ \bibnamefont {Wu}}, \bibinfo {author} {\bibfnamefont
  {Fanghang}\ \bibnamefont {Yu}}, \bibinfo {author} {\bibfnamefont {Jian}\
  \bibnamefont {Li}}, \bibinfo {author} {\bibfnamefont {Min}\ \bibnamefont
  {Shan}}, \bibinfo {author} {\bibfnamefont {Dan}\ \bibnamefont {Zhao}},
  \bibinfo {author} {\bibfnamefont {Shunjiao}\ \bibnamefont {Li}}, \bibinfo
  {author} {\bibfnamefont {Baolei}\ \bibnamefont {Kang}}, \bibinfo {author}
  {\bibfnamefont {Zhimian}\ \bibnamefont {Wu}}, \bibinfo {author}
  {\bibfnamefont {Yanbing}\ \bibnamefont {Zhou}}, \bibinfo {author}
  {\bibfnamefont {Kai}\ \bibnamefont {Liu}}, \bibinfo {author} {\bibfnamefont
  {Ziji}\ \bibnamefont {Xiang}}, \bibinfo {author} {\bibfnamefont {Jianjun}\
  \bibnamefont {Ying}}, \bibinfo {author} {\bibfnamefont {Zhenyu}\ \bibnamefont
  {Wang}}, \bibinfo {author} {\bibfnamefont {Tao}\ \bibnamefont {Wu}}, \ and\
  \bibinfo {author} {\bibfnamefont {Xianhui}\ \bibnamefont {Chen}},\ }\bibfield
   {title} {\enquote {\bibinfo {title} {Charge-density-wave-driven electronic
  nematicity in a kagome superconductor},}\ }\href {\doibase
  10.1038/s41586-022-04493-8 } {\bibfield  {journal} {\bibinfo  {journal}
  {Nature}\ }\textbf {\bibinfo {volume} {604}},\ \bibinfo {pages} {59--64}
  (\bibinfo {year} {2022})}\BibitemShut {NoStop}%
\bibitem [{\citenamefont {LaBollita}\ and\ \citenamefont
  {Botana}(2021)}]{Labollita2021}%
  \BibitemOpen
  \bibfield  {author} {\bibinfo {author} {\bibfnamefont {Harrison}\
  \bibnamefont {LaBollita}}\ and\ \bibinfo {author} {\bibfnamefont {Antia~S.}\
  \bibnamefont {Botana}},\ }\bibfield  {title} {\enquote {\bibinfo {title}
  {Tuning the van hove singularities in $a{\mathrm{v}}_{3}{\mathrm{sb}}_{5}$
  $(a=\mathrm{K},\mathrm{Rb},\mathrm{Cs})$ via pressure and doping},}\ }\href
  {\doibase 10.1103/PhysRevB.104.205129} {\bibfield  {journal} {\bibinfo
  {journal} {Phys. Rev. B}\ }\textbf {\bibinfo {volume} {104}},\ \bibinfo
  {pages} {205129} (\bibinfo {year} {2021})}\BibitemShut {NoStop}%
\bibitem [{\citenamefont {Tan}\ \emph {et~al.}(2021)\citenamefont {Tan},
  \citenamefont {Liu}, \citenamefont {Wang},\ and\ \citenamefont
  {Yan}}]{Tan2021d}%
  \BibitemOpen
  \bibfield  {author} {\bibinfo {author} {\bibfnamefont {Hengxin}\ \bibnamefont
  {Tan}}, \bibinfo {author} {\bibfnamefont {Yizhou}\ \bibnamefont {Liu}},
  \bibinfo {author} {\bibfnamefont {Ziqiang}\ \bibnamefont {Wang}}, \ and\
  \bibinfo {author} {\bibfnamefont {Binghai}\ \bibnamefont {Yan}},\ }\bibfield
  {title} {\enquote {\bibinfo {title} {Charge density waves and electronic
  properties of superconducting kagome metals},}\ }\href {\doibase
  10.1103/PhysRevLett.127.046401 } {\bibfield  {journal} {\bibinfo  {journal}
  {Phys. Rev. Lett.}\ }\textbf {\bibinfo {volume} {127}},\ \bibinfo {pages}
  {046401} (\bibinfo {year} {2021})}\BibitemShut {NoStop}%
\bibitem [{\citenamefont {Liang}\ \emph {et~al.}(2021)\citenamefont {Liang},
  \citenamefont {Hou}, \citenamefont {Zhang}, \citenamefont {Ma}, \citenamefont
  {Wu}, \citenamefont {Zhang}, \citenamefont {Yu}, \citenamefont {Ying},
  \citenamefont {Jiang}, \citenamefont {Shan}, \citenamefont {Wang},\ and\
  \citenamefont {Chen}}]{Liang2021c}%
  \BibitemOpen
  \bibfield  {author} {\bibinfo {author} {\bibfnamefont {Zuowei}\ \bibnamefont
  {Liang}}, \bibinfo {author} {\bibfnamefont {Xingyuan}\ \bibnamefont {Hou}},
  \bibinfo {author} {\bibfnamefont {Fan}\ \bibnamefont {Zhang}}, \bibinfo
  {author} {\bibfnamefont {Wanru}\ \bibnamefont {Ma}}, \bibinfo {author}
  {\bibfnamefont {Ping}\ \bibnamefont {Wu}}, \bibinfo {author} {\bibfnamefont
  {Zongyuan}\ \bibnamefont {Zhang}}, \bibinfo {author} {\bibfnamefont
  {Fanghang}\ \bibnamefont {Yu}}, \bibinfo {author} {\bibfnamefont {J.-J.}\
  \bibnamefont {Ying}}, \bibinfo {author} {\bibfnamefont {Kun}\ \bibnamefont
  {Jiang}}, \bibinfo {author} {\bibfnamefont {Lei}\ \bibnamefont {Shan}},
  \bibinfo {author} {\bibfnamefont {Zhenyu}\ \bibnamefont {Wang}}, \ and\
  \bibinfo {author} {\bibfnamefont {X.-H.}\ \bibnamefont {Chen}},\ }\bibfield
  {title} {\enquote {\bibinfo {title} {Three-dimensional charge density wave
  and surface-dependent vortex-core states in a kagome superconductor
  ${\mathrm{csv}}_{3}{\mathrm{sb}}_{5}$},}\ }\href {\doibase
  10.1103/PhysRevX.11.031026   } {\bibfield  {journal} {\bibinfo  {journal} {Phys.
  Rev. X}\ }\textbf {\bibinfo {volume} {11}},\ \bibinfo {pages} {031026}
  (\bibinfo {year} {2021})}\BibitemShut {NoStop}%
\bibitem [{\citenamefont {Yu}\ \emph {et~al.}(2021)\citenamefont {Yu},
  \citenamefont {Ma}, \citenamefont {Zhuo}, \citenamefont {Liu}, \citenamefont
  {Wen}, \citenamefont {Lei}, \citenamefont {Ying},\ and\ \citenamefont
  {Chen}}]{Yu2021cs}%
  \BibitemOpen
  \bibfield  {author} {\bibinfo {author} {\bibfnamefont {F.~H.}\ \bibnamefont
  {Yu}}, \bibinfo {author} {\bibfnamefont {D.~H.}\ \bibnamefont {Ma}}, \bibinfo
  {author} {\bibfnamefont {W.~Z.}\ \bibnamefont {Zhuo}}, \bibinfo {author}
  {\bibfnamefont {S.~Q.}\ \bibnamefont {Liu}}, \bibinfo {author} {\bibfnamefont
  {X.~K.}\ \bibnamefont {Wen}}, \bibinfo {author} {\bibfnamefont
  {B.}~\bibnamefont {Lei}}, \bibinfo {author} {\bibfnamefont {J.~J.}\
  \bibnamefont {Ying}}, \ and\ \bibinfo {author} {\bibfnamefont {X.~H.}\
  \bibnamefont {Chen}},\ }\bibfield  {title} {\enquote {\bibinfo {title}
  {Unusual competition of superconductivity and charge-density-wave state in a
  compressed topological kagome metal},}\ }\href {\doibase
  10.1038/s41467-021-23928-w   } {\bibfield  {journal} {\bibinfo  {journal}
  {Nature Communications}\ }\textbf {\bibinfo {volume} {12}},\ \bibinfo {pages}
  {3645} (\bibinfo {year} {2021})}\BibitemShut {NoStop}%
\bibitem [{\citenamefont {Jiang}\ \emph {et~al.}(2023)\citenamefont {Jiang},
  \citenamefont {Ma}, \citenamefont {Xia}, \citenamefont {Liu}, \citenamefont
  {Xiao}, \citenamefont {Liu}, \citenamefont {Yang}, \citenamefont {Ding},
  \citenamefont {Huang}, \citenamefont {Liu}, \citenamefont {Qiao},
  \citenamefont {Liu}, \citenamefont {Peng}, \citenamefont {Cho}, \citenamefont
  {Guo}, \citenamefont {Liu},\ and\ \citenamefont {Shen}}]{Jiang2022e}%
  \BibitemOpen
  \bibfield  {author} {\bibinfo {author} {\bibfnamefont {Zhicheng}\
  \bibnamefont {Jiang}}, \bibinfo {author} {\bibfnamefont {Haiyang}\
  \bibnamefont {Ma}}, \bibinfo {author} {\bibfnamefont {Wei}\ \bibnamefont
  {Xia}}, \bibinfo {author} {\bibfnamefont {Zhengtai}\ \bibnamefont {Liu}},
  \bibinfo {author} {\bibfnamefont {Qian}\ \bibnamefont {Xiao}}, \bibinfo
  {author} {\bibfnamefont {Zhonghao}\ \bibnamefont {Liu}}, \bibinfo {author}
  {\bibfnamefont {Yichen}\ \bibnamefont {Yang}}, \bibinfo {author}
  {\bibfnamefont {Jianyang}\ \bibnamefont {Ding}}, \bibinfo {author}
  {\bibfnamefont {Zhe}\ \bibnamefont {Huang}}, \bibinfo {author} {\bibfnamefont
  {Jiayu}\ \bibnamefont {Liu}}, \bibinfo {author} {\bibfnamefont {Yuxi}\
  \bibnamefont {Qiao}}, \bibinfo {author} {\bibfnamefont {Jishan}\ \bibnamefont
  {Liu}}, \bibinfo {author} {\bibfnamefont {Yingying}\ \bibnamefont {Peng}},
  \bibinfo {author} {\bibfnamefont {Soohyun}\ \bibnamefont {Cho}}, \bibinfo
  {author} {\bibfnamefont {Yanfeng}\ \bibnamefont {Guo}}, \bibinfo {author}
  {\bibfnamefont {Jianpeng}\ \bibnamefont {Liu}}, \ and\ \bibinfo {author}
  {\bibfnamefont {Dawei}\ \bibnamefont {Shen}},\ }\bibfield  {title} {\enquote
  {\bibinfo {title} {Observation of electronic nematicity driven by the
  three-dimensional charge density wave in kagome lattice kv3sb5},}\ }\href
  {\doibase 10.1021/acs.nanolett.3c01151 } {\bibfield  {journal} {\bibinfo
  {journal} {Nano Letters}\ }\textbf {\bibinfo {volume} {23}},\ \bibinfo
  {pages} {5625--5633} (\bibinfo {year} {2023})}\BibitemShut {NoStop}%
\bibitem [{\citenamefont {Ritz}\ \emph {et~al.}(2023)\citenamefont {Ritz},
  \citenamefont {Fernandes},\ and\ \citenamefont {Birol}}]{Ritz2022}%
  \BibitemOpen
  \bibfield  {author} {\bibinfo {author} {\bibfnamefont {Ethan~T.}\
  \bibnamefont {Ritz}}, \bibinfo {author} {\bibfnamefont {Rafael~M.}\
  \bibnamefont {Fernandes}}, \ and\ \bibinfo {author} {\bibfnamefont {Turan}\
  \bibnamefont {Birol}},\ }\bibfield  {title} {\enquote {\bibinfo {title}
  {Impact of sb degrees of freedom on the charge density wave phase diagram of
  the kagome metal ${\mathrm{csv}}_{3}{\mathrm{sb}}_{5}$},}\ }\href {\doibase
  10.1103/PhysRevB.107.205131 } {\bibfield  {journal} {\bibinfo  {journal}
  {Phys. Rev. B}\ }\textbf {\bibinfo {volume} {107}},\ \bibinfo {pages}
  {205131} (\bibinfo {year} {2023})}\BibitemShut {NoStop}%
\bibitem [{\citenamefont {Hu}\ \emph {et~al.}(2022{\natexlab{c}})\citenamefont
  {Hu}, \citenamefont {Wu}, \citenamefont {Ortiz}, \citenamefont {Han},
  \citenamefont {Plumb}, \citenamefont {Wilson}, \citenamefont {Schnyder},\
  and\ \citenamefont {Shi}}]{Hu2022t}%
  \BibitemOpen
  \bibfield  {author} {\bibinfo {author} {\bibfnamefont {Yong}\ \bibnamefont
  {Hu}}, \bibinfo {author} {\bibfnamefont {Xianxin}\ \bibnamefont {Wu}},
  \bibinfo {author} {\bibfnamefont {Brenden~R.}\ \bibnamefont {Ortiz}},
  \bibinfo {author} {\bibfnamefont {Xinloong}\ \bibnamefont {Han}}, \bibinfo
  {author} {\bibfnamefont {Nicholas~C.}\ \bibnamefont {Plumb}}, \bibinfo
  {author} {\bibfnamefont {Stephen~D.}\ \bibnamefont {Wilson}}, \bibinfo
  {author} {\bibfnamefont {Andreas~P.}\ \bibnamefont {Schnyder}}, \ and\
  \bibinfo {author} {\bibfnamefont {Ming}\ \bibnamefont {Shi}},\ }\bibfield
  {title} {\enquote {\bibinfo {title} {Coexistence of trihexagonal and
  star-of-david pattern in the charge density wave of the kagome superconductor
  $a{\mathrm{v}}_{3}{\mathrm{sb}}_{5}$},}\ }\href {\doibase
  10.1103/PhysRevB.106.L241106 } {\bibfield  {journal} {\bibinfo  {journal}
  {Phys. Rev. B}\ }\textbf {\bibinfo {volume} {106}},\ \bibinfo {pages}
  {L241106} (\bibinfo {year} {2022}{\natexlab{c}})}\BibitemShut {NoStop}%
\bibitem [{\citenamefont {Luo}\ \emph {et~al.}(2022{\natexlab{a}})\citenamefont
  {Luo}, \citenamefont {Zhao}, \citenamefont {Zhou}, \citenamefont {Yang},
  \citenamefont {Fang}, \citenamefont {Yang}, \citenamefont {Gao},
  \citenamefont {Zhou},\ and\ \citenamefont {Zheng}}]{Luo2022s}%
  \BibitemOpen
  \bibfield  {author} {\bibinfo {author} {\bibfnamefont {J.}~\bibnamefont
  {Luo}}, \bibinfo {author} {\bibfnamefont {Z.}~\bibnamefont {Zhao}}, \bibinfo
  {author} {\bibfnamefont {Y.~Z.}\ \bibnamefont {Zhou}}, \bibinfo {author}
  {\bibfnamefont {J.}~\bibnamefont {Yang}}, \bibinfo {author} {\bibfnamefont
  {A.~F.}\ \bibnamefont {Fang}}, \bibinfo {author} {\bibfnamefont {H.~T.}\
  \bibnamefont {Yang}}, \bibinfo {author} {\bibfnamefont {H.~J.}\ \bibnamefont
  {Gao}}, \bibinfo {author} {\bibfnamefont {R.}~\bibnamefont {Zhou}}, \ and\
  \bibinfo {author} {\bibfnamefont {Guo-qing}\ \bibnamefont {Zheng}},\
  }\bibfield  {title} {\enquote {\bibinfo {title} {Possible star-of-david
  pattern charge density wave with additional modulation in the kagome
  superconductor csv3sb5},}\ }\href {\doibase 10.1038/s41535-022-00437-7}
  {\bibfield  {journal} {\bibinfo  {journal} {npj Quantum Materials}\ }\textbf
  {\bibinfo {volume} {7}},\ \bibinfo {pages} {30} (\bibinfo {year}
  {2022}{\natexlab{a}})}\BibitemShut {NoStop}%
\bibitem [{\citenamefont {Mu}\ \emph {et~al.}(2022)\citenamefont {Mu},
  \citenamefont {Yin}, \citenamefont {Tu}, \citenamefont {Gong}, \citenamefont
  {Zheng}, \citenamefont {Lei}, \citenamefont {Li},\ and\ \citenamefont
  {Luo}}]{Mu_2022}%
  \BibitemOpen
  \bibfield  {author} {\bibinfo {author} {\bibfnamefont {Chao}\ \bibnamefont
  {Mu}}, \bibinfo {author} {\bibfnamefont {Qiangwei}\ \bibnamefont {Yin}},
  \bibinfo {author} {\bibfnamefont {Zhijun}\ \bibnamefont {Tu}}, \bibinfo
  {author} {\bibfnamefont {Chunsheng}\ \bibnamefont {Gong}}, \bibinfo {author}
  {\bibfnamefont {Ping}\ \bibnamefont {Zheng}}, \bibinfo {author}
  {\bibfnamefont {Hechang}\ \bibnamefont {Lei}}, \bibinfo {author}
  {\bibfnamefont {Zheng}\ \bibnamefont {Li}}, \ and\ \bibinfo {author}
  {\bibfnamefont {Jianlin}\ \bibnamefont {Luo}},\ }\bibfield  {title} {\enquote
  {\bibinfo {title} {Tri-hexagonal charge order in kagome metal csv3sb5
  revealed by 121sb nuclear quadrupole resonance},}\ }\href {\doibase
  10.1088/1674-1056/ac422c} {\bibfield  {journal} {\bibinfo  {journal} {Chinese
  Physics B}\ }\textbf {\bibinfo {volume} {31}},\ \bibinfo {pages} {017105}
  (\bibinfo {year} {2022})}\BibitemShut {NoStop}%
\bibitem [{\citenamefont {Luo}\ \emph {et~al.}(2022{\natexlab{b}})\citenamefont
  {Luo}, \citenamefont {Gao}, \citenamefont {Liu}, \citenamefont {Gu},
  \citenamefont {Wu}, \citenamefont {Yi}, \citenamefont {Jia}, \citenamefont
  {Wu}, \citenamefont {Luo}, \citenamefont {Xu}, \citenamefont {Zhao},
  \citenamefont {Wang}, \citenamefont {Mao}, \citenamefont {Liu}, \citenamefont
  {Zhu}, \citenamefont {Shi}, \citenamefont {Jiang}, \citenamefont {Hu},
  \citenamefont {Xu},\ and\ \citenamefont {Zhou}}]{Luo2022n}%
  \BibitemOpen
  \bibfield  {author} {\bibinfo {author} {\bibfnamefont {Hailan}\ \bibnamefont
  {Luo}}, \bibinfo {author} {\bibfnamefont {Qiang}\ \bibnamefont {Gao}},
  \bibinfo {author} {\bibfnamefont {Hongxiong}\ \bibnamefont {Liu}}, \bibinfo
  {author} {\bibfnamefont {Yuhao}\ \bibnamefont {Gu}}, \bibinfo {author}
  {\bibfnamefont {Dingsong}\ \bibnamefont {Wu}}, \bibinfo {author}
  {\bibfnamefont {Changjiang}\ \bibnamefont {Yi}}, \bibinfo {author}
  {\bibfnamefont {Junjie}\ \bibnamefont {Jia}}, \bibinfo {author}
  {\bibfnamefont {Shilong}\ \bibnamefont {Wu}}, \bibinfo {author}
  {\bibfnamefont {Xiangyu}\ \bibnamefont {Luo}}, \bibinfo {author}
  {\bibfnamefont {Yu}~\bibnamefont {Xu}}, \bibinfo {author} {\bibfnamefont
  {Lin}\ \bibnamefont {Zhao}}, \bibinfo {author} {\bibfnamefont {Qingyan}\
  \bibnamefont {Wang}}, \bibinfo {author} {\bibfnamefont {Hanqing}\
  \bibnamefont {Mao}}, \bibinfo {author} {\bibfnamefont {Guodong}\ \bibnamefont
  {Liu}}, \bibinfo {author} {\bibfnamefont {Zhihai}\ \bibnamefont {Zhu}},
  \bibinfo {author} {\bibfnamefont {Youguo}\ \bibnamefont {Shi}}, \bibinfo
  {author} {\bibfnamefont {Kun}\ \bibnamefont {Jiang}}, \bibinfo {author}
  {\bibfnamefont {Jiangping}\ \bibnamefont {Hu}}, \bibinfo {author}
  {\bibfnamefont {Zuyan}\ \bibnamefont {Xu}}, \ and\ \bibinfo {author}
  {\bibfnamefont {X.~J.}\ \bibnamefont {Zhou}},\ }\bibfield  {title} {\enquote
  {\bibinfo {title} {Electronic nature of charge density wave and
  electron-phonon coupling in kagome superconductor kv3sb5},}\ }\href {\doibase
  10.1038/s41467-021-27946-6} {\bibfield  {journal} {\bibinfo  {journal}
  {Nature Communications}\ }\textbf {\bibinfo {volume} {13}},\ \bibinfo {pages}
  {273} (\bibinfo {year} {2022}{\natexlab{b}})}\BibitemShut {NoStop}%
\bibitem [{\citenamefont {Kato}\ \emph {et~al.}(2022)\citenamefont {Kato},
  \citenamefont {Li}, \citenamefont {Kawakami}, \citenamefont {Liu},
  \citenamefont {Nakayama}, \citenamefont {Wang}, \citenamefont {Moriya},
  \citenamefont {Tanaka}, \citenamefont {Takahashi}, \citenamefont {Yao},\ and\
  \citenamefont {Sato}}]{Kato2022}%
  \BibitemOpen
  \bibfield  {author} {\bibinfo {author} {\bibfnamefont {Takemi}\ \bibnamefont
  {Kato}}, \bibinfo {author} {\bibfnamefont {Yongkai}\ \bibnamefont {Li}},
  \bibinfo {author} {\bibfnamefont {Tappei}\ \bibnamefont {Kawakami}}, \bibinfo
  {author} {\bibfnamefont {Min}\ \bibnamefont {Liu}}, \bibinfo {author}
  {\bibfnamefont {Kosuke}\ \bibnamefont {Nakayama}}, \bibinfo {author}
  {\bibfnamefont {Zhiwei}\ \bibnamefont {Wang}}, \bibinfo {author}
  {\bibfnamefont {Ayumi}\ \bibnamefont {Moriya}}, \bibinfo {author}
  {\bibfnamefont {Kiyohisa}\ \bibnamefont {Tanaka}}, \bibinfo {author}
  {\bibfnamefont {Takashi}\ \bibnamefont {Takahashi}}, \bibinfo {author}
  {\bibfnamefont {Yugui}\ \bibnamefont {Yao}}, \ and\ \bibinfo {author}
  {\bibfnamefont {Takafumi}\ \bibnamefont {Sato}},\ }\bibfield  {title}
  {\enquote {\bibinfo {title} {Three-dimensional energy gap and origin of
  charge-density wave in kagome superconductor kv3sb5},}\ }\href {\doibase
  10.1038/s43246-022-00255-1} {\bibfield  {journal} {\bibinfo  {journal}
  {Communications Materials}\ }\textbf {\bibinfo {volume} {3}},\ \bibinfo
  {pages} {30} (\bibinfo {year} {2022})}\BibitemShut {NoStop}%
\bibitem [{\citenamefont {Cho}\ \emph {et~al.}(2021)\citenamefont {Cho},
  \citenamefont {Ma}, \citenamefont {Xia}, \citenamefont {Yang}, \citenamefont
  {Liu}, \citenamefont {Huang}, \citenamefont {Jiang}, \citenamefont {Lu},
  \citenamefont {Liu}, \citenamefont {Liu}, \citenamefont {Li}, \citenamefont
  {Wang}, \citenamefont {Liu}, \citenamefont {Jia}, \citenamefont {Guo},
  \citenamefont {Liu},\ and\ \citenamefont {Shen}}]{Cho2021n}%
  \BibitemOpen
  \bibfield  {author} {\bibinfo {author} {\bibfnamefont {Soohyun}\ \bibnamefont
  {Cho}}, \bibinfo {author} {\bibfnamefont {Haiyang}\ \bibnamefont {Ma}},
  \bibinfo {author} {\bibfnamefont {Wei}\ \bibnamefont {Xia}}, \bibinfo
  {author} {\bibfnamefont {Yichen}\ \bibnamefont {Yang}}, \bibinfo {author}
  {\bibfnamefont {Zhengtai}\ \bibnamefont {Liu}}, \bibinfo {author}
  {\bibfnamefont {Zhe}\ \bibnamefont {Huang}}, \bibinfo {author} {\bibfnamefont
  {Zhicheng}\ \bibnamefont {Jiang}}, \bibinfo {author} {\bibfnamefont
  {Xiangle}\ \bibnamefont {Lu}}, \bibinfo {author} {\bibfnamefont {Jishan}\
  \bibnamefont {Liu}}, \bibinfo {author} {\bibfnamefont {Zhonghao}\
  \bibnamefont {Liu}}, \bibinfo {author} {\bibfnamefont {Jun}\ \bibnamefont
  {Li}}, \bibinfo {author} {\bibfnamefont {Jinghui}\ \bibnamefont {Wang}},
  \bibinfo {author} {\bibfnamefont {Yi}~\bibnamefont {Liu}}, \bibinfo {author}
  {\bibfnamefont {Jinfeng}\ \bibnamefont {Jia}}, \bibinfo {author}
  {\bibfnamefont {Yanfeng}\ \bibnamefont {Guo}}, \bibinfo {author}
  {\bibfnamefont {Jianpeng}\ \bibnamefont {Liu}}, \ and\ \bibinfo {author}
  {\bibfnamefont {Dawei}\ \bibnamefont {Shen}},\ }\bibfield  {title} {\enquote
  {\bibinfo {title} {Emergence of new van hove singularities in the charge
  density wave state of a topological kagome metal
  ${\mathrm{rbv}}_{3}{\mathrm{sb}}_{5}$},}\ }\href {\doibase
  10.1103/PhysRevLett.127.236401} {\bibfield  {journal} {\bibinfo  {journal}
  {Phys. Rev. Lett.}\ }\textbf {\bibinfo {volume} {127}},\ \bibinfo {pages}
  {236401} (\bibinfo {year} {2021})}\BibitemShut {NoStop}%
\bibitem [{\citenamefont {Denner}\ \emph {et~al.}(2021)\citenamefont {Denner},
  \citenamefont {Thomale},\ and\ \citenamefont {Neupert}}]{Denner2021}%
  \BibitemOpen
  \bibfield  {author} {\bibinfo {author} {\bibfnamefont {M.~Michael}\
  \bibnamefont {Denner}}, \bibinfo {author} {\bibfnamefont {Ronny}\
  \bibnamefont {Thomale}}, \ and\ \bibinfo {author} {\bibfnamefont {Titus}\
  \bibnamefont {Neupert}},\ }\bibfield  {title} {\enquote {\bibinfo {title}
  {Analysis of charge order in the kagome metal
  $a{\mathrm{v}}_{3}{\mathrm{sb}}_{5}$
  ($a=\mathrm{K},\mathrm{Rb},\mathrm{Cs}$)},}\ }\href {\doibase
  10.1103/PhysRevLett.127.217601} {\bibfield  {journal} {\bibinfo  {journal}
  {Phys. Rev. Lett.}\ }\textbf {\bibinfo {volume} {127}},\ \bibinfo {pages}
  {217601} (\bibinfo {year} {2021})}\BibitemShut {NoStop}%
\bibitem [{\citenamefont {Park}\ \emph {et~al.}(2021)\citenamefont {Park},
  \citenamefont {Ye},\ and\ \citenamefont {Balents}}]{Park2021}%
  \BibitemOpen
  \bibfield  {author} {\bibinfo {author} {\bibfnamefont {Takamori}\
  \bibnamefont {Park}}, \bibinfo {author} {\bibfnamefont {Mengxing}\
  \bibnamefont {Ye}}, \ and\ \bibinfo {author} {\bibfnamefont {Leon}\
  \bibnamefont {Balents}},\ }\bibfield  {title} {\enquote {\bibinfo {title}
  {Electronic instabilities of kagome metals: Saddle points and landau
  theory},}\ }\href {\doibase 10.1103/PhysRevB.104.035142} {\bibfield
  {journal} {\bibinfo  {journal} {Phys. Rev. B}\ }\textbf {\bibinfo {volume}
  {104}},\ \bibinfo {pages} {035142} (\bibinfo {year} {2021})}\BibitemShut
  {NoStop}%
\bibitem [{\citenamefont {Lin}\ and\ \citenamefont
  {Nandkishore}(2021)}]{Lin2021c}%
  \BibitemOpen
  \bibfield  {author} {\bibinfo {author} {\bibfnamefont {Yu-Ping}\ \bibnamefont
  {Lin}}\ and\ \bibinfo {author} {\bibfnamefont {Rahul~M.}\ \bibnamefont
  {Nandkishore}},\ }\bibfield  {title} {\enquote {\bibinfo {title} {Complex
  charge density waves at van hove singularity on hexagonal lattices:
  Haldane-model phase diagram and potential realization in the kagome metals
  $a{V}_{3}{\mathrm{sb}}_{5}$ ($a$=k, rb, cs)},}\ }\href {\doibase
  10.1103/PhysRevB.104.045122 } {\bibfield  {journal} {\bibinfo  {journal}
  {Phys. Rev. B}\ }\textbf {\bibinfo {volume} {104}},\ \bibinfo {pages}
  {045122} (\bibinfo {year} {2021})}\BibitemShut {NoStop}%
\bibitem [{\citenamefont {Christensen}\ \emph {et~al.}(2021)\citenamefont
  {Christensen}, \citenamefont {Birol}, \citenamefont {Andersen},\ and\
  \citenamefont {Fernandes}}]{Christensen2021}%
  \BibitemOpen
  \bibfield  {author} {\bibinfo {author} {\bibfnamefont {Morten~H.}\
  \bibnamefont {Christensen}}, \bibinfo {author} {\bibfnamefont {Turan}\
  \bibnamefont {Birol}}, \bibinfo {author} {\bibfnamefont {Brian~M.}\
  \bibnamefont {Andersen}}, \ and\ \bibinfo {author} {\bibfnamefont
  {Rafael~M.}\ \bibnamefont {Fernandes}},\ }\bibfield  {title} {\enquote
  {\bibinfo {title} {Theory of the charge density wave in
  $a{\mathrm{v}}_{3}{\mathrm{sb}}_{5}$ kagome metals},}\ }\href {\doibase
  10.1103/PhysRevB.104  . 214513} {\bibfield  {journal} {\bibinfo  {journal}
  {Phys. Rev. B}\ }\textbf {\bibinfo {volume} {104}},\ \bibinfo {pages}
  {214513} (\bibinfo {year} {2021})}\BibitemShut {NoStop}%
\bibitem [{\citenamefont {Jeong}\ \emph {et~al.}(2022)\citenamefont {Jeong},
  \citenamefont {Yang}, \citenamefont {Kim}, \citenamefont {Kim}, \citenamefont
  {Lee},\ and\ \citenamefont {Han}}]{Jeong2022}%
  \BibitemOpen
  \bibfield  {author} {\bibinfo {author} {\bibfnamefont {Min~Yong}\
  \bibnamefont {Jeong}}, \bibinfo {author} {\bibfnamefont {Hyeok-Jun}\
  \bibnamefont {Yang}}, \bibinfo {author} {\bibfnamefont {Hee~Seung}\
  \bibnamefont {Kim}}, \bibinfo {author} {\bibfnamefont {Yong~Baek}\
  \bibnamefont {Kim}}, \bibinfo {author} {\bibfnamefont {SungBin}\ \bibnamefont
  {Lee}}, \ and\ \bibinfo {author} {\bibfnamefont {Myung~Joon}\ \bibnamefont
  {Han}},\ }\bibfield  {title} {\enquote {\bibinfo {title} {Crucial role of
  out-of-plane sb $p$ orbitals in van hove singularity formation and electronic
  correlations in the superconducting kagome metal
  ${\mathrm{csv}}_{3}{\mathrm{sb}}_{5}$},}\ }\href {\doibase
  10.1103/PhysRevB.105.235145  } {\bibfield  {journal} {\bibinfo  {journal}
  {Phys. Rev. B}\ }\textbf {\bibinfo {volume} {105}},\ \bibinfo {pages}
  {235145} (\bibinfo {year} {2022})}\BibitemShut {NoStop}%
\bibitem [{\citenamefont {Tazai}\ \emph {et~al.}(2022)\citenamefont {Tazai},
  \citenamefont {Yamakawa}, \citenamefont {Onari},\ and\ \citenamefont
  {Kontani}}]{Rina2022}%
  \BibitemOpen
  \bibfield  {author} {\bibinfo {author} {\bibfnamefont {Rina}\ \bibnamefont
  {Tazai}}, \bibinfo {author} {\bibfnamefont {Youichi}\ \bibnamefont
  {Yamakawa}}, \bibinfo {author} {\bibfnamefont {Seiichiro}\ \bibnamefont
  {Onari}}, \ and\ \bibinfo {author} {\bibfnamefont {Hiroshi}\ \bibnamefont
  {Kontani}},\ }\bibfield  {title} {\enquote {\bibinfo {title} {Mechanism of
  exotic density-wave and beyond-migdal unconventional superconductivity in
  kagome metal av3sb5 (a = k, rb, cs)},}\ }\href {\doibase
  10.1126/sciadv.abl4108 } {\bibfield  {journal} {\bibinfo  {journal} {Science
  Advances}\ }\textbf {\bibinfo {volume} {8}},\ \bibinfo {pages} {eabl4108}
  (\bibinfo {year} {2022})}\BibitemShut {NoStop}%
\bibitem [{\citenamefont {Zhou}\ and\ \citenamefont {Wang}(2022)}]{Zhou2022f}%
  \BibitemOpen
  \bibfield  {author} {\bibinfo {author} {\bibfnamefont {Sen}\ \bibnamefont
  {Zhou}}\ and\ \bibinfo {author} {\bibfnamefont {Ziqiang}\ \bibnamefont
  {Wang}},\ }\bibfield  {title} {\enquote {\bibinfo {title} {Chern fermi
  pocket, topological pair density wave, and charge-4e and charge-6e
  superconductivity in kagom{\'e} superconductors},}\ }\href {\doibase
  10.1038/s41467-022-34832-2} {\bibfield  {journal} {\bibinfo  {journal}
  {Nature Communications}\ }\textbf {\bibinfo {volume} {13}},\ \bibinfo {pages}
  {7288} (\bibinfo {year} {2022})}\BibitemShut {NoStop}%
\bibitem [{\citenamefont {Wu}\ \emph {et~al.}(2022{\natexlab{b}})\citenamefont
  {Wu}, \citenamefont {Thomale},\ and\ \citenamefont {Raghu}}]{Wu2022sub}%
  \BibitemOpen
  \bibfield  {author} {\bibinfo {author} {\bibfnamefont {Yi-Ming}\ \bibnamefont
  {Wu}}, \bibinfo {author} {\bibfnamefont {Ronny}\ \bibnamefont {Thomale}}, \
  and\ \bibinfo {author} {\bibfnamefont {S.}~\bibnamefont {Raghu}},\
  }\href@noop {} {\enquote {\bibinfo {title} {Sublattice interference promotes
  pair density wave order in kagome metals},}\ } (\bibinfo {year}
  {2022}{\natexlab{b}}),\ \Eprint {http://arxiv.org/abs/2211.01388}
  {arXiv:2211.01388} \BibitemShut {NoStop}%
\bibitem [{\citenamefont {Li}\ \emph {et~al.}(2022{\natexlab{b}})\citenamefont
  {Li}, \citenamefont {Fabbris}, \citenamefont {Said}, \citenamefont {Sun},
  \citenamefont {Jiang}, \citenamefont {Yin}, \citenamefont {Pai},
  \citenamefont {Yoon}, \citenamefont {Lupini}, \citenamefont {Nelson},
  \citenamefont {Yin}, \citenamefont {Gong}, \citenamefont {Tu}, \citenamefont
  {Lei}, \citenamefont {Cheng}, \citenamefont {Hasan}, \citenamefont {Wang},
  \citenamefont {Yan}, \citenamefont {Thomale}, \citenamefont {Lee},\ and\
  \citenamefont {Miao}}]{Li2022pres}%
  \BibitemOpen
  \bibfield  {author} {\bibinfo {author} {\bibfnamefont {Haoxiang}\
  \bibnamefont {Li}}, \bibinfo {author} {\bibfnamefont {G.}~\bibnamefont
  {Fabbris}}, \bibinfo {author} {\bibfnamefont {A.~H.}\ \bibnamefont {Said}},
  \bibinfo {author} {\bibfnamefont {J.~P.}\ \bibnamefont {Sun}}, \bibinfo
  {author} {\bibfnamefont {Yu-Xiao}\ \bibnamefont {Jiang}}, \bibinfo {author}
  {\bibfnamefont {J.-X.}\ \bibnamefont {Yin}}, \bibinfo {author} {\bibfnamefont
  {Yun-Yi}\ \bibnamefont {Pai}}, \bibinfo {author} {\bibfnamefont {Sangmoon}\
  \bibnamefont {Yoon}}, \bibinfo {author} {\bibfnamefont {Andrew~R.}\
  \bibnamefont {Lupini}}, \bibinfo {author} {\bibfnamefont {C.~S.}\
  \bibnamefont {Nelson}}, \bibinfo {author} {\bibfnamefont {Q.~W.}\
  \bibnamefont {Yin}}, \bibinfo {author} {\bibfnamefont {C.~S.}\ \bibnamefont
  {Gong}}, \bibinfo {author} {\bibfnamefont {Z.~J.}\ \bibnamefont {Tu}},
  \bibinfo {author} {\bibfnamefont {H.~C.}\ \bibnamefont {Lei}}, \bibinfo
  {author} {\bibfnamefont {J.-G.}\ \bibnamefont {Cheng}}, \bibinfo {author}
  {\bibfnamefont {M.~Z.}\ \bibnamefont {Hasan}}, \bibinfo {author}
  {\bibfnamefont {Ziqiang}\ \bibnamefont {Wang}}, \bibinfo {author}
  {\bibfnamefont {Binghai}\ \bibnamefont {Yan}}, \bibinfo {author}
  {\bibfnamefont {R.}~\bibnamefont {Thomale}}, \bibinfo {author} {\bibfnamefont
  {H.~N.}\ \bibnamefont {Lee}}, \ and\ \bibinfo {author} {\bibfnamefont
  {H.}~\bibnamefont {Miao}},\ }\bibfield  {title} {\enquote {\bibinfo {title}
  {Discovery of conjoined charge density waves in the kagome superconductor
  csv3sb5},}\ }\href {\doibase 10.1038/s41467-022-33995-2 } {\bibfield
  {journal} {\bibinfo  {journal} {Nature Communications}\ }\textbf {\bibinfo
  {volume} {13}},\ \bibinfo {pages} {6348} (\bibinfo {year}
  {2022}{\natexlab{b}})}\BibitemShut {NoStop}%
\bibitem [{\citenamefont {Tsirlin}\ \emph {et~al.}(2022)\citenamefont
  {Tsirlin}, \citenamefont {Fertey}, \citenamefont {Ortiz}, \citenamefont
  {Klis}, \citenamefont {Merkl}, \citenamefont {Dressel}, \citenamefont
  {Wilson},\ and\ \citenamefont {Uykur}}]{Tsirlin2022s}%
  \BibitemOpen
  \bibfield  {author} {\bibinfo {author} {\bibfnamefont {Alexander~A.}\
  \bibnamefont {Tsirlin}}, \bibinfo {author} {\bibfnamefont {Pierre}\
  \bibnamefont {Fertey}}, \bibinfo {author} {\bibfnamefont {Brenden~R.}\
  \bibnamefont {Ortiz}}, \bibinfo {author} {\bibfnamefont {Berina}\
  \bibnamefont {Klis}}, \bibinfo {author} {\bibfnamefont {Valentino}\
  \bibnamefont {Merkl}}, \bibinfo {author} {\bibfnamefont {Martin}\
  \bibnamefont {Dressel}}, \bibinfo {author} {\bibfnamefont {Stephen~D.}\
  \bibnamefont {Wilson}}, \ and\ \bibinfo {author} {\bibfnamefont {Ece}\
  \bibnamefont {Uykur}},\ }\bibfield  {title} {\enquote {\bibinfo {title}
  {{Role of Sb in the superconducting kagome metal CsV$_3$Sb$_5$ revealed by
  its anisotropic compression}},}\ }\href {\doibase
  10.21468/SciPostPhys.12.2.049 } {\bibfield  {journal} {\bibinfo  {journal}
  {SciPost Phys.}\ }\textbf {\bibinfo {volume} {12}},\ \bibinfo {pages} {049}
  (\bibinfo {year} {2022})}\BibitemShut {NoStop}%
\bibitem [{\citenamefont {Frassineti}\ \emph {et~al.}(2023)\citenamefont
  {Frassineti}, \citenamefont {Bonf\`a}, \citenamefont {Allodi}, \citenamefont
  {Garcia}, \citenamefont {Cong}, \citenamefont {Ortiz}, \citenamefont
  {Wilson}, \citenamefont {De~Renzi}, \citenamefont
  {Mitrovi\ifmmode~\acute{c}\else \'{c}\fi{}},\ and\ \citenamefont
  {Sanna}}]{Frass2022}%
  \BibitemOpen
  \bibfield  {author} {\bibinfo {author} {\bibfnamefont {Jonathan}\
  \bibnamefont {Frassineti}}, \bibinfo {author} {\bibfnamefont {Pietro}\
  \bibnamefont {Bonf\`a}}, \bibinfo {author} {\bibfnamefont {Giuseppe}\
  \bibnamefont {Allodi}}, \bibinfo {author} {\bibfnamefont {Erick}\
  \bibnamefont {Garcia}}, \bibinfo {author} {\bibfnamefont {Rong}\ \bibnamefont
  {Cong}}, \bibinfo {author} {\bibfnamefont {Brenden~R.}\ \bibnamefont
  {Ortiz}}, \bibinfo {author} {\bibfnamefont {Stephen~D.}\ \bibnamefont
  {Wilson}}, \bibinfo {author} {\bibfnamefont {Roberto}\ \bibnamefont
  {De~Renzi}}, \bibinfo {author} {\bibfnamefont {Vesna~F.}\ \bibnamefont
  {Mitrovi\ifmmode~\acute{c}\else \'{c}\fi{}}}, \ and\ \bibinfo {author}
  {\bibfnamefont {Samuele}\ \bibnamefont {Sanna}},\ }\bibfield  {title}
  {\enquote {\bibinfo {title} {Microscopic nature of the charge-density wave in
  the kagome superconductor ${\mathrm{rbv}}_{3}{\mathrm{sb}}_{5}$},}\ }\href
  {\doibase 10.1103/PhysRevResearch.5.L012017} {\bibfield  {journal} {\bibinfo
  {journal} {Phys. Rev. Res.}\ }\textbf {\bibinfo {volume} {5}},\ \bibinfo
  {pages} {L012017} (\bibinfo {year} {2023})}\BibitemShut {NoStop}%
\bibitem [{\citenamefont {Li}\ \emph {et~al.}(2023)\citenamefont {Li},
  \citenamefont {Liu}, \citenamefont {Kim},\ and\ \citenamefont
  {Kee}}]{Li2023kagome}%
  \BibitemOpen
  \bibfield  {author} {\bibinfo {author} {\bibfnamefont {Heqiu}\ \bibnamefont
  {Li}}, \bibinfo {author} {\bibfnamefont {Xiaoyu}\ \bibnamefont {Liu}},
  \bibinfo {author} {\bibfnamefont {Yong~Baek}\ \bibnamefont {Kim}}, \ and\
  \bibinfo {author} {\bibfnamefont {Hae-Young}\ \bibnamefont {Kee}},\
  }\bibfield  {title} {\enquote {\bibinfo {title} {Origin of
  $\ensuremath{\pi}$-shifted three-dimensional charge density waves in the
  kagom\'e metal ${A\mathrm{V}}_{3}{\mathrm{sb}}_{5}$ $(a=\mathrm{Cs},
  \mathrm{Rb}, \mathrm{K})$},}\ }\href {\doibase 10.1103/PhysRevB.108.075102}
  {\bibfield  {journal} {\bibinfo  {journal} {Phys. Rev. B}\ }\textbf {\bibinfo
  {volume} {108}},\ \bibinfo {pages} {075102} (\bibinfo {year}
  {2023})}\BibitemShut {NoStop}%
\bibitem [{\citenamefont {Le}\ \emph {et~al.}(2023)\citenamefont {Le},
  \citenamefont {Pan}, \citenamefont {Xu}, \citenamefont {Liu}, \citenamefont
  {Wang}, \citenamefont {Lou}, \citenamefont {Wang}, \citenamefont {Yao},
  \citenamefont {Wu},\ and\ \citenamefont {Lin}}]{Le2023c}%
  \BibitemOpen
  \bibfield  {author} {\bibinfo {author} {\bibfnamefont {Tian}\ \bibnamefont
  {Le}}, \bibinfo {author} {\bibfnamefont {Zhiming}\ \bibnamefont {Pan}},
  \bibinfo {author} {\bibfnamefont {Zhuokai}\ \bibnamefont {Xu}}, \bibinfo
  {author} {\bibfnamefont {Jinjin}\ \bibnamefont {Liu}}, \bibinfo {author}
  {\bibfnamefont {Jialu}\ \bibnamefont {Wang}}, \bibinfo {author}
  {\bibfnamefont {Zhefeng}\ \bibnamefont {Lou}}, \bibinfo {author}
  {\bibfnamefont {Zhiwei}\ \bibnamefont {Wang}}, \bibinfo {author}
  {\bibfnamefont {Yugui}\ \bibnamefont {Yao}}, \bibinfo {author} {\bibfnamefont
  {Congjun}\ \bibnamefont {Wu}}, \ and\ \bibinfo {author} {\bibfnamefont
  {Xiao}\ \bibnamefont {Lin}},\ }\href@noop {} {\enquote {\bibinfo {title}
  {Evidence for chiral superconductivity in kagome superconductor csv3sb5},}\ }
  (\bibinfo {year} {2023}),\ \Eprint {http://arxiv.org/abs/2309.00264}
  {arXiv:2309.00264  [cond-mat.supr-con]} \BibitemShut {NoStop}%
\bibitem [{\citenamefont {Wenzel}\ \emph {et~al.}(2023)\citenamefont {Wenzel},
  \citenamefont {Tsirlin}, \citenamefont {Capitani}, \citenamefont {Chan},
  \citenamefont {Ortiz}, \citenamefont {Wilson}, \citenamefont {Dressel},\ and\
  \citenamefont {Uykur}}]{Wenzel2023p}%
  \BibitemOpen
  \bibfield  {author} {\bibinfo {author} {\bibfnamefont {Maxim}\ \bibnamefont
  {Wenzel}}, \bibinfo {author} {\bibfnamefont {Alexander~A.}\ \bibnamefont
  {Tsirlin}}, \bibinfo {author} {\bibfnamefont {Francesco}\ \bibnamefont
  {Capitani}}, \bibinfo {author} {\bibfnamefont {Yuk~T.}\ \bibnamefont {Chan}},
  \bibinfo {author} {\bibfnamefont {Brenden~R.}\ \bibnamefont {Ortiz}},
  \bibinfo {author} {\bibfnamefont {Stephen~D.}\ \bibnamefont {Wilson}},
  \bibinfo {author} {\bibfnamefont {Martin}\ \bibnamefont {Dressel}}, \ and\
  \bibinfo {author} {\bibfnamefont {Ece}\ \bibnamefont {Uykur}},\ }\bibfield
  {title} {\enquote {\bibinfo {title} {Pressure evolution of electron dynamics
  in the superconducting kagome metal csv3sb5},}\ }\href {\doibase
  10.1038/s41535-023-00577-4 } {\bibfield  {journal} {\bibinfo  {journal} {npj
  Quantum Materials}\ }\textbf {\bibinfo {volume} {8}},\ \bibinfo {pages} {45}
  (\bibinfo {year} {2023})}\BibitemShut {NoStop}%
\bibitem [{\citenamefont {Scammell}\ \emph {et~al.}(2023)\citenamefont
  {Scammell}, \citenamefont {Ingham}, \citenamefont {Li},\ and\ \citenamefont
  {Sushkov}}]{Scammell2023-wl}%
  \BibitemOpen
  \bibfield  {author} {\bibinfo {author} {\bibfnamefont {Harley~D}\
  \bibnamefont {Scammell}}, \bibinfo {author} {\bibfnamefont {Julian}\
  \bibnamefont {Ingham}}, \bibinfo {author} {\bibfnamefont {Tommy}\
  \bibnamefont {Li}}, \ and\ \bibinfo {author} {\bibfnamefont {Oleg~P}\
  \bibnamefont {Sushkov}},\ }\bibfield  {title} {\enquote {\bibinfo {title}
  {Chiral excitonic order from twofold van hove singularities in kagome
  metals},}\ }\href@noop {} {\bibfield  {journal} {\bibinfo  {journal} {Nature
  Communications}\ }\textbf {\bibinfo {volume} {14}},\ \bibinfo {pages} {605}
  (\bibinfo {year} {2023})}\BibitemShut {NoStop}%
\bibitem [{\citenamefont {Tsirlin}\ \emph {et~al.}(2023)\citenamefont
  {Tsirlin}, \citenamefont {Ortiz}, \citenamefont {Dressel}, \citenamefont
  {Wilson}, \citenamefont {Winnerl},\ and\ \citenamefont
  {Uykur}}]{Tsirlin2023}%
  \BibitemOpen
  \bibfield  {author} {\bibinfo {author} {\bibfnamefont {Alexander~A.}\
  \bibnamefont {Tsirlin}}, \bibinfo {author} {\bibfnamefont {Brenden~R.}\
  \bibnamefont {Ortiz}}, \bibinfo {author} {\bibfnamefont {Martin}\
  \bibnamefont {Dressel}}, \bibinfo {author} {\bibfnamefont {Stephen~D.}\
  \bibnamefont {Wilson}}, \bibinfo {author} {\bibfnamefont {Stephan}\
  \bibnamefont {Winnerl}}, \ and\ \bibinfo {author} {\bibfnamefont {Ece}\
  \bibnamefont {Uykur}},\ }\bibfield  {title} {\enquote {\bibinfo {title}
  {Effect of nonhydrostatic pressure on the superconducting kagome metal
  ${\mathrm{csv}}_{3}{\mathrm{sb}}_{5}$},}\ }\href {\doibase
  10.1103/PhysRevB.107.174107    }   {\bibfield  {journal} {\bibinfo  {journal}
  {Phys. Rev. B}\ }\textbf {\bibinfo {volume} {107}},\ \bibinfo {pages}
  {174107} (\bibinfo {year} {2023})}\BibitemShut {NoStop}%
\bibitem [{sup()}]{supp}%
  \BibitemOpen
  \href@noop {} {}\bibinfo {note} {See Supplemental Materials for details on symmetry constraints on the effective Hamiltonian, and the discussion on Ginzburg-Landau analysis.}\BibitemShut
  {Stop}%
\end{thebibliography}

%

\appendix

\setcounter{equation}{0}
\setcounter{figure}{0}
\setcounter{table}{0}
\makeatletter
\renewcommand{\theequation}{S\arabic{equation}}
\renewcommand{\thefigure}{S\arabic{figure}}
\renewcommand{\bibnumfmt}[1]{[S#1]}
\renewcommand{\citenumfont}[1]{S#1}

\newpage

\begin{widetext}

\section{Supplementary materials}

\section{Symmetry constraints on the wave function structure at momentum $M$}

The mirror symmetries $m'$ and $m''$ perpendicular to the kagome plane put strong constraints on the form of wave function at momentum $M$. Suppose both vH1 and vH2 are even under $m''$ and vH1 (vH2) is odd (even) under $m'$, and the orbital at each kagome site is even (odd) under $m''(m')$ as in Fig.\ref{fig_orbitsign}(a), where the red (blue) color denotes regions of orbitals with positive (negative) amplitude. This case corresponds to the vHS in AV$_3$Sb$_5$ made of $d_{xz},d_{yz}$ orbitals. Then the wave function components corresponding A,B,C sublattices for vH2 at $M_C$ will take the form $(b,-b,0)$ as explained in the main text. The explanation for vH1 is as follows. Consider momentum $M_c$, the wave function of vH1 is odd under $m_x$ and even under $m_y$, where $m_x$ flips $x$ to $-x$. This leads to the sign structure shown in Fig.\ref{fig_orbitsign}(a), where the signs denote the region of orbitals with positive/negative amplitude. However, this sign structure is not allowed for a wave function with momentum $M_C$, because a wave function with momentum $M_C$ should have opposite signs at sites A and A', as seen by $e^{iM_C\cdot (\mathbf r_A-\mathbf r_{A'})}=e^{iM_C\cdot (\mathbf r_B-\mathbf r_{B'})}=-1$. Therefore the wave function amplitude has to vanish at sublattice A and B. This leads to the form of wave function $(0,0,b')$ for vH1 at $M_C$ in Fig.1(c) in the main text.

\begin{figure}[h]
\centering
\includegraphics[width=3.5 in]{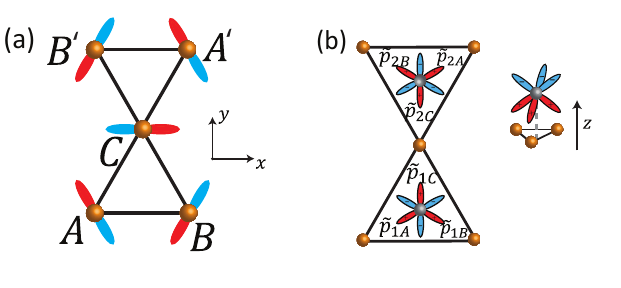}
\caption{ Illustration of the orbitals used in the tight-binding model.
}
\label{fig_orbitsign}
\end{figure}

The above discussion assumes the orbital is even (odd) under $m''(m')$. In the following we show that this assumption can be released such that the conclusion is applicable to orbitals with any symmetries under $m'$ and $m''$. Denote $\phi_{m'}$ and $\phi_{m''}$ as the eigenvalues of the orbital under $m'$ and $m''$ and denote $u_{m'}$ and $u_{m''}$ as the eigenvalues of the wave function at vHS under $m'$ and $m''$. $(\phi_{m'},\phi_{m''})=(-1,+1)$ corresponds to the orbitals in Fig.\ref{fig_orbitsign}(a) discussed above. $(\phi_{m'},\phi_{m''})=(+1,-1)$ corresponds to the orbitals in Fig.\ref{fig_orbitsign}(a) rotated by 90 degree. $(\phi_{m'},\phi_{m''})=(+1,+1)$ denotes orbitals with s-wave symmetry. $(\phi_{m'},\phi_{m''})=(-1,-1)$ denotes orbitals with symmetry of $d_{xy}$. Since $\phi_{m'},\phi_{m''},u_{m'},u_{m''}$ can be $\pm 1$ independently, it leads to 16 different cases. We enumerate these 16 cases in Fig.\ref{fig_orbitclass}. We focus on vHS at momentum $M_C$ without loss of generality and write down the allowed form of wave function components at A,B,C sublattices for each case. Each figure shows the orbital orientation consistent with mirror symmetries, where the $\pm$ signs denote the region of orbital with positive/negative amplitude. Some of the symmetry combinations give wave function structure of $(0,0,0)$, which means this orbital type cannot give rise to a vHS at momentum $M$ with the given mirror eigenvalues. Take the combination $(\phi_{m'},\phi_{m''},u_{m'},u_{m''})=(+1,+1,-1,-1)$ for example. The sign structure in Fig.\ref{fig_orbitclass} is consistent with the mirror eigenvalues. However, it contradicts with the property of a wave function at momentum $M_C$ which requires that the wave function amplitude should have opposite signs between sites A and A' due to $e^{iM_C\cdot (\mathbf r_A-\mathbf r_{A'})}=-1$. Hence this symmetry combination is not allowed.

From Fig.\ref{fig_orbitclass} we find that, no matter what symmetry does the orbital has, the wave functions at vH1 and vH2 will take the form of Fig.1(c) as long as the following conditions are satisfied: (1) vH1 and vH2 have opposite eigenvalues for $m'$ and same eigenvalues for $m''$. (2) vH1 and vH2 are made of the same orbital type. These conditions result in the wave functions in Fig.1(c), which is crucial for the emergence of loop current order.

\begin{figure*}[ht]
\centering
\includegraphics[width=4.8 in]{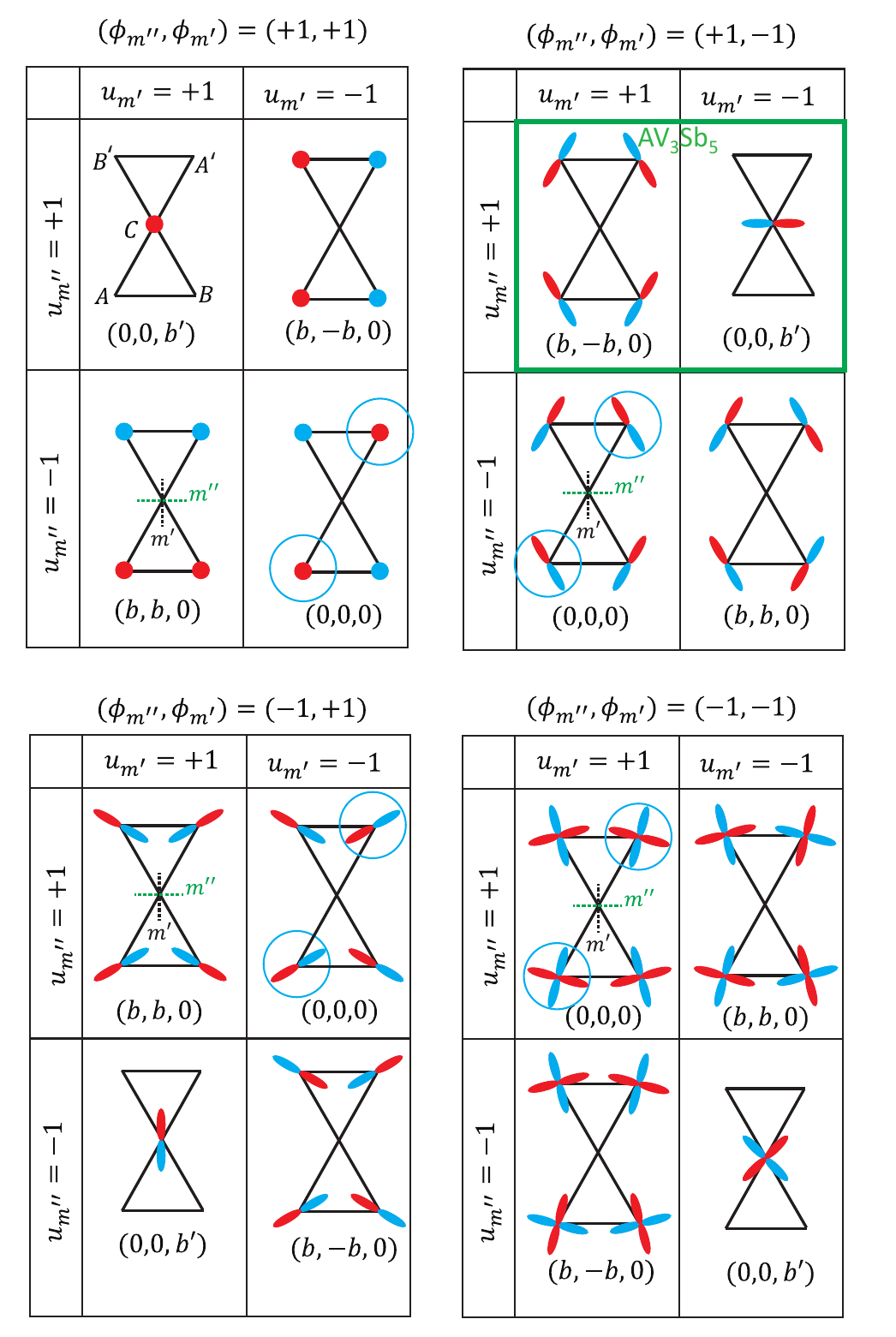}
\caption{ Orbital arrangement for the wave function of vHS at momentum $M_C$ under different symmetry conditions. The allowed form of wave function components at sublattice A,B,C are shown in the parentheses. The red (blue) color denotes regions of orbitals with positive (negative) amplitude. The wave function amplitude at sites marked by blue circles have the same sign, which contradicts with the requirement of being at momentum $M_C$ and hence it is not allowed. The green box labels the symmetry conditions relevant to AV$_3$Sb$_5$.
}
\label{fig_orbitclass}
\end{figure*}

\section{Details on derivation of effective model $H_{\rm{eff}}$}

More details on the derivation of $H_{\rm{eff}}$ in the main text are given as follows. We start from the Hamiltonian $H(k)$ of the 9-band model~\cite{Li2023kagome}, with basis chosen as $\{\tilde \Phi\}=\{\tilde d_A,\tilde d_B,\tilde d_C,\tilde p_{1A}^-,\tilde p_{1B}^-,\tilde p_{1C}^-,\tilde p_{2A}^-,\tilde p_{2B}^-,\tilde p_{2C}^- \}$. Here $\tilde d$ orbitals are:
\bea
\tilde d_A&=&-\frac{1}{2}d_{A,xz}+\frac{\sqrt{3}}{2}d_{A,yz}\nonumber\\
\tilde d_B&=&-\frac{1}{2}d_{B,xz}-\frac{\sqrt{3}}{2}d_{B,yz} \nonumber\\
\tilde d_C&=&d_{C,xz}.
\eea
The orbitals $\tilde p_{\sigma\alpha}^-$ with $\sigma=1,2$, $\alpha=A,B,C$ are superpositions of $p$ orbitals at the four out-of-plane Sb sites with negative eigenvalue of $M_z$. $\tilde p_{\sigma\alpha}^-\equiv(\tilde p_{\sigma\alpha}-M_z\tilde p_{\sigma\alpha})/\sqrt{2}$, and $\tilde p_{\sigma\alpha}$ is shown in Fig.\ref{fig_orbitsign}(b) with $\sigma=1,2$ representing the two Sb sites above the kagome plane, which are explicitly written as:
\bea
\tilde p_{1A}&=&-\frac{1}{\sqrt{2}}p_{1x}-\frac{1}{\sqrt{6}}p_{1y}-\frac{1}{\sqrt{3}}p_{1z},\ \tilde p_{2A}=\frac{1}{\sqrt{2}}p_{2x}+\frac{1}{\sqrt{6}}p_{2y}-\frac{1}{\sqrt{3}}p_{2z}, \nonumber\\
\tilde p_{1B}&=&\frac{1}{\sqrt{2}}p_{1x}-\frac{1}{\sqrt{6}}p_{1y}-\frac{1}{\sqrt{3}}p_{1z},\ \tilde p_{2B}=-\frac{1}{\sqrt{2}}p_{2x}+\frac{1}{\sqrt{6}}p_{2y}-\frac{1}{\sqrt{3}}p_{2z} \nonumber\\
\tilde p_{1C}&=&\sqrt{\frac{2}{3}}p_{1y}-\frac{1}{\sqrt{3}}p_{1z},\ \tilde p_{2C}=-\sqrt{\frac{2}{3}}p_{2y}-\frac{1}{\sqrt{3}}p_{2z}
\label{tildep}
\eea

The hopping amplitudes and onsite potentials are obtained from DFT, which leads to the 9-band model $H(k)$. We choose the convention of Fourier transform as $c^\dagger_{k,\rho}=\frac{1}{\sqrt{N_c}}\sum_R c^\dagger_{R,\rho}e^{ik\cdot (R+r_\rho)}$, where $R$ labels unit cells, $\rho=1,...,9$ denotes the orbitals in $\{\tilde \Phi\}$ and $r_\rho$ is the location within the unit cell. Then $H(k)$ has a threefold rotation symmetry $C_3$ given by:
\bea
C_3=\begin{pmatrix}
1 & 0 & 0  \\
0 & 1 & 0  \\
0 & 0 & 1  
\end{pmatrix}\otimes\begin{pmatrix}
0 & 0 & 1  \\
1 & 0 & 0  \\
0 & 1 & 0  
\end{pmatrix},\ \ \ \ C_3H(k)C_3^\dagger=H(C_3 k)
\eea
$H(k)$ can be diagonalized by $U(k)^\dagger H(k)U(k)=\Lambda(k)$ with $\Lambda_{11}\ge \Lambda_{22}\ge...\ge\Lambda_{99}$. At momentum $M$, the energy of vH1 (vH2) is $\Lambda_{44}$ ($\Lambda_{33}$). With CDW orders connecting different $M$ points, the mean field Hamiltonian near the $M$ points is written as
\bea
H_{MF}(k)=\begin{pmatrix}
H(k+M_A) & D^{AB}(k) & D^{CA}(k)^\dagger \\
D^{AB}(k)^\dagger & H(k+M_B)   & D^{BC}(k) \\
D^{CA}(k) & D^{BC}(k)^\dagger & H(k+M_C)
\end{pmatrix}.
\eea
Here $k$ denotes the small deviation from $M$ points, and $k$ is taken to be inside a small hexagon with side length $k_{cut}\ll |G|$, where $G$ is reciprocal lattice vector. $D^{\alpha\beta}$ is a $9\times 9$ matrix from mean field doupling of the interaction. It has nonzero elements only in the first $3\times 3$ diagonal block corresponding to $\tilde d$ orbitals, which is denoted as $\overline{D}^{\alpha\beta}$. Consider the threefold symmetric CDW phase with $\Delta_{AB}=\Delta_{BC}=\Delta_{CA}\equiv \Delta$, the explicit forms of $\overline{D}^{\alpha\beta}(k)$ are:
\bea
\overline{D}^{AB}(k)=-2 \cos \frac{k\cdot d_{AB}}{2}\begin{pmatrix}
0 & \Delta & 0  \\
\Delta^* & 0 & 0  \\
0 & 0 & 0  
\end{pmatrix},\ \overline{D}^{BC}(k)=-2 \cos \frac{k\cdot d_{BC}}{2}\begin{pmatrix}
0 & 0 & 0  \\
0 & 0 & \Delta  \\
0 & \Delta^* & 0  
\end{pmatrix},\ \overline{D}^{CA}(k)=-2 \cos \frac{k\cdot d_{CA}}{2}\begin{pmatrix}
0 & 0 & \Delta^*  \\
0 & 0 & 0  \\
\Delta & 0 & 0  
\end{pmatrix}.
\eea
To construct the $6\times 6$ effective Hamiltonian $H_{\rm{eff}}$ near vH1 and vH2, We perform a unitary transformation to define $\tilde H(k)=U^\dagger H_{MF}(k) U$, where $U$ is given by
\bea
U=\begin{pmatrix}
U(M_A) & 0 & 0 \\
0 & U(M_B)   & 0 \\
0 & 0 & U(M_C) 
\end{pmatrix}.
\eea
It is essential to fix $U(M_B)=C_3U(M_A),U(M_C)=C_3U(M_B)$ to eliminate the gauge ambiguity between different $M$ points. The $k$-independent form of $C_3$ matrix is important in this gauge fixing procedure. The diagonal elements of $\tilde H(k)$ at $k=0$ represent the energies of the original Hamiltonian at momentum $M$, which also include vH1 and vH2. To construct an effective Hamiltonian that describes the vHS, we only keep the columns and rows in $\tilde H(k)$ corresponding to vH1 and vH2. More precisely, because vH1 and vH2 are the fourth and the third largest eigenvalues, $H_{\rm{eff}}$ is a $6\times 6$ matrix obtained from the $3,4,12,13,21,22$-th columns and rows of $\tilde H(k)$. Therefore $H_{\rm{eff}}$ captures the influence of CDW order on vHS as well as the hybridization between vH1 and vH2 due to the deviation from momentum $M$. We make a further approximation by keeping only the leading order Taylor expansion of $H_{\rm{eff}}$. Let $u_n(k)$ be the 9-component wave function of $H(k)$ at vH$n$ with $n=1,2$. By choosing the basis $\{u_1(M_A),u_1(M_B),u_1(M_C),u_2(M_A),u_2(M_B),u_2(M_C)\}$, we get the following form of $H_{\rm{eff}}$:
\bea
&&H_{\rm{eff}}(\mathbf k,\Delta)=\begin{pmatrix}
\epsilon_1 &  s_1 \Delta & s_1 \Delta^* &  \lambda^* k_1 &0 &0\\
s_1 \Delta^* & \epsilon_1  & s_1 \Delta &0 &  \lambda^* k_2 & 0\\
s_1 \Delta & s_1 \Delta^* & \epsilon_1 &0 &0 &  \lambda^* k_3 \\
 \lambda k_1 &0 & 0 & \epsilon_2 & s_2 \Delta^* & s_2 \Delta\\ 
0&  \lambda k_2  & 0  & s_2 \Delta & \epsilon_2 & s_2 \Delta^*\\
0  & 0 &  \lambda k_3 & s_2 \Delta^* & s_2 \Delta & \epsilon_2
\end{pmatrix}, \nonumber\\
&&\ \ \ \ \ \ \ \ \ \ \ \ \ \ \  \equiv\begin{pmatrix}
P_1 & Q^\dagger \\
Q & P_2  
\end{pmatrix}. 
\label{Heff}
\eea
This is the effective Hamiltonian introduced in the main text.

\section{Constraints on the effective Hamiltonian from the sublattice structure of wave functions}

The symmetry and the structure of wave functions at vH1 and vH2 in Fig.1(c) in the main text imposes strong constraints on the form of effective Hamiltonian in Eq.\eqref{Heff}. In particular, it enforces the off-diagonal block $Q$ to be a diagonal matrix. This can be seen by explicitly writing down the elements of $Q$: $Q(k)_{\alpha\beta}=\sum_{ij}u_{1i}(M_\alpha)^* D^{\alpha\beta}_{ij}(k) u_{2j}(M_\beta)$, where $u_{nj}(M_\alpha)$ is the $j$-th element of $u_n(M_\alpha)$ with $\alpha,\beta=A,B,C=1,2,3$. Then from Fig.1(c) the off-diagonal element of $Q$ vanishes identically because it meets a zero in the wave function at either vH1 or vH2. Explicitly, one can show that:
\bea
Q_{1,2}=u_1(M_A)^\dagger D^{AB}(0) u_2(M_B)=(b'^*\ 0\ 0)\begin{pmatrix}
0 & -2\Delta & 0  \\
-2\Delta^* & 0 & 0  \\
0 & 0 & 0  
\end{pmatrix} \begin{pmatrix}
-b   \\
0  \\
b 
\end{pmatrix}=b'^*\times\left(-2\Delta\right)\times 0=0
\label{Qzero}
\eea
Here we only wrote down the first three components corresponding to the $\tilde d$ orbitals because $D^{\alpha\beta}$ has nonzero elements only at $\tilde d$ orbitals. The other off-diagonal elements of $Q$ vanish in a similar way as Eq.\eqref{Qzero}. Furthermore, the diagonal elements of $Q$ should vanish when $k=0$ because $\epsilon_1$ and $\epsilon_2$ are exact eigenvalues at $k=0$ in the absence of CDW. Therefore, $Q$ is a diagonal matrix with the leading order diagonal elements linear in $k$. Another important consequence of the sublattice structure of wave function in Fig.1(c) is that the coefficient of the real part of $\Delta$ has opposite sign between blocks $P_1$ and $P_2$. This can be seen from the explicit expression $(P_n)(k)_{\alpha\beta}=\sum_{ij}u_{ni}(M_\alpha)^* D^{\alpha\beta}_{ij}(k) u_{nj}(M_\beta)$ for $n=1,2$. At momentum $M$ represented by $k=0$ it gives:
\bea
(P_1)_{1,2}&=&u_1(M_A)^\dagger D^{AB}(0) u_1(M_B)=(b'^*\ 0\ 0)\begin{pmatrix}
0 & -2\Delta & 0  \\
-2\Delta^* & 0 & 0  \\
0 & 0 & 0  
\end{pmatrix} \begin{pmatrix}
0   \\
b'  \\
0 
\end{pmatrix}=-2|b'|^2\Delta\equiv s_1\Delta \\
(P_2)_{1,2}&=&u_2(M_A)^\dagger D^{AB}(0) u_2(M_B)=(0\ b^*\ -b^*)\begin{pmatrix}
0 & -2\Delta & 0  \\
-2\Delta^* & 0 & 0  \\
0 & 0 & 0  
\end{pmatrix} \begin{pmatrix}
-b   \\
0  \\
b 
\end{pmatrix}=2|b|^2\Delta^* \equiv s_2\Delta^*
\eea
Therefore $s_1=-2|b'|^2<0$ and $s_2=2|b|^2>0$.


\section{Other types of loop current orders}

In the main text we focused on the loop current order with a pattern given in Fig.2(a) denoted as LCBO. There are other possible LCO patterns with the same periodicity, such as the (3,-1,-1,-1) phase~\cite{Lin2021c} in Fig.\ref{fig_LCOP}(a), denoted as LCO$'$. In LCO$'$ the direction of current is clockwise in one of the hexagons in the extended unit cell and counterclockwise in the other three hexagons, and the magnitude of current in the clockwise hexagon is three times as large as the current in the other hexagons. We demonstrate that LCO$'$ has higher free energy than the LCBO phase considered in the main text, hence LCO$'$ is energetically unfavored.

\begin{figure}[h]
\centering
\includegraphics[width=6.5 in]{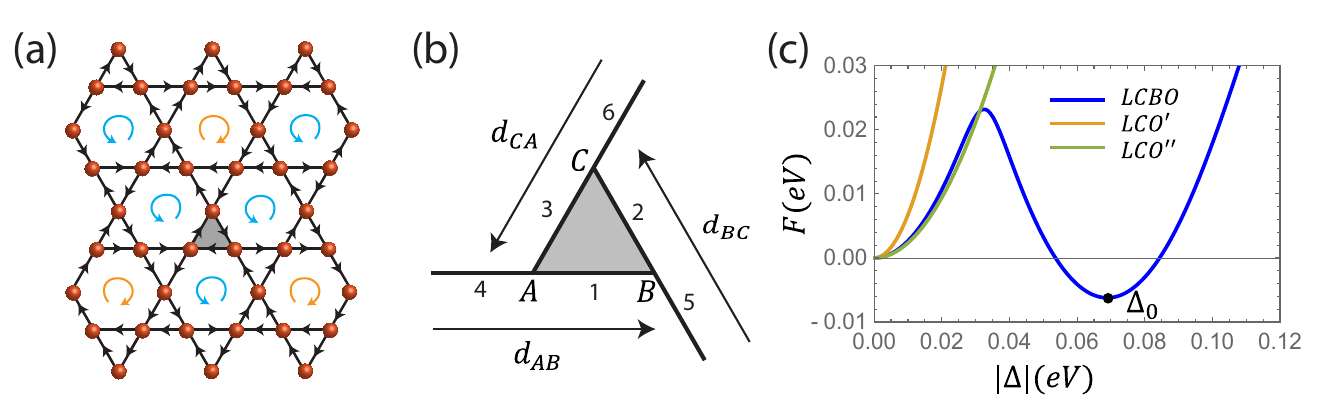}
\caption{ (a): Illustration of Loop current orders LCO$'$ and LCO$''$. Each hexagon has either clockwise or counterclockwise currents. For LCO$'$, the currents at the edges of the clockwise hexagons are three times as large as those in counterclockwise hexagons. For LCO$''$, the current on all bonds have the same magnitude. (b): Notations for distinct bonds in the shaded triangle in (a). (c): Free energy for LCBO, LCO$'$ and LCO$''$ under the same interaction strength. LCBO is energetically favored as the ground state.
}
\label{fig_LCOP}
\end{figure}

To characterize the order parameter in LCO$'$, we choose the origin inside the shaded upper triangle in Fig.\ref{fig_LCOP}(a) and denote the six distinct bonds in the original unit cell by 1$\sim$6 as in Fig.\ref{fig_LCOP}(b). Let $\Delta_l( R)$ denote the current on the $l$-th bond ($l=1\sim 6$) in unit cell $ R$. Due to the $2\times 2$ periodicity, $\Delta_l( R)$ can be decomposed as $\Delta_l( R)=\sum_Q\Delta_{l,Q}\cos(Q\cdot R)$ where $Q$ is summed over $0,Q_{AB},Q_{BC},Q_{CA}$. Different LCO patterns correspond to different choices of $\Delta_{l,Q}$. Denote the current at the edges of counterclockwise hexagons as $\Delta$. The values of $\Delta_{l,Q}$ for LCO$'$ is given by:
\begin{table}[h]
\centering
\begin{tabular}{ |c|c|c|c|c|c|c|  }
 \hline
  & 1 &  2 & 3 &4 &5 &6   \\
 \hline
 $\Gamma$& 0&0 &0 &0 &0 &0 \\
 \hline
 $Q_{AB}$& -$\Delta$&$\Delta$ &$\Delta$ &$\Delta$ &-$\Delta$ &$\Delta$ \\
 \hline
 $Q_{BC}$& $\Delta$&-$\Delta$ &$\Delta$ &$\Delta$ &$\Delta$ &-$\Delta$ \\
 \hline
 $Q_{CA}$& $\Delta$&$\Delta$ &-$\Delta$ &-$\Delta$ &$\Delta$ &$\Delta$ \\
 \hline
\end{tabular}
\caption{ Values of $\Delta_{l,Q}$ with $l=1\sim 6,Q\in\{0,Q_{AB},Q_{BC},Q_{CA}\}$ corresponding to LCO$'$. $\Gamma$ denotes $Q=0$.   }
\label{LCOptb}
\end{table}

The contribution from LCO to the Hamiltonian is given by:
\bea
\hat H_{\Delta}&=&-\sum_{l=1}^6\sum_{Q\in\{0,Q_{AB},Q_{BC},Q_{CA}\}}\Delta_{l,Q}\hat O_{l,Q}+\frac{N_c}{4}\sum_{j=1}^{24}\frac{|\Delta_j|^2}{V}  \label{Hdelt}\\
\hat O_{1,Q}&=&\sum_R \left(ic^\dagger_{R,A}c_{R,B} +h.c.\right)\cos(Q\cdot R) \nonumber\\
\hat O_{2,Q}&=&\sum_R \left(ic^\dagger_{R,B}c_{R,C} +h.c.\right)\cos(Q\cdot R)\nonumber\\
\hat O_{3,Q}&=&\sum_R \left(ic^\dagger_{R,C}c_{R,A} +h.c.\right)\cos(Q\cdot R)\nonumber\\
\hat O_{4,Q}&=&\sum_R \left(ic^\dagger_{R,A}c_{R-d_{AB},B} +h.c.\right)\cos(Q\cdot R)\nonumber\\
\hat O_{5,Q}&=&\sum_R \left(ic^\dagger_{R,B}c_{R-d_{BC},C} +h.c.\right)\cos(Q\cdot R)\nonumber\\
\hat O_{6,Q}&=&\sum_R \left(ic^\dagger_{R,C}c_{R-d_{CA},A} +h.c.\right)\cos(Q\cdot R)
\eea
Here $R$ is summed over original unit cells, $N_c$ is the number of original unit cells and $\Delta_j$ is the current on the 24 distinct bonds in the extended unit cell. For LCO$'$, there are six bonds with $\Delta_j=3\Delta$ and 18 bonds with $\Delta_j=\Delta$.

The free energy of LCO$'$ can be computed via Eq.\eqref{Hdelt} with the coefficients in Table.\ref{LCOptb}. Using the effective Hamiltonian of LCO$'$ with parameters $\epsilon_1=6.16eV,\epsilon_2=6.4eV,\mu=6.28eV,V=0.9eV,T=7.8meV,b=0.52,b'=0.96$, the free energy of LCO$'$ as a function of the magnitude of current $|\Delta|$ is plotted in Fig.\ref{fig_LCOP}(c). It shows that LCO$'$ has a much higher free energy compared to LCBO at the same interaction strength. LCBO has a nonzero global minimum $\Delta_0$ corresponding to Fig.2(a) discussed in the main text, whereas the LCO$'$ only has a trivial minimum at zero. Therefore LCO$'$ is energetically unfavored compared to LCBO.

The large free energy of LCO$'$ mainly comes from the non-uniform distribution of currents, i.e., a bond with current $3\Delta$ contribute a term $9|\Delta|^2/V$ to the free energy, which is nine times as large as a bond with current $\Delta$. We can also consider another pattern of loop current order in which the direction of currents in each bond is the same as LCO$'$ but all the bonds have the same magnitude of current. Denote this phase as LCO$''$ and let the current on each bond be $\pm\Delta$. The values of $\Delta_{l,Q}$ for LCO$''$ is given by:
\begin{table}[h]
\centering
\begin{tabular}{ |c|c|c|c|c|c|c|  }
 \hline
  & 1 &  2 & 3 &4 &5 &6   \\
 \hline
 $\Gamma$& $\frac{\Delta}{2}$&$\frac{\Delta}{2}$ &$\frac{\Delta}{2}$ &$\frac{\Delta}{2}$ &$\frac{\Delta}{2}$ &$\frac{\Delta}{2}$ \\
 \hline
 $Q_{AB}$& -$\frac{\Delta}{2}$&$\frac{\Delta}{2}$ &$\frac{\Delta}{2}$ &$\frac{\Delta}{2}$ &-$\frac{\Delta}{2}$ &$\frac{\Delta}{2}$ \\
 \hline
 $Q_{BC}$& $\frac{\Delta}{2}$&-$\frac{\Delta}{2}$ &$\frac{\Delta}{2}$ &$\frac{\Delta}{2}$ &$\frac{\Delta}{2}$ &-$\frac{\Delta}{2}$ \\
 \hline
 $Q_{CA}$& $\frac{\Delta}{2}$&$\frac{\Delta}{2}$ &-$\frac{\Delta}{2}$ &-$\frac{\Delta}{2}$ &$\frac{\Delta}{2}$ &$\frac{\Delta}{2}$ \\
 \hline
\end{tabular}
\caption{ Values of $\Delta_{l,Q}$ with $l=1\sim 6,Q\in\{0,Q_{AB},Q_{BC},Q_{CA}\}$ corresponding to LCO$''$. $\Gamma$ denotes $Q=0$.   }
\label{LCOpptb}
\end{table}

The free energy of LCO$''$ is also plotted in Fig.\ref{fig_LCOP}(c). It is higher than LCBO, and it only has a trivial minimum at zero. Therefore the energetically favored phase is LCBO.

\section{Ginzburg-Landau analysis and the effect of dispersion near vHS}

Many studies computed the CDW phase diagram via a Ginzburg-Landau (GL) approach by expanding the free energy as a power series of order parameter $\Delta$. Since the GL approach is a perturbation method for small $\Delta$ near the Fermi surface, the phase diagram obtained by the GL method is sensitive to the dispersion near the vHS. In the following, we will show that the CDW order we found is beyond the perturbation theory for small $\Delta$, hence it cannot be described by GL free energy and is not sensitive to the dispersion near the vHS.

The free energy of the system described by $H_{\rm{eff}}$ is given by:
\be
F(\Delta)=-\frac{1}{N}T\sum_{n,\mathbf{k}}\rm Tr\log[-G_{n,\mathbf k}^{-1}(\Delta)]+\frac{6|\Delta|^2}{V},
\ee
where $G_{n,\mathbf k}(\Delta)=\frac{1}{i\omega_n-H_{\rm{eff}}(\mathbf k,\Delta)+\mu}$ and the factor of 6 comes from the six bonds in each original unit cell. The GL free energy $F_{GL}(\Delta)$ is obtained by expanding the free energy to powers of $\Delta$. The most straightforward way to demonstrate the inability of $F_{GL}$ to capture our CDW solution is by considering the $\lambda=0$ limit. In this case, the free energy in the complex $\Delta$ plane has degenerate minima at $\Delta=-|\Delta|$ and $\Delta=|\Delta|e^{\pm \frac{\pi}{3}i}$ corresponding to CBO and LCBO respectively, where $|\Delta|$ is determined by interaction strength. At these free energy minima, the eigenvalues of $H_{\rm{eff}}-\mu$ can be written down analytically by 
\bea
&&E_1=\epsilon_2-\mu-4|b|^2 |\Delta|,\ E_2=E_3=\epsilon_1-\mu-2|b'|^2 |\Delta|,\nonumber\\
&&E_4=E_5=\epsilon_2-\mu+2|b|^2 |\Delta|,\ E_6=\epsilon_1-\mu+4|b'|^2 |\Delta|.
\eea
The free energy is given by
\be
F(\Delta)=-\frac{1}{4}T\sum_{n=1}^6\log\left(1+e^{-\frac{E_n}{T}}\right)+\frac{6|\Delta|^2}{V},
\label{Fex}
\ee
and the GL free energy $F_{GL}$ is the Taylor expansion of Eq.\eqref{Fex}. We fix the complex phase of $\Delta$ to be $\frac{\pi}{3}$ and plot the exact and GL free energy as a function of $\Delta$ in Fig.\ref{fig_FGL}(a) with parameters $\epsilon_1=6.16eV,\epsilon_2=6.4eV,\mu=6.28eV,V=0.9eV,T=7.8meV,b=0.52,b'=0.96$, where we have kept $O(|\Delta|^4)$ terms in $F_{GL}$. The global minimum $\Delta_0$ corresponds to our CDW solution. Fig.\ref{fig_FGL}(a) shows that $F$ and $F_{GL}$ agree with each other when $|\Delta|$ is small, but the global minimum $\Delta_0$ does not appear in $F_{GL}$. We also find that the coefficients of $O(|\Delta|^6),O(|\Delta|^8),...,O(|\Delta|^{20})$ terms in $F_{GL}$ are all negative, hence even if we add higher power terms of $|\Delta|$ into $F_{GL}$, the global minimum at $\Delta_0$ will not show up either. Therefore, the LCBO phase we found cannot be described by GL free energy which is valid only at small $|\Delta|$.

We can estimate the magnitude of $\Delta$ required to realize LCBO. From Eq.(4) in the main text, the ground state is LCBO when $4(|b|^2+|b'|^2)|\Delta|>\epsilon_2-\epsilon_1$, which implies $|\Delta|>\frac{\epsilon_2-\epsilon_1}{4(|b|^2+|b'|^2)}\sim (\epsilon_2-\epsilon_1)/4\sim 60meV$. This value is comparable with the magnitude of CDW order parameter $|\Delta|\sim 60meV$ observed in optical conductivity experiments~\cite{Uykur2022,wilson2023k}.

The finite magnitude of $\Delta$ makes the CDW phases less sensitive to the detailed dispersion near the vHS. To demonstrate this, we add to the effective Hamiltonian the following saddle-type term obtained from the dispersion of the tight-binding model:
\bea
&&\delta H(\mathbf k)= diag\{\ g_1(C_3\mathbf k),g_1(C_3^{-1}\mathbf k),g_1(\mathbf k),g_2(C_3\mathbf k),g_2(C_3^{-1}\mathbf k),g_2(\mathbf k) \ \},\\ 
&&g_1(\mathbf k)=-c_1 k_x^2+c_2 k_y^2,\ \ g_2(\mathbf k)=c_3 k_x^2-c_4 k_y^2 \\
&&(c_1,c_2,c_3,c_4)=(0.46,1.19,0.5,0.39)\ eV\cdot a^2
\eea
Here $a$ is the lattice constant, $diag$ denotes diagonal matrix and $C_3$ is threefold rotation. The new phase diagram obtained by $H_{\rm{eff}}+\delta H$ is shown in Fig.\ref{fig_FGL}(b). It demonstrates that the addition of dispersion near the vHS only results in a slight shift of phase boundaries, but no significant change occurs.

\begin{figure}
\centering
\includegraphics[width=5 in]{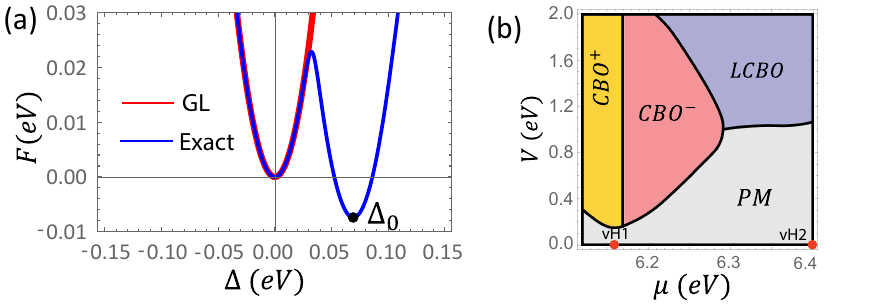}
\caption{ (a): Comparison between the exact free energy and GL free energy at $\lambda=0$ with the complex phase of $\Delta$ fixed at $\frac{\pi}{3}$. (b): Phase diagram of $H_{\rm{eff}}+\delta H$. The other parameters are the same as Fig.4(a) in the main text.
}
\label{fig_FGL}
\end{figure}

\end{widetext}

\end{document}